\documentclass[preprint]{elsarticle}
\usepackage{latexsym}
\usepackage{graphicx}

\usepackage{caption}
\usepackage{subcaption}

\usepackage{multirow}
\usepackage{hyperref}
\hypersetup{
	colorlinks=true,
	linkcolor=blue,
	filecolor=magenta,
	urlcolor=cyan,
}

\input epsf
\date{}

\newtheorem{exmpl}{Example}
\newdefinition{dfn}{Definition}

\newtheorem{thm}{Theorem}
\newtheorem{lem}[thm]{Lemma}
\newtheorem{prop}[thm]{Proposition}
\newdefinition{rmk}{Remark}
\newproof{pf}{Proof}
\newproof{pot}{Proof of Theorem \ref{thm2}}

\newcommand{\CN}{{\cal N}}
\newcommand{\CD}{{\cal D}}
\newcommand{\CV}{{\cal V}}
\newcommand{\CS}{{\cal S}}

\begin{document}

\title{Efficient query evaluation techniques over large amount of distributed linked data
}

\author[addr1]{Eleftherios Kalogeros}

\author[addr1]{Manolis Gergatsoulis}

\author[addr1]{Matthew Damigos}

\author[addr2]{\\ Christos Nomikos}


\address[addr1]{
Database \& Information Systems Group (DBIS),\\
Laboratory on Digital Libraries and Electronic Publishing,\\
Department of Archives, Library Science and Museology, Ionian University\\
Ioannou Theotoki 72, 49100 Corfu, Greece\\
{\normalsize{\tt \{kalogero, manolis\}@ionio.gr}} \ \ \  {\normalsize{\tt mgdamig@gmail.com}}}
\address[addr2]{
Department of Computer Science and Engineering, University of Ioannina, Greece\\
P.O Box 1186, 45110 Ioannina, Greece\\
{\normalsize{\tt cnomikos@cs.uoi.gr}}}
%

\begin{abstract}
As RDF becomes more widely established and the amount of linked data is rapidly increasing, the efficient querying of large amount of data becomes a significant challenge. In this paper, we propose a family of algorithms for querying large amount of linked data in a distributed manner. These query evaluation algorithms are independent of the way the data is stored, as well as of the particular implementation of the query evaluation. We then use the MapReduce paradigm to present a distributed implementation of these algorithms and experimentally evaluate them, although the algorithms could be straightforwardly translated into other distributed processing frameworks.
	We also investigate and propose multiple query decomposition approaches of  Basic Graph Patterns (subclass of SPARQL queries) that are used to improve the overall performance of the distributed query answering. A deep analysis of the effectiveness of these decomposition algorithms is also provided.
\end{abstract}

\begin{keyword}Linked Data \sep Graph Querying \sep Big Data \sep Map-Reduce \sep Distributed Processing \sep Cloud Computing \sep Semantic Web\end{keyword}

\maketitle

\section{Introduction}
\label{sec:Intro}

Linked data has become a widely-established approach for publishing and sharing semantically-meaningful information through distributed and interrelated data. RDF is the standard model Linked data is built upon. As RDF data is rapidly increasing, the efficient querying of large amount of Linked data becomes a significant challenge in many business and research areas, such as bioinformatics and cheminformatics, and digital libraries \cite{ExperimentalComparisonSurvey, ozsu2016survey}.

Both centralized (e.g., \cite{sequeda2013ultrawrap,spanos2012bringing}) and distributed (e.g., \cite{HadoopRDF2,SHARD}) processing of RDF data has been extensively investigated in the past, where SPARQL \cite{sparql, semanticsSparql} is mainly used as query language.
To process large amount of RDF data in a distributed manner, parallel processing frameworks are considered \cite{ExperimentalComparisonSurvey,review,SPARQL2Flink}. Apache Hadoop \cite{hadoop} (the open source alternative of Google's MapReduce \cite{dean2008mapreduce}), Spark \cite{zaharia2010spark, spark} and Flink \cite{Flink} are three widely-used programming frameworks for distributed processing. Although Apache Spark and Flink typically improve and outperform Hadoop/MapReduce, mainly due to in-memory processing, from algorithmic perspective, they all handle distributed processing in a similar manner; i.e., define workflows of tasks running in parallel and determine the way the data is reshuffled in order to be properly and efficiently joined.

In addition to the query evaluation approaches, a variety of effective storage schemes has been used to improve query answering over RDF data, such as the use of relational databases (e.g., \cite{SPARQLtoSQL, spanos2012bringing}) and NoSQL databases (e.g., \cite{Bigdatasurvey}). In distributed environ\-ments, the proper partitioning of the RDF data into a distributed repository (either file system or distributed NoSQL database) can significantly improve the perfor\-mance of query answering \cite{RDFpartitioningSurvey,RdfStorage2,RDFintheClouds}. Following this approach, most of the distributed based methods and systems utilize efficient partitioning of the data across a cluster of machines in order to ensure efficient query processing through minimizing the communication cost and improving parallel execution \cite{RDFpartitioningSurvey}. To take advantage of the selected partitioning during query answering, certain approaches for decomposing the given query, and creating a proper query plan (consisting of multiple steps of distributed processing) are proposed \cite{HadoopRDF,CliqueSquare}.

In this work, we present three distributed evaluation algorithms for querying large amount of RDF data. The main idea behind these algorithms is described as follows: a) The data graph is decomposed into a set of (possibly overlapping) data graph segments stored in different nodes of a cluster of commodity machines. b) The query graph $Q$ is also decomposed into a set of (possibly overlapping) subqueries. c) Subqueries are applied to each data graph segment, in isolation, and intermediate results are computed. d) The intermediate results are appropria\-tely combined to obtain the answers of the query $Q$. Note that the algorithms are independent of the way the data is stored as well as of the particular implementation of the query evaluation. We then use the MapReduce paradigm to present a distributed implementation of these algorithms and experimentally evaluate them, although the algorithms could be straightforwardly translated into Spark jobs and/or Flink dataflows.

This paper consolidates our previous  work presented in \cite{obd, globe, graphq} into a single unified framework for distributed evaluation of Basic Graph Pattern queries (subclass of SPARQL queries), and extend this framework by proposing multiple query decomposition algori\-thms that could be used by a wide-variety of query evaluation approaches. In particular, we investigate decomposition approaches a) which are based on producing subqueries
of special forms (with or without replication of query triples), and b) that take into account certain replication of the distributed data.

The paper is organized as follows.
In Section~\ref{sec:related}, related work is presented and discussed.
In Section~\ref{sec:Framework}, the framework of our work is defined.
More specifically, after presenting some preliminary definitions in Subsection~\ref{subsec:DataAndQueries}, we introduce the concept of (data and query) \emph{graph decomposition} in Subsection~\ref{subsec:dataQueryDecomposition}. Then, we define the concept of \emph{partial embeddings} in Subsection~\ref{subsec:partial} and distinguish special forms of queries in Subsection~\ref{subsec:special}.

In Section~\ref{sec:query-eval-approaches}, we presented three query evaluation approaches. More specifically, in Subsection~\ref{subsec:QEJPE-algorithm}, we present a query evaluation approach which is based on the concept of partial embeddings. In Subsection~\ref{subsec:QE-using-STARS}, we present an approach  which is based on the decomposition of queries into subqueries of a specific form called \emph{generalized star queries}. Finally, in Subsection~\ref{subsec:QE-redundancy}, we present an approach which is based on the idea that replication in data decomposition can be taken into account to efficiently answer queries.
In Subsection~\ref{subsec:query-decomp-algorithms}, we present a set of query decomposition algorithms.

In Section~\ref{sec:implementations}, we presenta set of query evaluation algorithms, which implement the approaches presented in Section~\ref{sec:query-eval-approaches}.
Experimental evaluation results of the algorithms are presented in Section~\ref{sec:experiments}.
Finally, Section~\ref{sec:conclusion} concludes the paper.

\section{Related work}
\label{sec:related}

The problem of efficiently querying linked data has been widely investigated, for both centralized (single-machine) \cite{sequeda2013ultrawrap,spanos2012bringing} (e.g., systems such as RDF-3X \cite{RDF3X1,RDF3X2} and Hexastore \cite{Hexastore}) and distributed environments (e.g., \cite{HadoopRDF2,SHARD}).
	Processing large amount of linked data into a single machine has significant limitations, since it lacks scalability \cite{RDFpartitioningSurvey}. To handle this problem, a variety of distributed methods for storing and processing linked data has been proposed \cite{Bigdatasurvey}. Most of the approaches proposed in the literature to handle scalability of answering SPARQL queries over big linked datasets \cite{ExperimentalComparisonSurvey} focus on two aspects, distributed storage of linked data and distributed processing of SPARQL queries.  Typically, the proper partitioning of the data into a distributed repository (either file system or distributed NoSQL database) can significantly improve the performance of query answering \cite{RDFpartitioningSurvey,RdfStorage2,RDFintheClouds} (e.g., Random Partitioning \cite{globe,graphq}, Hash Partitioning \cite{SHARD,HadoopRDF, HAQWA}, Graph Partitioning \cite{nhop, METIS}, and Semantic Partitioning \cite{SHAPE}).

	To process and query large amount of linked data, distributed processing frameworks, such as MapReduce \cite{hadoop} or Apache Spark \cite{spark}, are used. Apart from these approaches, there is a noteworthy amount of related work focusing on utilizing distributed NoSQL databases \cite{rya,AMADA,MAPSIN,CumulusRDF} to store the linked data graph and answer the given queries. In these cases, the query evaluation is achieved either through translating the given SPARQL query into the query language supported by the NoSQL database \cite{dsparq}, or by using a distributed processing framework to implement the overall query execution plan \cite{CliqueSquare} (in such a case, the NoSQL database mainly used as a storage layer ensuring proper data partitioning).

 In this context, Afrati et al. \cite{Afrati1} proposed an approach for optimizing joins in MapReduce by choosing the appropriate map-key and shares. This approach is extended in \cite{Afrati2} to data graphs, where the cost of evaluating queries on data graphs using one round of Map-Reduce is investigated, and an approach of translating the query patterns into conjunctive queries is proposed. The communication cost is minimized using the techniques of the approach proposed in \cite{Afrati3}. Such an approach could be used for answering conjunctive SPARQL queries \cite{neumann2009scalable, picalausa2011real, vidal2010efficiently}.
	
	A method of answering SPARQL Basic Graph Pattern using traditional multi-way join into MapReduce, instead of multiple individual joins, is also presented in \cite{iterativeMapReduce}, where certain joining keys are selected to avoid unnecessary iterations. This approach can be used for every type of partitioning of the RDF data.
	
	SHARD \cite{SHARD} is built on top of Hadoop, and uses the Hadoop distributed files system (HDFS) to store data in native text files.
It uses subject hash partitioning to decompose the RDF data graph; all the triples with the same subject are stored in the same line of the text file.
	For the execution, one MapReduce job is created for every query triple, while an additional job is used, at the end, to remove duplicated results and apply the required projection. Hence, assuming an $n$-triple query pattern, $n+1$ jobs are required, and all the data graph is scanned $n$ times.
	
	HadoopRDF \cite{HadoopRDF2} uses predicate hash partition method to distribute the data graph; similar to the vertical partitioning approach applied by SW-Store \cite{SW-Store2}.
	In general, the number of the data fragments is equal to the number of the distinct predicates.
The query evaluation is performed through a sequence of MapReduce jobs and is optimized using a heuristic and a greedy approach.
	
	CliqueSquare \cite{CliqueSquare} presents a method that generates highly parallelizable query plans for BGP queries, which rely on n-ary equality joins with minimum amount of MapReduce stages. CliqueSquare uses a data partitioning scheme that permits first-level joins can be evaluated locally at each node.
	The triples that share the same value in subject, predicate or object are placed on the same node. This partition ensures that queries sharing the same variable (like star queries) can be evaluated locally.
	
	H2RDF \cite{H2RDF} uses the Apache HBase \cite{hbase} to store data triples. Three RDF indices on subject, predicate and object (spo, pos and osp combinations) are materialized and stored to HBase in the form of key-value pairs.
	Different strategies are used to execute joins and answer the given query. H2RDF+ \cite{H2RDF2} extends H2RDF by considering three more indices (ops, osp and sop).
	Furthermore, MapReduce Merge Join algorithm is used to join query triples  that share the same variable and the MapReduce Sort-Merge Join algorithm is used for joining the intermediate results.
	
	PigSPARQL \cite{PigSPARQL} is yet another approach which uses Hadoop-based implementation of vertical partitioning of the data stored into HDFS. It implements a translation from SPARQL to Pig Latin \cite{PigLatin}. In the system RAPID+ \cite{Rapid}, an alternative query algebra, called the Nested Triple Group Algebra, is used as an extension of Apache Pig, to improve  the performance of SPARQL query processing over MapReduce.
	
	The authors in \cite{nhop} proposed a graph partitioning schema, which resembles the s-decomposition partitioning defined in this work. In particular, the data is partitioned in such a way that the vertices that are relatively close to each other are included in the same segment. In this context, the following main methodologies are investigated and proposed: the undirected and the directed n-hop guarantee. The former focuses on initially partitioning the vertices and then assigning the triple-paths of length $n$ that start from a vertex that is already included in the segment.
	The latter is similar to the directed one but considers any undirected path of length $n$. In both cases, a graph practitioner tool which is based on METIS is used for partitioning the vertices of the RDF graph into disjoint partitions so that the minimum number of edges is cut.
	The queries are also decomposed in such a way that the subqueries generated can be computed locally, in each cluster node. MapReduce is used for the joins of the intermediate results of subqueries.
	Although s-decomposition partitioning approach is similar to 1-hop undirected guarantee (or hash partitioning), the n-hop guarantee of the data graph may cause data explosion especially in coherent data graphs if $n> 2$.

	SHAPE \cite{SHAPE} proposes a semantic hash partitioning which is based on the similarity of the URI hierarchy of the vertices. The vertices with same URI prefixes are placed in the same partition. After a simple hash partitioning is used, a replication of only a set of necessary triples is performed, using a k-hop semantic hash partitioning and context-aware filters. The system also uses a RDF-3X triple store in each data node. Query processing and the joins of the intermediate results is based in MapReduce.
	
	The papers \cite{obd,graphq,globe} focus on both decomposing queries and partition the RDF data, where the data is stored into MySQL and the framework used to evaluate the queries is MapReduce. SPARQL to SQL translations is used for query processing, and MapReduce is used to apply  the joins.
	
	D-SPARQ \cite{dsparq} uses the document database MongoDB \cite{mongodb} to store and index data using subject hash partition. A single MapReduce job is then used to import data into the document database and to collect statistical information for query optimization process based on join reordering. All triples sharing the same subject value are stored in the same document (JSON) file.
	
	Another approach which is based on the MapReduce is Sempala \cite{Sempala}, which applies SPARQL-to-SQL translation on top of Hadoop. It uses Impala \cite{Impala} as a distributed SQL processing engine. Sempala uses a unified vertical partitioning (single property table) in order to boost the star-shaped queries.
	
	In \cite{EfficientSubgraphMatching}, the authors proposed a MapReduce algorithm, called StarMR, which is based on star
	decomposition for answering subgraph matching queries. The StarMR algorithm is improved with two optimization strategies. The first applies an RDF property filtering approach and the second one postpones any Cartesian product operation. RDF graph is stored in a distributed adjacency list.
	
	\cite{DataIntensiveQuery} uses a partitioning method over the predicate value and the type of objects to store the RDF data. Query processing
	is performed using MapReduce and the algorithm proposed applies a number of the jobs that depends on the form of the given query.
	
	Apache Spark anf Flink have been used to improve the performance of SPARQL query evaluation over big RDF data \cite{review}. SPARQLGX \cite{SPARQLGX} uses a vertical partition approach, where the triples are partitioned according to their predicate values.  The query evaluation is performed by initially filtering the triples matching a query triple,  in each segment, and then, by applying a sequence of join operations through a query plan which is generated according to predefined statistics. The authors in \cite{SPARQL2Flink} propose an approach for translating SPARQL queries into Apache Flink \cite{Flink} programs for querying RDF data, as well as investigate the semantic correspondence between Apache Flink's subset transformations and the SPARQL Algebra operators.
	
	S2RDF \cite{S2RDF} also proposes a vertical-like partitioning, called Extended Vertical Partitioning (ExtVP), which is based on semi-joins reductions (i.e., a certain number of semi-joins are applied between the vertical partitioning tables and their results are materialized for improving the overall performance). To evaluate queries over ExtVP, an approach of applying a certain partitioning of the query triples (in order to achieve parallel/local computation) and utilizing Spark SQL is followed.

	HAQWA \cite{HAQWA} proposes a hash-based partitioning over the subject values of the RDF triples. This ensures local computation of subject-centric star queries (a subclass of generalized star queries). To extend the supported queries, the query is decomposed into subqueries and missing triples of each subquery are replicated. The overall computation process is managed through a Spark application.
	
	The authors in~\cite{NAC17} analyze the query evaluation plan of a BGP expression on Spark and proposes a joins plan for efficiently evaluating BGPs over a large RDF graph. Considering an initially hash-based partitioning of the data (e.g., the triples are partitioned by their subject), the authors propose a hybrid method to find a query plan. The approach uses a cost-driven combination of partitioned/cascade and broadcast joins over Spark.
	
	Apart from the previous approaches, it's worth mentioning the approaches S2X \cite{S2X}, Spar(k)ql  \cite{Sparkql} and \cite{kassaie2017sparql}, which focus on evaluating SPARQL queries using the Spark GraphX library.
	Sparklify \cite{Sparklify} applies a SPARQL-to-SQL rewriter for translating SPARQL queries into Spark executable code.
	
	In \cite{SemanticDataQueryingSPARK}, a property table scheme is built on top of HBase storage system and
	a vertical partitioning scheme on top of Cassandra storage system. Query processing is based on SPARQL query translation to SparkSQL for both HBase and Cassandra storage schemas.
	
	As mentioned previously, multiple approaches that use NoSQL platforms to store RDF data and answer SPARQL queries have been proposed in the literature \cite{RDFintheClouds}. Representative examples include the distributed systems Rya \cite{rya}, AMADA \cite{AMADA}, MAPSIN \cite{MAPSIN} and CumulusRDF \cite{CumulusRDF} which use NoSQL \cite{NoSQL} databases to store RDF data and provide efficient query processing using three different indices SPO, POS and OSP (S for subject, P for predicate and O for object values). More specifically, Rya uses Apache Accumulo \cite{accumulo}, AMADA use Amazon DynamoDB \cite{dynamodb}, MAPSIN Apache HBase and CumulusRDF Apache Cassandra \cite{cassandra}.  MAPSIN (Map-Side Index Nested Loop Join) joins are performed in the map phase, so shuffle and reduce phase are not required. The proposed algorithm is optimized for the efficient processing of multiway joins.

\section{Definition of the framework}
\label{sec:Framework}

\subsection{Preliminaries}
\label{subsec:DataAndQueries}

Let $U_{so}$ and $U_p$ be two countably infinite disjoint sets of URI references, $L$ be a countably infinite set of (plain) literals\footnote{In this paper we do not consider typed literals} and $V$ be a countably infinite set of variables.
In the following, we define two types of graphs, \emph{data graphs} and \emph{query graphs}. The former describes the data model
and the latter determines the form of the query expressions over the stored data.

\begin{dfn}
\label{def:datagraph}
A triple $(s, p, o) \in  U_{so} \times U_p \times (U_{so} \cup L)$ is called a \emph{data triple}.
In a data triple $t = (s,p,o)$, $s$ is called the \emph{subject}, $p$ the \emph{predicate} and $o$ the \emph{object} of $t$.
A \emph{data graph} $G$ is a non-empty set of data triples. A data graph $G'$ is a \emph{subgraph} of a data graph $G$ if $G' \subseteq G$.
\end{dfn}

\begin{dfn}
\label{def:querygraph}
A triple $(s, p, o) \in  (U_{so} \cup V) \times U_p \times (U_{so} \cup L \cup V)$ is called a \emph{query triple}.
In a query triple $q = (s,p,o)$, $s$ is called the \emph{subject}, $p$ the \emph{predicate} and $o$ the \emph{object} of $q$.
A \emph{query graph} (or simply a \emph{query}) $Q$ is a nonempty set of query triples.
The \emph{output pattern} $O(Q)$ of a query graph $Q$ is the tuple $(X_1, \dots, X_n)$, with $n \geq 0$, of all the variables
appearing in $Q$.
A query $Q$ is said to be a \emph{Boolean query} if $n=0$.
A query graph $Q'$ is a \emph{subquery} of a query graph $Q$ if $Q' \subseteq Q$.
\end{dfn}

\begin{dfn}
Let $G$ be a data or query graph. A \emph{directed path} (or simply \emph{path}) in  $G$ is a sequence of triples
$(v_0, p_1, v_1), (v_1, p_2, v_2), \dots, (v_{n-1}, p_n, v_n)$ in $G$, where $n \geq 1$.
The \emph{length} of the path is the number $n$ of triples in the path. A finite directed path always has
a \emph{start node} which corresponds to the subject of its first triple, and an \emph{end node} which is the object of its last triple.
A \emph{cycle} is a path in which the start node and the end node are the same.
A path with no repeated nodes (i.e. without cycles) is called a \emph{simple path}.
\end{dfn}

The set of \emph{nodes} of a data graph $G$ (resp. a query graph $Q$), denoted by ${\cal N}(G)$ (resp. ${\cal N}(Q)$), is the set of elements of $U_{so} \cup L$ (resp. $U_{so} \cup L \cup V$) that occur in the triples of $G$ (resp. $Q$).
The set of \emph{edge labels} of a data graph $G$ (resp. a query graph $Q$), denoted by ${\cal E}(G)$ (resp. ${\cal E}(Q)$), is the set of elements of $U_{p}$ that occur in the triples of $G$ (resp. $Q$).
Finally, the set of variables in a query $Q$ is denoted by ${\cal V}(Q)$.

Notice that the data graphs defined above correspond to ground RDF graphs defined in~\cite{GHM11-520}.
Notice also that query graphs correspond to Basic Graph Patterns (BGP) SPARQL queries.
In this paper,
we do not allow queries with variables in the place of predicates.
However, the query evaluation algorithms proposed in this paper can be easily extended
to allow such variables.

Data and query graphs are graphically represented as follows: A node (subject or object), which is a URI or a variable, is represented as a rounded rectangle, while an object which is a literal is represented by a rectangle. Each triple $(s, p, o)$ is represented
by a labeled edge $s$ $\stackrel{p}{\longrightarrow}$ $o$ connecting the nodes $s$ and $o$.
%

In this paper, we use strings with initial lowercase letters
to represent elements in $U_{p}$ (i.e., URIs corresponding to predicates), while strings with
initial uppercase letters denote elements in $U_{so}$ (i.e., URIs
corresponding to objects and subjects).
Literals are represented as strings enclosed in double quotes.
Finally, we assume that  variables are represented by strings whose first symbol is the question mark symbol (?).

\begin{exmpl}
\label{ex:datagraph}
Fig.~\ref{fig:datagraph}(a) depicts a data graph showing information about three journal papers, their authors and the relationships between the authors. Fig.~\ref{fig:datagraph}(b)
shows a query graph.

\begin{figure*}[htb]
\centering
\includegraphics[width=0.8\textwidth]{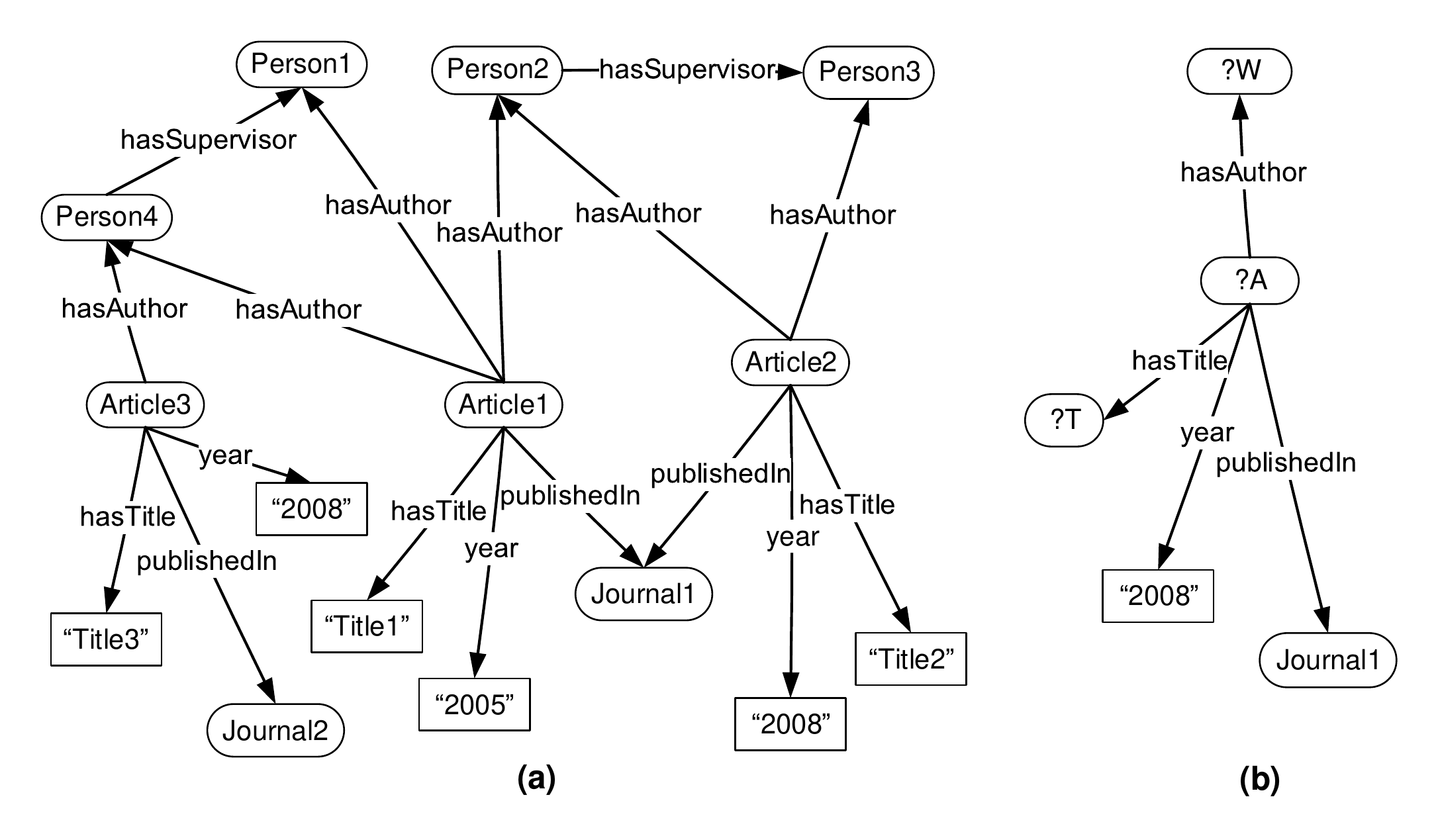}
\caption{(a) A data graph, and (b) a query graph
\label{fig:datagraph}}
\end{figure*}
\hfill$\Box$
\end{exmpl}

\begin{dfn}
\label{def:embedding}
A \emph{(total) embedding} of a query graph $Q$ in a data graph $G$ is a total mapping $e: {\cal{N}}(Q) \rightarrow {\cal{N}}(G)$ with the following properties:
\begin{enumerate}
\item
For each node $v \in  {\cal{N}}(Q)$, if $v$ is not a variable then $e(v) = v$.
\item
For each triple $(v_1, p, v_2) \in Q$, the triple  $(e(v_1), p, e(v_2))$ is in $G$.
\end{enumerate}

The tuple $(e(X_1), \dots, e(X_n))$, where $(X_1,\dots,X_n)$ is the output pattern
of $Q$, is said to be an \emph{answer} to the query $Q$.
\end{dfn}

\begin{exmpl}
Fig.~\ref{fig:rdf-embedding}
depicts an embedding of the query
graph $Q$ in data graph $G$, where $Q$ and $G$
are the graphs appearing in Fig.~\ref{fig:datagraph}.
\begin{figure*}[htb]
\centering
\includegraphics[width=0.85\textwidth]{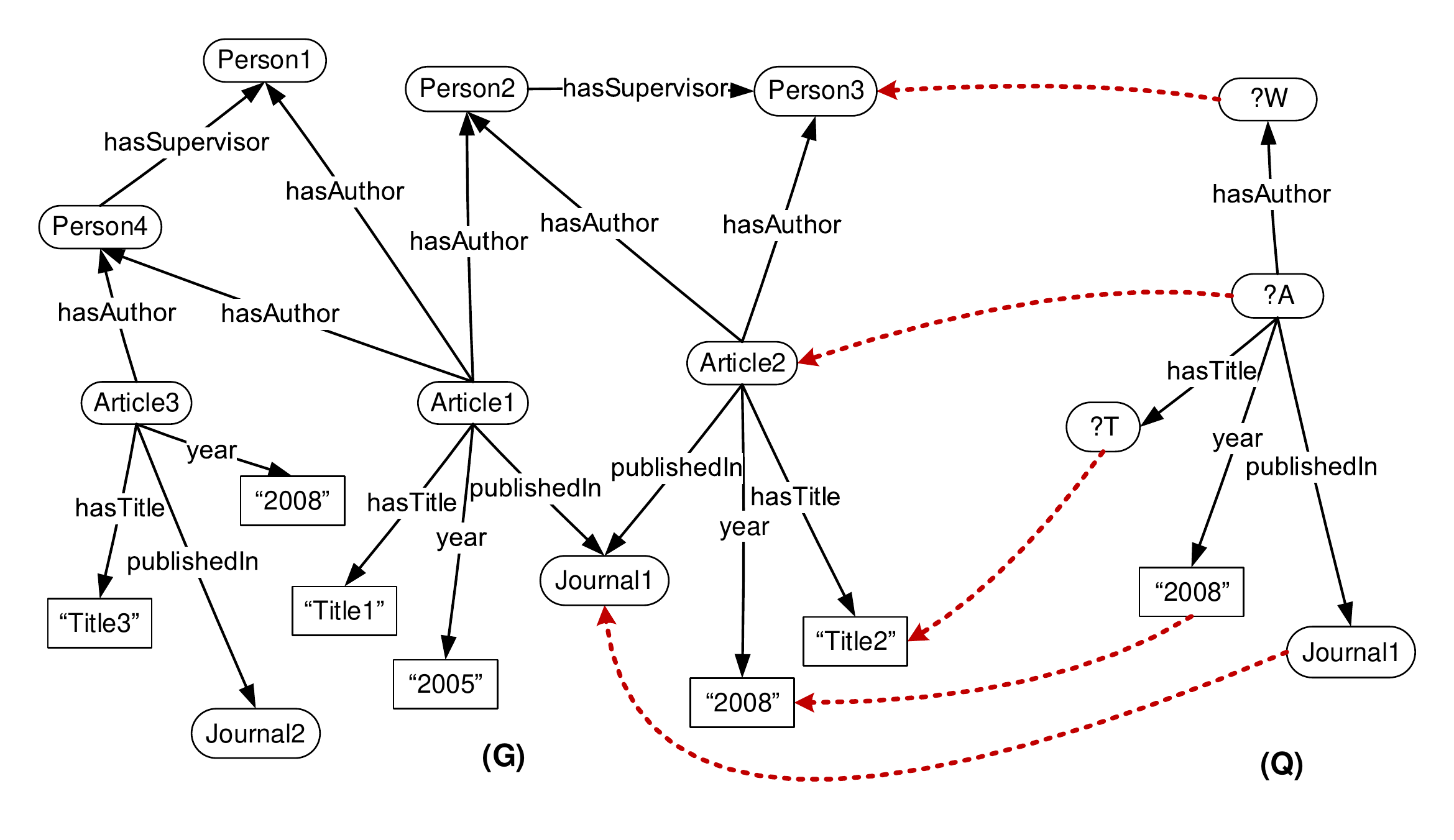}
\caption{An embedding of the query graph $Q$ in the data graph $G$.\label{fig:rdf-embedding}}
\end{figure*}
The answer obtained
by this embedding
is $(?A, ?W, ?T) = (Article2, Person3, ``Title2")$. Notice that
a second embedding exists giving  the answer $(?A, ?W, ?T)$ = $(Article2$, $Person2$, $``Title2")$.
\hfill$\Box$
\end{exmpl}

\subsection{Data and query graph decomposition}
\label{subsec:dataQueryDecomposition}

A crucial problem, when we use a cluster of computer nodes to evaluate queries, is how to distribute the data in the computers of the cluster as well as how to compute the queries on the distributed data.
In this section we define the concept of \emph{data} and  \emph{query graph decomposition}.

\begin{dfn}
\label{def:graphpartition}\label{def:graphdecomposition}
A \emph{data (resp. query) graph decomposition} of a data (resp. query) graph $F$
is an $m$-tuple of data graphs ${\cal D}_F=(F_1, \dots, F_m)$, where $m \geq 1$,  such that:

\begin{enumerate}
\item $F_i \subseteq F$, for $i = 1, \dots, m$, and
\item $\bigcup_i F_i = F$.
\end{enumerate}

Each data (resp. query) graph $F_i$ in a data (resp. query) graph decomposition is called a \emph{data (resp. query) graph segment}.
When, in a data/query graph decomposition, for all the pairs $i$, $j$, with $1 \leq i < j \leq m$, it also holds $F_i \cap F_j = \emptyset$,
i.e. data (resp. query) graph segments are disjoint of each other,  then the data (resp. query) graph decomposition is said to be \emph{non-redundant} and the
graph (resp. query) segments obtained form a partition of the triples of data (resp. query) graph $F$,
called \emph{$m$-triple partition} of  $F$.
\end{dfn}

Notice that, $F_i \neq \emptyset$ for $i = 1, \dots, m$, since, because of Definitions~\ref{def:datagraph} and ~\ref{def:querygraph}, a data/query graph is nonempty.

It should be noted that, in a data or a query graph decomposition, a triple is (in general) allowed  to participate in multiple data or query graph segments.
At first sight, this redundancy seems to burden the system with the extra cost of storing and managing or evaluating more data.
However, as it is shown in subsequent sections (see for example Section~\ref{subsec:QE-redundancy}), if it is used  appropriately it may lead to more efficient computation of the query answers, due to proper parallelization of the query execution.

\begin{dfn}
\label{def:border-node}
Let ${\cal D}_F =(F_1, \dots, F_m)$, with $m \geq 1$,  be a data (resp. query) graph decomposition  of a data graph $F$, and $F_i$, $F_j$, with $i \neq j$, be two data (resp. query) graph segments in ${\cal D}_F$.
A \emph{border node} $v$ of $F_i$ and $F_j$, is a node that belongs to ${\cal N}(F_i) \cap {\cal N}(F_j) - L$.
By ${\cal B}(F_i, F_j)$ we denote the set of border nodes of $F_i$ and $F_j$, while,
by ${\cal B}(F_i)$, we denote the set $\bigcup_{(1 \leq j \leq m) \wedge (j \neq i)} {\cal B}(F_i, F_j)$. Finally, by ${\cal B}(F)$ we denote the set of all border nodes of $F$ i.e. ${\cal B}(F) = \bigcup_{1 \leq i \leq m}  {\cal B}(F_i)$.
\end{dfn}

Notice that, according to the above definition, literals that occur in more that one segments
in ${\cal D}_F$, are not considered to be border nodes.

\begin{dfn}
Let ${\cal D}_Q =  (Q_1, \dots, Q_n)$, where $n \geq 1$, be a query decomposition of a query graph $Q$.
A node $n \in {\cal B}(Q)$ is said to be a \emph{common border node} if  $n \in  {\cal B}(Q_i)$ for each $Q_i$ in ${\cal D}_Q$.
The set of common border nodes in $Q$ is denoted as ${\cal CB}(Q)$.
\end{dfn}

\begin{exmpl}
A data graph decomposition ${\cal D}_G$
(more specifically a 3-triple partition)
of the data graph $G$ of Fig.~\ref{fig:datagraph}(a) appears in Fig.~\ref{fig:rdf-apartition}.
\begin{figure*}[htb]
\centering
\includegraphics[width=0.9\textwidth]{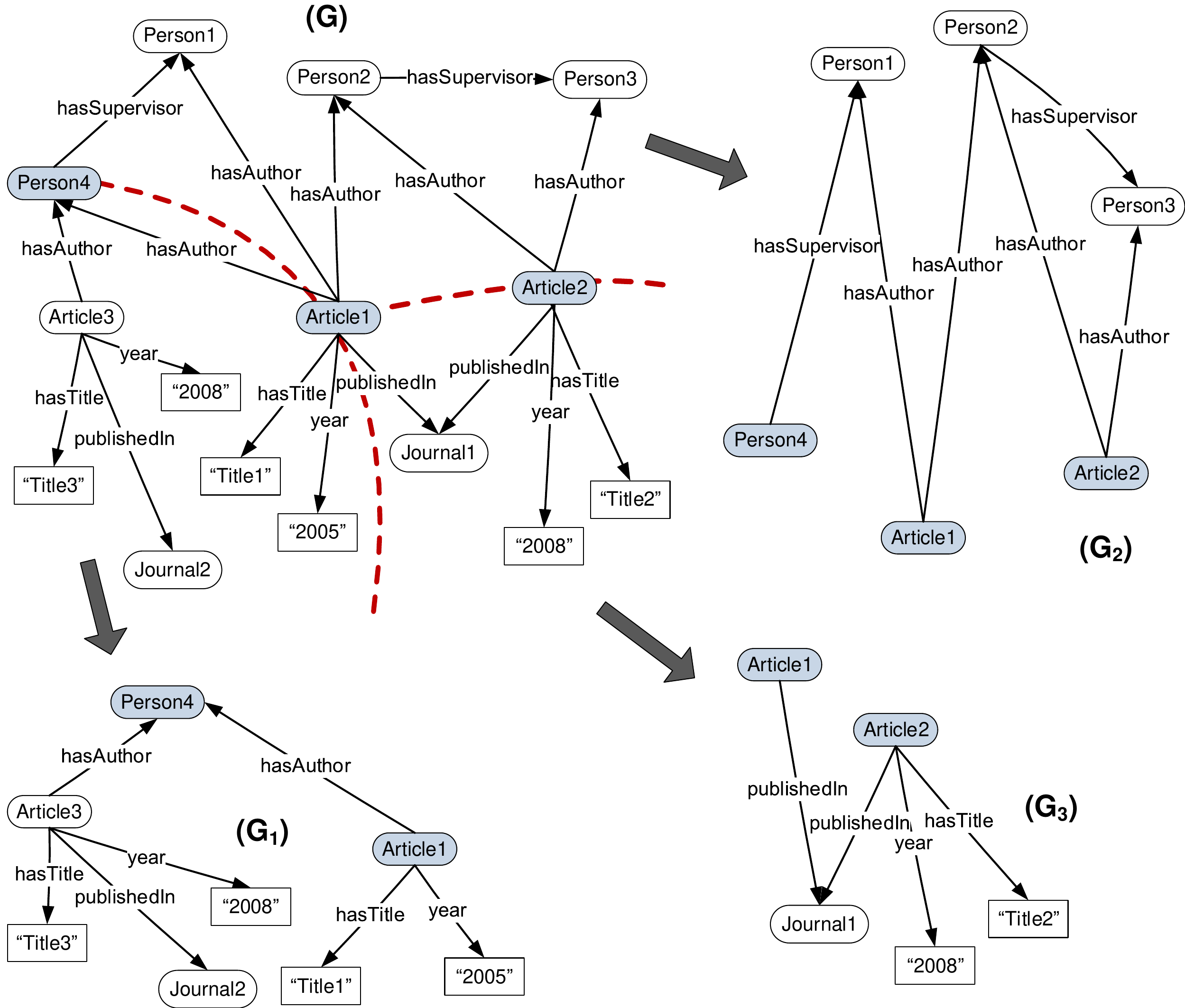}
\caption{3-triple partition of the data graph $G$ of Fig.~\ref{fig:datagraph}(a).\label{fig:rdf-apartition}}
\end{figure*}
The dark nodes correspond to the border nodes between the data graph segments, that is:\\
 ${\cal B}(G_1) = \{Person4, Article1\}$\\
 ${\cal B}(G_2) = \{Person4, Article1, Article2\}$\\
 ${\cal B}(G_3) = \{Article1, Article2\}$.

\noindent
A decomposition ${\cal D}_Q$ of a query $Q$ into a 3 query graph segments (subqueries) $Q_1$, $Q_2$, and $Q_3$ is illustrated in Fig.~\ref{fig:rdf-query-decomp}.
The border nodes between the query graph segments are:\\
 ${\cal B}(Q_1) = \{n1, n2\} = \{?P1, ?A\}$\\
 ${\cal B}(Q_2) = \{n2, n3\} = \{?A, ?P2\}$\\
 ${\cal B}(Q_3) = \{n1, n3\} = \{?P1, ?P2\}$.

\noindent
 while the set ${\cal CB}(Q)$ of common border nodes in $Q$ is empty.

\noindent
Notice that the query decomposition appearing in Fig.~\ref{fig:rdf-query-decomp} is non-redundant.

\begin{figure*}[htb]
\centering
\includegraphics[width=0.8\textwidth]{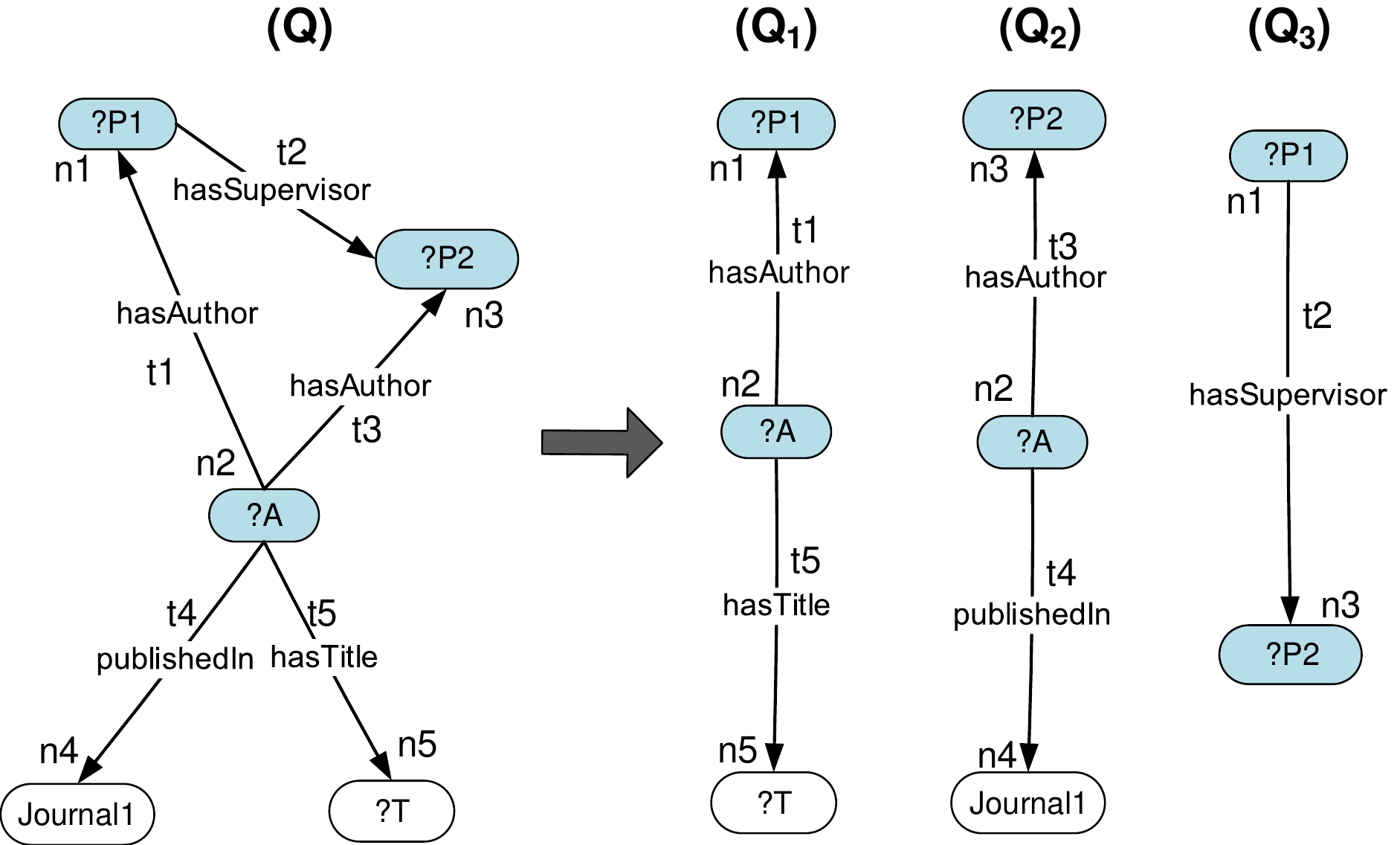}
\caption{Query decomposition.\label{fig:rdf-query-decomp}}
\end{figure*}
\hfill$\Box$
\end{exmpl}

\begin{exmpl}\label{ex:query-datadecomp}
Consider the query graph $Q$ appearing in the left part of Fig.~\ref{fig:rdf-query-decomp}.
$Q$ represents the query:
\emph{``Find an article (variable $?A$) and its title (variable $?T$) published in \emph{Journal1},
which has as authors a person (variable $?P1$) and his supervisor (variable $?P2$)''}.
It is easy to see that the evaluation of this query on the data graph $G$
depicted in Fig.~\ref{fig:rdf-apartition} returns the answers:

Answer 1: $(?P1, ?A, ?P2, ?T)$ = $(Person4, Article1, Person1, ``Title1")$.

Answer 2: $(?P1, ?A, ?P2, ?T)$ = $(Person2, Article2, Person3, ``Title2")$.

Notice, however, that,
we cannot evaluate $Q$ on a single data graph segment in ${\cal D}_G$  depicted in Fig.~\ref{fig:rdf-apartition}.
Instead, all these graph  segments are needed in order to compute the answers to this specific query
$Q$ as each of them contains part of the data needed to answer the query $Q$.
\hfill$\Box$
\end{exmpl}

Query decomposition will be proved very useful in the subsequent sections in query evaluation.
The general idea behind the algorithms that will be presented is that, in order to find the answers to a query $Q$, it suffices
to decompose $Q$ into a tuple of subqueries $(Q_1, \dots, Q_m)$,
find the answers (or partial answers) of $Q_1$, \dots, $Q_m$ and then combine appropriately  these answers to construct the answers to the query $Q$.

\subsection{Partial embeddings}
\label{subsec:partial}

When a query $Q$ is evaluated over a data graph segment $G_i$ of a data graph $G$,
it is likely that no embedding of $Q$ in $G_i$ exists.
However, this does not necessarily mean that there is no embedding of $Q$ in $G$, at all.
Instead, it is possible that ''part'' of an embedding of $Q$ in $G$ has images in $G_i$, while other ''parts''of the embeddings have images in other data graph segments of $G$.
Then, to obtain the embedding of $Q$ in $G$, we have to combine appropriately  these ''partial embeddings''.
This situation is formulated as follows:


\begin{dfn}
\label{def:partial-embedding}
A \emph{partial embedding} of a query graph $Q$ in a data graph $G$ is a partial mapping $e: {\cal{N}}(Q) \rightarrow {\cal{N}}(G)$ such that for every node
$v \in {\cal{N}}(Q)$ for which $e(v)$ is defined, the following properties hold:
\begin{enumerate}
\item
if $v$ is not a variable, then $e(v) = v$.
\item 
if $v$ is a variable, then there exists a node $u\in  {\cal{N}}(Q)$ for which
$e(u)$ is defined and an edge label $p \in {\cal{E}}(Q)$, such that
$(v, p, u) \in Q$ and $(e(v), p, e(u)) \in G$ or $(u, p, v) \in Q$ and
$(e(u), p, e(v)) \in G$.
\end{enumerate}
A partial embedding is said to be \emph{non-trivial} if there exists
a triple $(v_1, p, v_2) \in Q$ such that both $e(v_1)$ and $e(v_2)$ are defined and the triple $(e(v_1), p, e(v_2))$ belongs to $G$.
In other words, a non-trivial partial embedding is a partial embedding that maps at least one edge of $Q$ in $G$.
\end{dfn}

In essence, a partial embedding represents a mapping from a subset of nodes and edges of $Q$ to a
given data graph $G$. In other words, partial embeddings represent partial answers to $Q$, provided that, they can be appropriately ``combined''
with other ``compatible'' partial embeddings to give complete answers to the query $Q$.

The intuition behind Condition (2) is that when $e(v)$  is defined for a variable $v$ of a query $Q$, then there is a triple $t$ in $Q$
such that the variable is either the subject or the object of $t$, and $t$
is mapped, through $e$, to a triple in the data graph $G$. Notice that, as we will prove in the next section,  no answers are lost by imposing this condition to the definition of partial embeddings, while it substantially restricts the search space for computing partial embeddings.

It is easy to see that a total embedding $e$ of $Q$ in $G$ is also a partial embedding of $Q$ in $G$.
Moreover, a total embedding of a subquery of $Q$, corresponds to a partial embedding of $Q$.

\begin{dfn}
\label{def:compatible}
Two partial mappings $e_1: D_1 \rightarrow R_1$ and $e_2: D_2 \rightarrow R_2$ are said to be \emph{compatible} if
for every node $v \in  D_1 \cap D_2$ such that $e_1(v)$ and $e_2(v)$ are defined, it is $e_1(v)= e_2(v)$.
\end{dfn}

\begin{dfn}
\label{def:join}
Let $e_1: D_1 \rightarrow R_1$ and $e_2: D_2 \rightarrow R_2$ be two compatible partial mappings.
The \emph{join} of $e_1$ and $e_2$ is the partial mapping $e: D_1 \cup D_2 \rightarrow R_1 \cup R_2$ defined as follows:
\[ e(v) = \left\{ \begin{array}{ll}
                      e_1(v) & \mbox{if $e_1(v)$ is defined} \\
                      e_2(v) & \mbox{if $e_2(v)$ is defined and $e_1(v)$ is undefined} \\
                      \textup{undefined} & \mbox{if both $e_1(v)$ and $e_2(v)$ are undefined} \\
                   \end{array}
                   \right. \]
\end{dfn}

Note that, the above definitions apply also to total embeddings as they are partial mappings.
Notice also that in the first case of Definition~\ref{def:join}, $e_2(v)$ may be defined or not. If it is defined, then
the compatibility of the two partial mappings (embeddings) implies that $e_2(v)=e_1(v)$.
It is trivial to prove that the join of two compatible partial embeddings is a partial embedding
and that the join operation is commutative and associative. Therefore, we can refer
to the partial embedding resulting by the join of $n$ mutually compatible partial embeddings without ambiguity.

It should also be noted that if $Q'$ is a subquery of a query $Q$ and $e$ is a total embedding of $Q'$
in a data graph $G$, then $e$ is a partial embedding of $Q$ in $G$.

\begin{exmpl}
\label{ex:emb-com-join}
The mappings $e_1$, with $e_1(?A) = Article2$ and $e_1(?T) = Title2$, and
$e_2$, with $e_2(?A) = Article2$ and $e_2(?W) = Person3$, are partial embeddings of the query graph of Fig.~\ref{fig:datagraph}(b) in the data graph of Fig.~\ref{fig:datagraph}(a). $e_1$ and $e_2$ are compatible and their join is the partial embedding $e_3$, with $e_3(?A) = Article2$, $e_3(?T) = Title2$ and $e_3(?W) = Person3$.
\hfill$\Box$
\end{exmpl}

\subsection{Special forms of queries}
\label{subsec:special}

In this subsection we define several forms of queries. We begin by defining the \emph{path queries}:

\begin{dfn}
A query $Q$ is said to be \emph{a path query of length $n$}, with $ n \geq 1$, if it is of the form
$(v_0, p_1, v_1), (v_1, p_2, v_2), \dots, (v_{n-1}, p_n, v_n)$.
\end{dfn}

We now define the \emph{generalized star queries} as follows:

\begin{dfn}
\label{def:srar-query-def}
A query $Q$ is called a \emph{generalized star query} if there exists a node $c \in {\cal N}(Q)$,
called the \emph{central node} of $Q$  and denoted as $C(Q)$,
such that for every
triple $t = (u, p, v) \in Q$ it is either $u = c$ or $v = c$.
\end{dfn}

We now define three special forms of generalized star queries, namely  \emph{subject star queries}, \emph{object star queries} and
\emph{subject-object star queries} (\emph{s-query},  \emph{o-query}, and \emph{so-query} for sort respectively).

\begin{dfn}
A generalized star query $Q$ is said to be a \emph{subject star query} (resp. \emph{object star query})
if for every triple $t \in Q$ the central node $c$ of $Q$ is the subject (resp. object) of $t$.
\end{dfn}

\begin{dfn}
A generalized star query $Q$ is said to be a \emph{subject-object star query}
if for every triple $t \in Q$,   the central node $c$ of $Q$ is either the subject or the object of $t$ and
there is a triple $t' \in Q$, such that $c$ is the subject of $t'$.
\end{dfn}

The interest in the above special forms of queries lies in that these queries are
easier to evaluate (are evaluated more efficiently and are amenable to parallelization) than
the general  query graphs.
Besides, as we will see in the next section, we can easily decompose a huge data graph into a set of graph segments such that a query (as those defined above) can be evaluated independently on each graph segment.
These special classes of queries have the following property:
for every query graph $Q$ there exist
a (non-redundant) decomposition into s-queries (or o-quaries, or so-queries, or path queries of length 1).
This follows trivially from the fact that every query that consists of a single triple belongs to each of these classes of queries.

The algorithms for query evaluation proposed in this paper are based on these observations.
In the following sections we will use two of the above classes of queries, namely, generalized star queries and subject-object star queries (so-star queries).

\section{Query evaluation approaches}
\label{sec:query-eval-approaches}

In this section we present three procedures for query evaluation. All procedures are based on the idea that the query is decomposed on a set of subqueries which are evaluated on data segments that are obtained from decomposing the data set using various approaches.  Finally, we present algorithms for decomposing a query into so-queries.

\subsection{Query evaluation using partial embeddings}
\label{subsec:QEJPE-algorithm}

The query evaluation approach presented below, called  QEJPE-algorithm, is based on the  idea of computing (possibly in a distributed manner)
partial embeddings of subqueries (query graph segments) of a query $Q$ over data segments of a decomposed data graph $G$
and combining these partial embeddings to obtain (total) embeddings of the initial query $Q$ in the data graph $G$.
To narrow down the search space for finding partial embeddings we introduce the concept of \emph{useful partial embeddings}:

\begin{dfn}
\label{def:useful}
Let ${\cal D}_G = (G_1, \dots, G_m)$, with $m \geq 1$, be a data graph decomposition  of a data graph $G$ and
let $e$ be a partial embedding of a query graph $Q$ in some $G_i$. Then $e$ is called a \emph{useful partial embedding of Q in $G_i$}
if the following conditions hold:
\begin{enumerate}
\item 
$e$ is non-trivial.
\item 
$e$ is defined for all the nodes in $({\cal N}(Q) \cap {\cal N}(G_i))$.
\item 
for each triple $(v, p, u) \in Q$, if $e(v)$ is defined and $e(v) \notin ({\cal B}(G_i) \cup L)$, then  $e(u)$ is also defined and $(e(v), p, e(u))$ is a triple in $G_i$.
\item 
for each triple $(v, p, u) \in Q$, if $e(u)$ is defined and $e(u) \notin ({\cal B}(G_i) \cup L)$, then  $e(v)$ is also defined and $(e(v), p, e(u))$ is a triple in $G_i$.
\end{enumerate}
\end{dfn}

Notice that, according to the above definition, if $v$ is a non-variable node
of the query graph
$Q$  that maps to a non-border node of $G_i$,
then
the second property implies that $e(v)$ is defined, and the third and fourth properties
enforce every triple that contains $v$ to be mapped in $G_i$.
More generally, the edges which start from or end to a node
that maps to a non-border node in a data graph segment $G_i$ should also have images
that belong entirely to $G_i$ otherwise the partial embedding cannot be used to construct a query answer.

\begin{lem}
\label{lem:Gdecomp}
Let ${\cal D}_G = (G_1, \dots, G_m)$, with $m \geq 1$, be a (redundant or non-redundant) data graph decomposition  of a data graph $G$ and let $Q$ be a query graph.
Then the following statements are equivalent:
\begin{enumerate}
\item $e$ is a total embedding of $Q$ in $G$.
\item there exist mutually compatible useful partial embeddings $e_1, \dots, e_k$ of $Q$ in $G_{i_1}, \dots, G_{i_k}$, respectively,
for some $i_1, \dots, i_k$ with $1 \leq i_1 < \dots < i_k \leq m$, that satisfy the following properties:
\begin{enumerate}
\item for every triple $(v, p, u) \in Q$ there exists some $j$ for which $e_j(v)$, $e_j(u)$ are defined
and $(e_j(v), p, e_j(u)) \in G_{i_j}$.
\item the join of $e_1, \dots, e_k$ is $e$.
\end{enumerate}
\end{enumerate}
\end{lem}

\begin{pf}
Assume that (1) holds, that is, $e$ is an embedding of $Q$ in $G$.
Let $Q_i = \{ (v,p,u) \in Q \mid (e(v),p,e(u)) \in G_i \}$, $1 \leq i \leq m$,
and let $I$ be the set of indices for which $Q_i$ is non-empty, that is, $I = \{ i \mid Q_i \neq \emptyset \}$.
Since the query graph $Q$ is non-empty, $I$ must be also non-empty. Suppose that $|I|=k$ and let
$i_1, \dots, i_k$ be the elements of $I$ in increasing order.

For every $j$, $1 \leq j\leq k$, define the following mapping $e_j$ from $Q$ to $G_{i_j}$:
\[ e_j(v) = \left\{
       \begin{array}{ll}
          v & \mbox{if $v$ is a non-variable node in ${\cal N}(Q)\cap {\cal N}(G_{i_j})$} \\
          e(v) & \mbox{if $v$ is a variable node in ${\cal N}(Q_{i_j})$} \\
           \textup{undefined} & \mbox{otherwise} \\
        \end{array}
              \right. \]
It is not hard to see that
$e_j$ is a partial embedding of $Q$ in $G_{i_j}$ and that the join of $e_1, \dots, e_k$ is exactly $e$. Thus, property (2b) holds. In order to prove that property (2a) holds,
consider a triple $(v, p, u) \in Q$. Then, $(v, p, u) \in Q_{i_j}$ for some $j$, which implies that $v,u \in {\cal N}(Q_{i_j})$ and $e(v),e(u) \in {\cal N}(G_{i_j})$.
From the definition of $e_j$ it follows that $e_j(v)=e(v)$ and $e_j(u)=e(u)$, which implies that $(e_j(v), p, e_j(u)) = (e(v), p, e(u))$, which is in
$G_{i_j}$ by the definition of $Q_{i_j}$.

It remains to prove that $e_j$ is useful. The fact that $e_j$  is non-trivial is straightforward, since $Q_{i_j}$ is non empty. Moreover, it
obviously satisfies condition (2) of Definition~\ref{def:useful}.

In order to prove that $e_j$ satisfies condition (3) of  Definition~\ref{def:useful},  consider a triple $(v, p, u) \in Q$ such that $e_j(v)$ is defined and $e_j(v) \notin ({\cal B}(G_{i_j}) \cup L)$.
Since $e$ is an embedding of $Q$ in $G$, it must be $(e(v),p,e(u)) \in G$.
Moreover, $e(v)=e_j(v)$ which implies that $e(v)$ is not a border node of $G_{i_j}$ nor an element of $L$.
Therefore, $e(v)$ appears only in $G_{i_j}$, which implies that $(e(v),p,e(u))$ must be a triple in $G_{i_j}$.
Hence, $(v,p,u) \in Q_{i_j}$, which implies that $u \in {\cal N}(Q_{i_j})$
and by the definition of $e_j$ it is $e_j(u)=e(u)$ (i.e. $e_j(u)$ is defined).
The fact that $(e_j(v),p,e_j(u))$ is a triple in $G_{i_j}$ is now clear, since it equals $(e(v),p,e(u))$. The proof for condition (4) of  Definition~\ref{def:useful},   is similar.

For the other direction, assume that (2) holds.
We first show that $e$ (the join of $e_1, \dots, e_k$) is a total mapping from ${\cal N}(Q)$ to  ${\cal N}(G)$. Notice that $e_1, \dots, e_k$ are compatible.
Suppose that $v \in {\cal N}(Q)$, that is, $v$ appears in some triple of the form $(u,p,v)$ or $(v,p,u)$ in $Q$. Then, $e_j(v)$ is defined for some $j$ (by property (2a)),
which implies (using the definition of join) that $e(v)$ is also defined.

We next show that $e$ is an embedding of $Q$ in $G$. Let $v$ be a non-variable element in ${\cal N}(Q)$. From the definition of join, it must be $e(v)=e_j(v)$ for some $j$.
Since $e_j$ is a useful partial embedding, it is $e_j(v)=v$. Therefore, it holds $e(v)=v$.

Finally, consider a triple $(v,p,u) \in Q$. By property (2a), there exists some $j$ such that $(e_j(v), p, e_j(u)) \in G_{i_j}$ which implies that
$(e(v), p, e(u)) \in G$ (since $e(v)=e_j(v)$, $e(u)=e_j(u)$,  and $G_{i_j} \subseteq G$). \qed
\end{pf}

\begin{lem}\label{lem:Qdecomp}
Let ${\cal D}_Q = (Q_1, \dots, Q_n)$, with $n \geq 1$, be a query decomposition of a query graph $Q$ and $G$ be a data graph.
Then $e$ is a total embedding of $Q$ in $G$ if and only if
there exist mutually compatible total embeddings $e_1, \dots, e_n$ of $Q_1, \dots, Q_n$ in $G$
such that the join of $e_1, \dots, e_n$ is $e$.
\end{lem}

\begin{pf}
For the one direction, assume that $e$ is a total embedding of $Q$ in $G$. For every $i$ define
$e_i$ to be the restriction of $e$ in ${\cal N}(Q_i)$ (that is, $e_i: {\cal N}(Q_i) \rightarrow {\cal N}(G)$, with $e_i(v)=e(v)$).
Obviously $e_i$ is a total mapping. Furthermore, for every non-variable element $v \in {\cal N}(Q_i)$ it is $e_i(v)=e(v)=v$ and for every triple
$(v_1,p,v_2) \in Q_i$ it is $(e_i(v_1),p,e_i(v_2)) = (e(v_1),p,e(v_2)) \in G$, which implies that $e_i$ is actually an embedding of $Q_i$ in $G$.

Moreover, for every $i,j$ with $i \neq j$, if $v \in {\cal N}(Q_i) \cap {\cal N}(Q_j)$ then it is $e_i(v) = e_j(v) = e(v)$, which implies that $e_i$ and $e_j$ are compatible.
Therefore, the join $e'$ of $e_1, \dots, e_n$ exists. It remains to show that $e'=e$.
Consider an arbitrary $v \in {\cal N}(Q)$. Then $v$ appears in some triple $t \in Q$. Since ${\cal D}_Q$ is a decomposition of $Q$, there exists some $i$ such that
$t \in Q_i$. Thus, $v \in {\cal N}(Q_i)$, which implies that $e_i(v)$ is defined. From the definition of join, it follows that $e'(v) = e_i(v)$, which implies $e'(v) = e(v)$.

For the other direction, assume that $e_1, \dots, e_n$ are compatible total embeddings of $Q_1, \dots, Q_n$ in $G$ and let $e$ be their join.
Using the same argument as above, we can prove that for every $v \in {\cal N}(Q)$ there exists some $i$ such that $e_i(v)$ is defined, which implies that $e(v)$ is also defined.
Thus, $e$ is a total mapping.

We next show that $e$ is an embeding of $Q$ in $G$. Consider any non-variable element $v \in {\cal N}(G)$.
Since $e$ is total, $e(v)$ is defined. From the definition of join, there must be some $i$ such that $e(v)=e_i(v)$, which implies $e(v)=v$ (since $e_i$ is an embedding).

Finally, let $(v_1,p,v_2)$ be a triple in $Q$. Since ${\cal D}_Q$ is a decomposition of $Q$, $(v_1,p,v_2)$ belongs to some $Q_i$. Since $e_i$ is a total embedding of $Q_i$ in $G$,
it holds $(e_i(v_1),p,e_i(v_2)) \in G$, which implies $(e(v_1),p,e(v_2)) \in G$. \qed
\end{pf}

\begin{thm}
\label{th:theoremTotal}
Let ${\cal D}_Q =(Q_1, \dots, Q_n)$, with $n \geq 1$, be a query decomposition of a query graph $Q$ and
${\cal D}_G = (G_1, \dots, G_m)$, with $m \geq 1$,  be  a data graph decomposition
of a data graph $G$.
Then the following statements are equivalent:
\begin{enumerate}
\item $e$ is a total embedding of $Q$ in $G$.
\item for every $j$, with $1 \leq j \leq n$, there exist useful partial embeddings
$e_{j,1}, \dots, e_{j,k_j}$ of $Q_j$ in $G_{i_{j,1}}, \dots, G_{i_{j,k_j}}$ for
some $i_{j,1}, \dots, i_{j,k_j}$ with $1 \leq i_{j,1} < \dots < i_{j,k_j} \leq m$ that satisfy the following properties:
\begin{enumerate}
\item for every $j$, with $1 \leq j \leq n$, and every triple $(v, p, u) \in Q_j$ there exists some $\ell$ such that $e_{j,\ell}(v)$, $e_{j,\ell}(u)$ are defined
and $(e_{j,\ell}(v), p, e_{j,\ell}(u)) \in G_{i_{j,\ell}}$.
\item for every $j_1,j_2,\ell_1,\ell_2$, with $1 \leq j_1 \leq j_2 \leq n$, $1 \leq \ell_1 \leq k_{j_1}$, $1 \leq \ell_2 \leq k_{j_2}$,
the partial embeddings $e_{j_1,\ell_1}$ and $e_{j_2,\ell_2}$ are compatible.
\item the join of $e_{j,\ell_j}$ for all $j \in \{1,\dots, n\}$ and all $\ell_j \in \{1,\dots, k_j\}$ is $e$.
\end{enumerate}
\end{enumerate}
\end{thm}

\begin{pf}
For the one direction, assume that (1) holds, that is, $e$ is an embedding of $Q$ in $G$.
From Lemma~\ref{lem:Qdecomp}  we conclude that
there are mutually compatible total embeddings $e_1, \dots, e_n$ of $Q_1, \dots, Q_n$ in $G$, such that the join of
$e_1, \dots, e_n$ is $e$. Now, from Lemma~\ref{lem:Gdecomp} we conclude that, for each $Q_j$, there exist mutually compatible useful partial embeddings
$e_{j,1}, \dots, e_{j,k_j}$ of $Q_j$ in $G_{i_{j,1}}, \dots, G_{i_{j,{k_j}}}$ such that property (a) holds and the join of
$e_{j,1}, \dots, e_{j,k_j}$ is $e_j$.
In order to show that propery (b) holds, suppose for the sake of contradiction, that $e_{j_1,\ell_1}$ and $e_{j_2,\ell_2}$ are not compatible, for
some $j_1,j_2,\ell_1,\ell_2$. Then, there exists some $v$ such that $e_{j_1,\ell_1}(v) \neq e_{j_2,\ell_2}(v)$. Since $e_{j_1}(v) = e_{j_1,\ell_1}(v)$
and $e_{j_2}(v) = e_{j_2,\ell_2}(v)$, the total embeddings $e_{j_1}$ and $e_{j_2}$ must also be incompatible, which is a contradiction. Therefore, property (b) holds.
Finally, property (c) holds since for all $j$, the join of
$e_{j,1}, \dots, e_{j,k_j}$ is $e_j$ and the join of $e_1, \dots, e_n$ is $e$.

For the other direction, assume that (2) holds.
From Lemma~\ref{lem:Gdecomp}, it follows that for every $j$, the join $e_j$ of the partial embeddings $e_{j,1}, \dots, e_{j,k_j}$ is
a total embedding of $Q_i$ in $G$. Moreover, the resulting embeddings $e_1, \dots, e_n$ are mutually compatible, since we have assumed that
$e_{j_1,\ell_1}$ and $e_{j_2,\ell_2}$ are compatible for all $j_1,j_2,\ell_1,\ell_2$.
Now, from Lemma~\ref{lem:Qdecomp} it follows that the join $e$ of $e_1, \dots, e_n$ is a total embedding of $Q$ in $G$.
\qed
\end{pf}

Theorem~\ref{th:theoremTotal} implies a generic query evaluation strategy,
named \emph{Query Evaluation by Joining Partial Embeddings (QEJPE) strategy} consisting of four steps.
The algorithm assumes an arbitrary decomposition of the data graph $G$ into a tuple ${\cal D}_G$ of data graph segments $G_1, \dots, G_m$, with $m \geq 1$, stored into a cluster of computer nodes.

\begin{quote}
\begin{itemize}
\item [\textbf{Step 1:}] Decompose the query $Q$ into a tuple  ${\cal D}_Q$ of subqueries $Q_1, \dots, Q_n$, with $n \geq 1$.

\item [\textbf{Step 2:}] Compute  all possible useful partial embeddings of each subquery $Q_j$ over each data graph segment $G_i$ of $G$.

\item [\textbf{Step 3:}] For each subquery $Q_j$, collect all the partial embeddings of $Q_j$ obtained in Step 2 and join them to get total embeddings of $Q_j$.

\item [\textbf{Step 4:}] To construct the total embeddings (i.e. answers) of $Q$, join the total embeddings obtained in Step 3 by using one embedding for each subquery, in all possible ways.
\end{itemize}
\end{quote}

Notice that the above generic query evaluation strategy has several interesting properties: a) it is independent of the way the data graph is decomposed and the way the data graph segments obtained by this decomposition are stored in the nodes of the cluster, b) it is independent of the way the query graph is decomposed, and c) it is independent of the algorithm used to compute (partial) embeddings.

In Subsection~\ref{subsec:QEJPE-algorithm-impl},  we present an implementation of this strategy on a cluster of commodity computers based on the Map-Reduce programming framework.

\subsection{Query evaluation by decomposing queries into generalized stars}
\label{subsec:QE-using-STARS}


In this section we present another approach, called eval-STARS algorithm, for evaluating queries over linked data.
The algorithm is based on assumptions similar to these on which the QEJPE-algorithm, presented in Section~\ref{subsec:QEJPE-algorithm}, is based.
The main difference is that we now impose subqueries obtained from the decomposition of a user query $Q$ to be in the form called \emph{generalized star queries}. Besides, the algorithm is based on evaluation of total embeddings of the subqueries instead of partial embeddings.
 
Recall that, as we proved in Lemma~\ref{lem:Qdecomp},
in order to compute the answers to a given query in a data graph $G$,
we can decompose the query into a tuple of subqueries,
compute the embeddings of the subqueries in $G$ (which may be a more efficient task due to
the simpler or special form of the subqueries) and then join these embedding to obtain the desired result.
However, given a target class of queries $C$, it may not be always possible to decompose an arbitrary query $Q$
into subqueries that belong to $C$. For example, if $C$ is the class of path queries of length 3, in other words if the subqueries must be of the form
$\{(u,p,v),(v,p',w),(w,p'',z)\}$, then it can be proved that it is not possible to decompose every user query into a set of path queries of length 3.
Nevertheless, if the target class $C$ is the class of generalized star queries, then for every query $Q$ there exist
a (non-redundant) decomposition of $Q$ into a tuple of generalized star subqueries. This trivially follows from the fact that
every query that consists of a single triple is also a generalized star query (with either the subject or
the object being the central node). We next present a more general result, relating
the decomposition of a query graph  $Q$ into generalized star subqueries, to the \emph{node covers} of this query graph.

\begin{dfn}
\label{def:node-cov}
Let $Q$ be a query graph. A set of nodes $V \subseteq {\cal N}(Q) - L$ is called a {\em node cover} of $Q$
if for every triple $(s,p,o) \in Q$, it holds either $s \in V$ or $o \in V$.
\end{dfn}

\begin{lem}\label{lem:nc-decom}
Let $Q$ be a query graph and $V=\{v_1, \dots, v_k \}$ be a node cover of $Q$.
For each $v_i \in V$ define the generalized star query $Q_{v_i} = \{ t \in Q \mid t = (s,p,v_i) \} \cup \{ t \in Q \mid t = (v_i,p,o) \textup{ and } o \notin V \}$.
Then ${\cal D}_Q = (Q_{v_1}, \dots, Q_{v_k})$ is a non-redundant decomposition of $Q$.
\end{lem}

\begin{pf}
It easy to see that $(Q_{v_1}, \dots, Q_{v_k})$ forms a decomposition of $Q$ since:

 (1) by construction $Q_{v_i} \subseteq Q$, for $i = 1, \dots, k$, and

 (2) $\bigcup_i  Q_{v_i} = Q$, since for every triple $t = (s, p , o ) \in Q$, either $s \in V$ or $o \in V$.
 If $o \in V$ then, by construction $t \in Q_{o}$. Otherwise (i.e. if $s \in V$ and $o \not\in V$) then $t \in  Q_{s}$.

We will now prove (by contradiction) that ${\cal D}_Q = (Q_{v_1}, \dots, Q_{v_k})$ is   non-redundant.
Assume that ${\cal D}_Q$ is redundant.
Then there exists a triple $t = (s, p, o) \in Q$ such that $t$ belongs to two different subqueries in ${\cal D}_Q$.
It is easy to see that these subqueries should be $Q_s$ and $Q_o$ and $s, o \in V$.
However, since $t \in Q_s$, then, by construction, $o \not\in V$, which contradicts with the fact that $s, o \in V$.
\qed
\end{pf}

Therefore, if a set of nodes is a node cover of a query $Q$, then its elements are the central nodes of
the generalized star subqueries in a non-redundant decomposition of $Q$. It turns out that the converse also holds.

\begin{lem}\label{lem:decom-nc}
Let $Q$ be a query graph, let ${\cal D}_Q = (Q_{1}, \dots, Q_{k})$ be a decomposition of $Q$
such that $Q_{1}, \dots, Q_{k}$ are generalized star queries
and let $c_1,\dots,c_k$ be their central nodes. Then $\{c_1,\dots,c_k\}$ is a node cover
of $Q$.
\end{lem}

\begin{pf}
 It immediately follows from Definitions~\ref{def:graphdecomposition}, \ref{def:srar-query-def} and~\ref{def:node-cov}.
 \qed
\end{pf}

\begin{exmpl}
In Fig.~\ref{fig:rdf-query-star-decomp} we see a decomposition of
the query $Q$ into three generalized star queries $Q_1$, $Q_2$ and $Q_3$,
which is obtained by the construction of
Lemma~\ref{lem:nc-decom}, using the node cover $\{n_4,n_2,n_5\}$
of $Q$.
\begin{figure*}[htb]
\centering
\includegraphics[width=0.95\textwidth]{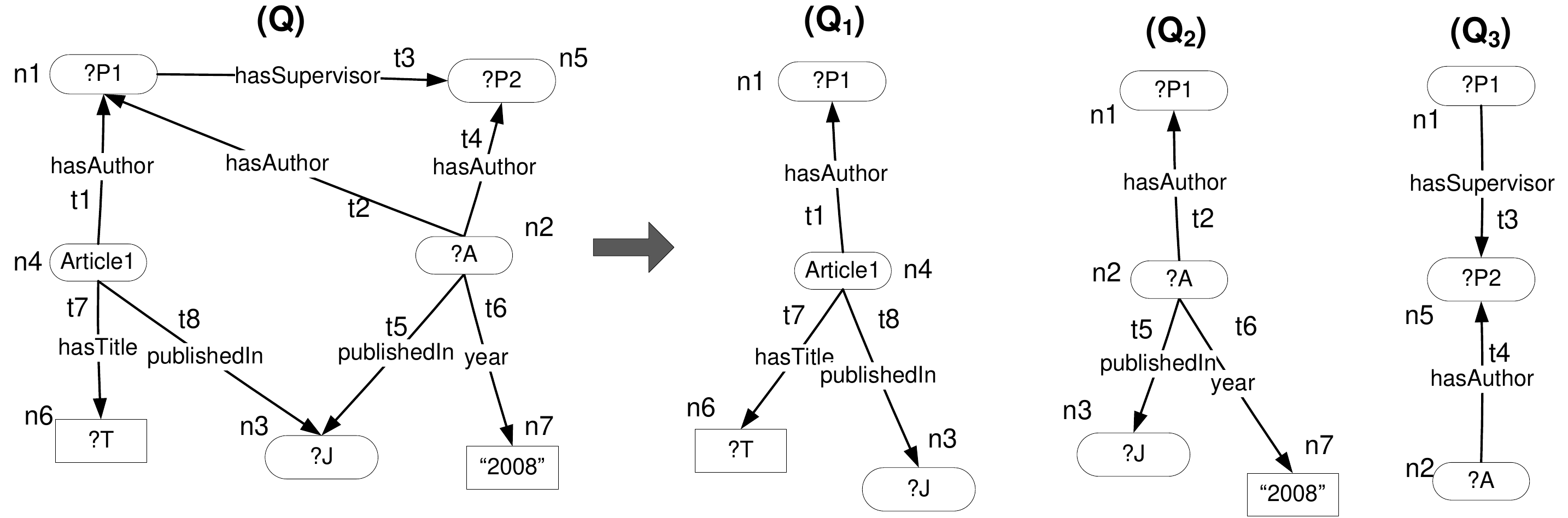}
\caption{Query decomposition into star queries.\label{fig:rdf-query-star-decomp}}
\end{figure*}
\hfill$\Box$
\end{exmpl}

Following the discussion above we can be specialize the generic query evaluation strategy (QEJPE strategy) presented in Subsection~\ref{subsec:QEJPE-algorithm}, obtaining in this way a new algorithm called \emph{eval-STARS algorithm}.
As in the case of QEJPE strategy we assume an arbitrary decomposition of the data graph $G$  into a tuple ${\cal D}_G$ of data graph segments $G_1, \dots, G_m$, with $m \geq 1$, stored into a cluster of computer nodes.

The  \emph{eval-STARS algorithm} algorithm consists of the following steps:

\begin{quote}
\begin{itemize}
\item [\textbf{Step 1:}] Decompose the query $Q$ into a tuple of generalized star subqueries ${\cal D}_Q = (Q_1, \dots, Q_n)$, with $n \geq 1$.

\item [\textbf{Step 2:}] Compute all possible embeddings of each triple in $Q$ over each data graph segment $G_i$ of $G$.

\item [\textbf{Step 3:}] For each subquery $Q_j$, collect the embeddings of all the triples in $Q_j$ and join compatible embeddings in all possible ways to compute the total embeddings of $Q_j$ in $G$.

\item [\textbf{Step 4:}] To construct the total embeddings (i.e. answers) of $Q$, join the total embeddings obtained in Step 3 by using one embedding for each subquery, in all possible ways.
\end{itemize}
\end{quote}

Note that eval-STARS algorithm applies two query decomposition processes. Initially, the given query is decomposed into generalized star queries (Step 1 of eval-STARS) and each star query is further decomposed (Step 2 of eval-STARS) into its triples. On the contrary, QEJPE applies a single decomposition (Step 1 of QEJPE). Following this stepwise approach of two decompositions, in fact, we achieve the construction of the total embeddings in two phases, where each phase gathers the compatible partial embeddings and join them together (i.e., it applies Lemma~\ref{lem:Qdecomp}  twice). Although this extra decomposition could be thought of as a redundant step, in parallel computation (see Section~\ref{sec:implementations}), such an approach brings a significant performance improvement and facilitates the distribution of both the intermediate data and the computation.

\subsection{Query evaluation by data decomposition using replication}
\label{subsec:QE-redundancy}


In this section we propose a query evaluation approach, called \emph{QE-with-Redundancy},
which uses a specific form of replication in the data graph decomposition to efficiently answer queries. More specifically:

\begin{enumerate}
\item [(a)]
Data graphs are decomposed into data graph segments in which replication of the data triples  is allowed.
Data triples are replicated in such a way that all the answers to a special form of queries,
namely \emph{subject-object star queries}, can be obtained from a single data segment.
The partition of the data graph is specified by an arbitrary partition of the data nodes, while data segments consist of the in- and out-edges of each block of nodes.
Therefore, triples containing nodes that are in two different blocks occur in both segments of the data graph
corresponding to these blocks.
This redundancy, as we show, ensures that the subject-object star subqueries can be easily evaluated over each segment, independently.

\item [(b)] Each query posed by the user is decomposed into a tuple of subject-object star subqueries.
\end{enumerate}

In the evaluation strategy presented in this section, our aim is to construct the embeddings of a query $Q$ in a data graph $G$, by appropriately
combining embeddings (i.e. joining compatible embeddings) of so-subqueries of $Q$ over the proper sugbgaphs of $G$.

The following lemma refers to the compatibility of embeddings:

\begin{lem}
\label{lem:border-compatible}
Let ${\cal D}_Q = (Q_1, \dots, Q_n)$, with $n \geq 1$, be a query decomposition of a query graph $Q$ and ${\cal D}_G = (G_1, \dots, G_m)$, with $m \geq 1$, be a data graph decomposition of  a data graph $G$.
Let $e_{Q_i}$ and  $e_{Q_j}$ be two embeddings of the subqueries $Q_i$ and $Q_j$ respectively, with $1 \leq i \not= j \leq n$, on two (not necessarily different) graph segments $D_k$ and $D_l$ in ${\cal D}_G$. Let ${\cal B}(Q_i, Q_j)$ be the border nodes of
$Q_i$, $Q_j$. Then $e_{Q_i}$ and  $e_{Q_j}$ are compatible if and only if for each node $v \in {\cal B}(Q_i, Q_j)$, it holds that $e_{Q_i}(v) = e_{Q_j}(v)$.
\end{lem}

\begin{pf}
It immediately follows from Definitions~\ref{def:compatible} and ~\ref{def:border-node}.
\qed
\end{pf}

In the following definition  we present a decomposition scheme for a data graph $G$, called \textit{star-oriented decomposition} (or simply \emph{s-decomposition}).

 \begin{dfn}\label{def:stargraphpartition}
A \emph{star-oriented decomposition} (or \emph{s-decomposition} for short) of a data graph $G$ is a tuple of graphs ${\cal D}_G = (G_1, \dots, G_m)$, with $m \geq 1$, if ${\cal N_P}=(N_1, \dots, N_m)$ is a partition of the nodes in ${\cal N}(G)- L$ and
for each $i$, with $1 \leq i \leq m$, $G_i = \{t \mid  t=(s, p, o)\ and\ t \in G\ and\ s \in N_i\ or\ o \in N_i\}$.
Subgraphs $G_1, \dots, G_m$ are called \emph{s-graph segments}.
A node in ${\cal N}(G_i) - L - N_i$ is called a \emph{replicated node} in $G_i$.
A \emph{replicated triple} $t = (s, p, o)$ in a s-graph segment $G_i$ is a data triple in $G_i$ such that ether $s$ or $o$ is a replicated node.
\end{dfn}

In the following, the set of replicated nodes in a s-graph segment $G_i$ is denoted by ${\cal R}_N(G_i)$.
The replicated nodes of a data graph $G$  is ${\cal R}_N(G) = \bigcup_i{\cal R}_N(G_i)$.
Similarly, the set of replicated triples in a s-graph segment $G_i$ is denoted by ${\cal R}_t(G_i)$.
Finally, replicated triples of a data graph $G$  is ${\cal R}_t(G) =  \bigcup_i {\cal R}_t(G_i)$.

\begin{exmpl} 
Fig.~\ref{fig:rdfStarpartition} shows an s-decomposition of the data graph $G$ of Fig.~\ref{fig:datagraph}(a),
which is based on the following partition of the set of nodes in ${\cal N}(G)-L$:\\
$N_1 = \{Article1,  Article3, Journal2, Person4\}$\\
$N_2 = \{Person1,  Person2, Person3\}$\\
$N_3 = \{Article2,  Journal1\}.$\\

\begin{figure*}[htb]
\centering
\includegraphics[width=0.99\textwidth]{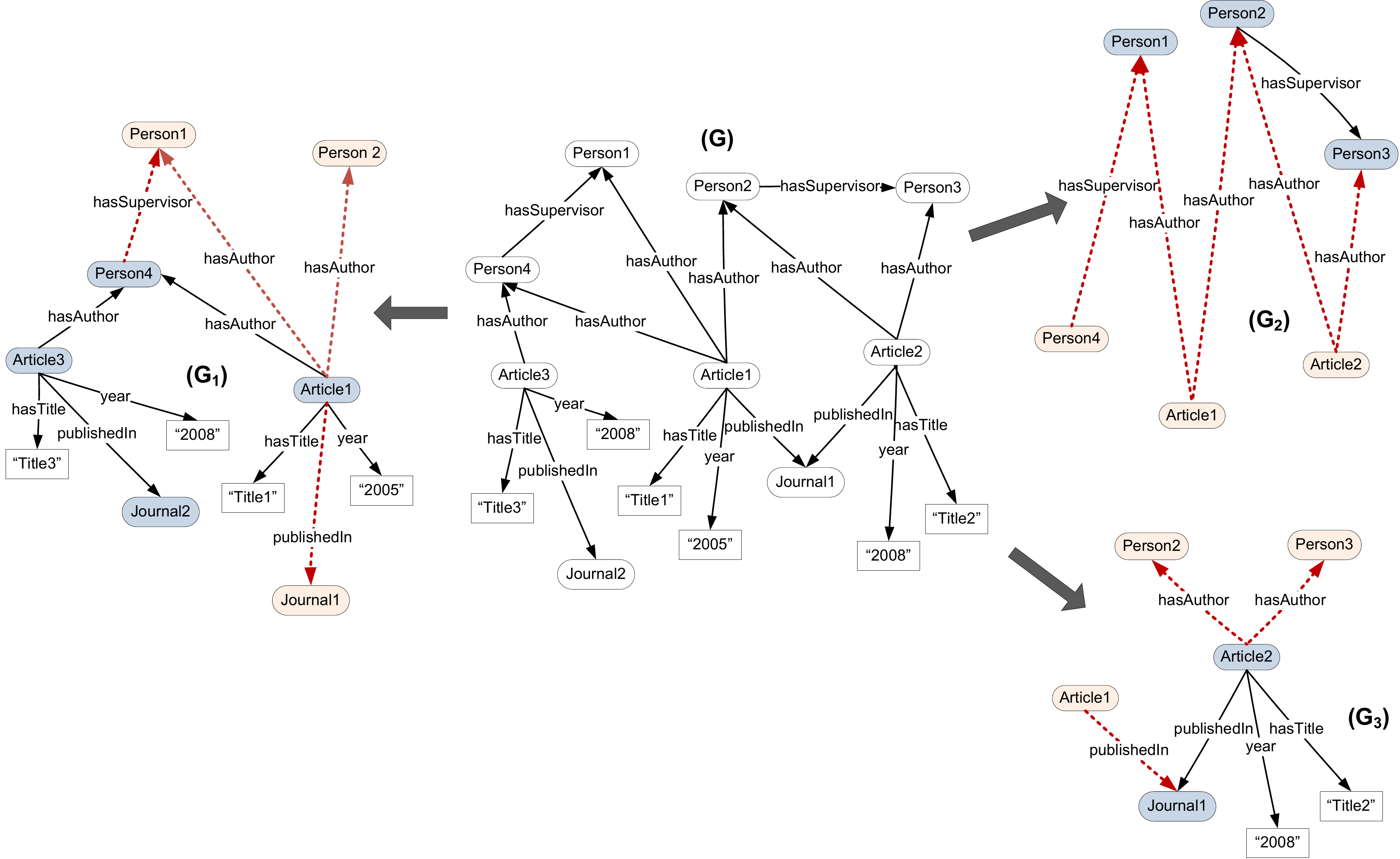}
\caption{An s-decomposition of the data graph $G$ of Fig.~\ref{fig:datagraph}.\label{fig:rdfStarpartition}}
\end{figure*}
The
grey colored nodes in the segments $G_1$, $G_2$, and $G_3$ correspond to the nodes in $N_1$, $N_2$ and $N_3$, respectively, while the pink colored nodes are the replicated nodes. Finally,  the dashed lines in the graph segments correspond to replicated data triples.
%
Consider now the query graph $Q$ appearing in the right part of Fig.~\ref{fig:datagraph}.
It is easy to see that we cannot obtain the solution described in Example~\ref{ex:datagraph} by finding an embedding of $Q$ in a single graph segment of $G$ appearing in Fig.~\ref{fig:rdfStarpartition} (as such an embedding does not exist).
\hfill
\end{exmpl}

The following lemma presents some interesting properties of the star-oriented decomposition of a data graph.

\begin{lem}
\label{lem:properties-of-s-decomp}
Let  ${\cal D}_G = (G_1, \dots, G_m)$, with $m \geq 1$, be an \emph{s-decomposition} of a data graph $G$  based on the partition ${\cal N_P}=(N_1, \dots, N_m)$ of the nodes in ${\cal N}(G)- L$.
Then the following hold:

\begin{enumerate}
\item
$({\cal N}(G_i) - L) \supseteq N_i$, for each $i$, with $1 \leq i \leq m$.
\item
$\bigcup_{i \leq m} {\cal N}(G_i) = {\cal N}(G)$
\item
$\bigcup_{i \leq m} G_i = G$
\item
Consider a node $s \in {\cal R}_N(G_i)$. Then for each triple $t = (s, p, o) \in G_i$ it holds that $o \in N_i$ and $t \in {\cal R}_t(G_i)$.
\item
Consider a node $o \in {\cal R}_N(G_i)$. Then for each triple $t = (s, p, o) \in G_i$ it holds that $s \in N_i$ and $t \in {\cal R}_t(G_i)$.
\item
For each node $v \in {\cal R}_N(G_i)$, with  $1 \leq i \leq m$
there exists an index $j$, with $1 \leq j \leq m$ and $i \neq j$, such that $v \in N_j$.
\item
For each triple $t \in {\cal R}_t(G_i)$, with $1 \leq i \leq m$,
there exists an index $j$, with $1 \leq j \leq m$ and $i \neq j$, such that $t \in G_j$. 
\end{enumerate}
\end{lem}

\begin{pf}
\hfill\\
\noindent\underline{Proof of 1:} It immediately follows from Defitition~\ref{def:stargraphpartition}.

\noindent\underline{Proof of 2:}
Let $v$ be a node in $\bigcup_{i \leq m} {\cal N}(G_i)$. Then, there exist a triple $(s,p,o) \in G_i$,
for some $i$, such that $v=s$ or $v=o$. Since $G_i \subseteq G$, we have $(s,p,o) \in G$, which implies that
$s,o \in {\cal N}(G)$. Therefore, $v \in {\cal N}(G)$.

Now let $v$ a node in ${\cal N}(G)$. Then, there exist a triple $(s,p,o) \in G$,
such that $v=s$ or $v=o$. Since $s$ is the subject of this triple, it must be $s \in {\cal N}(G) - L$,
which implies that $s \in N_i$, for some $i$. Therefore, $(s,p,o) \in G_i$ and thus $s,o \in {\cal N}(G_i)$.
Consequently, $v \in \bigcup_{i \leq m} {\cal N}(G_i)$.

\noindent\underline{Proof of 3:} From Definition~\ref{def:stargraphpartition}  we conclude that $\bigcup_{i \leq m} G_i \subseteq G$. To prove the inverse let $t = (s, p, o)$ be a triple in $G$. Then $s \in ({\cal N}(G) - L)$. Thus $s \in N_i$ for some $i$ with $1 \leq i \leq n$. Hence, by construction of the s-segments, $t \in G_i$ and therefore $t \in \bigcup_{i \leq m} G_i$. Therefore, $G \subseteq \bigcup_{i \leq m} G_i$.

\noindent\underline{Proof of 4:}
It immediately follows form Definition~\ref{def:stargraphpartition}.

\noindent\underline{Proof of 5:}
It immediately follows form Definition~\ref{def:stargraphpartition}.

\noindent\underline{Proof of 6:}
As $v \in {\cal R}_N(G_i)$, from Definition~\ref{def:stargraphpartition} we conclude that
$v \not\in N_i$. But as $v$ is a node in ${\cal N}(G) - L$, the node $v$ should belong to another set $N_j$, with $j \not= i$,  of the partition of the nodes in ${\cal N}(G) - L$.

\noindent\underline{Proof of 7:}
Assume that $t$  is of the form $t = (s,p,o)$. As $t \in {\cal R}_t(G_i)$, from Definition~\ref{def:stargraphpartition} we see that either $s$ or $o$ is a replicated node in ${\cal R}_N(G_i)$. Assume that the replicated node is $s$. Then, from (6) we conclude that there is an index $j \neq i$ such that $s \in N_j$. Then,  from Definition~\ref{def:stargraphpartition}, we conclude that $t \in G_j$. In a similar way we reach the same conclusion by assuming that $o$ is the replicated node.
\qed
\end{pf}

The following theorem relates the embeddings of the so-queries obtained from graph segments to the embeddings of the query $Q$ on the graph $G$.


\begin{thm}\label{th:QandGdecompandeval}
Let ${\cal D}_Q = (Q_1, \dots, Q_n)$,  with $n \geq 1$, be a query decomposition of a query graph $Q$, such that each $Q_i$, with $1 \leq i \leq n$, is an so-query.
Let also ${\cal D}_G = (G_1, \dots, G_m)$, with $m \geq 1$, be an s-decomposition of a data graph $G$.
Then $e$ is a total embedding of $Q$ in $G$ if and only if $e$ is the join of $e_1, \dots, e_n$, where $e_1, \dots, e_n$ are mutually compatible embeddings such that for each $i$, with $1 \leq i \leq n$, $e_i$ is a total embedding of $Q_i$ in some segment $G_j$, where $1 \leq j \leq m$.
\end{thm}

\begin{pf}
For the one direction, assume that $e$ is a (total) embedding of $Q$ in $G$.
For every $i$, with $1 \leq i \leq n$, define $e_i$ to be the restriction of $e$ in $Q_i$
(that is, $e_i: {\cal N}(Q_i) \rightarrow {\cal N}(G)$, with $e_i(v)=e(v)$ for every node $v \in {\cal N}(Q_i)$).
Obviously $e_i$ is a total mapping.
Furthermore, for every non-variable element $v \in {\cal N}(Q_i)$ it is $e_i(v)=e(v)=v$ and for every triple
$(v_1,p,v_2) \in Q_i$ it is $(e_i(v_1),p,e_i(v_2)) = (e(v_1),p,e(v_2)) \in G$, which implies that $e_i$ is actually an embedding of $Q_i$ in $G$.

As $Q_i$ is an so-query, let $C(Q_i)$ be the central node of $Q_i$ and $e(C(Q_i)) \in N(G)$ the image of $C(Q_i)$ in $G$.
From Definition~\ref{def:stargraphpartition} we conclude that  $e(C(Q_i)) \in  N_j$, for some $N_j$ with $1 \leq j \leq m$ and that
$e_i$ is an embedding of $Q_i$ in $G_j$.

We next prove that the embeddings $e_i$, with $1 \leq i \leq n$ are mutually compatible and their join is $e$.
By construction, for every $i, j$ with $i \neq j$, if $v \in {\cal N}(Q_i) \cap {\cal N}(Q_j)$ then it is $e_i(v) = e_j(v) = e(v)$, which implies that $e_i$ and $e_j$ are compatible.
Therefore, the join $e'$ of $e_1, \dots, e_n$ exists.
It remains to show that $e'=e$.
Consider an arbitrary $v \in {\cal N}(Q)$. Then $v$ appears in some triple $t \in Q$. Since ${\cal D}_Q$ is a decomposition of $Q$, there exists some $i$ such that
$t \in Q_i$. Thus, $v \in {\cal N}(Q_i)$, which implies that $e_i(v)$ is defined. From the definition of join, it follows that $e'(v) = e_i(v)$, which implies $e'(v) = e(v)$.


For the other direction, assume that for each $i$, with $1 \leq i \leq m$,
there is an embedding $e_i$ for the subquery $Q_i$ in some graph segment $G_j$. Assume also that $e_1, \dots, e_n$ are mutually compatible embeddings and let $e$ be their join.
We will prove that $e$ is an embedding of $Q$ in $G$.
It is easy to see that for every $v \in {\cal N}(Q)$ there exists some $i$ such that $e_i(v)$ is defined, which implies that $e(v)$ is also defined.
Thus, $e$ is a total mapping.

We next show that $e$ is an embeding of $Q$ in $G$. Consider any non-variable element $v \in {\cal N}(G)$. Since $e$ is total, $e(v)$ is defined. From the definition of join, there must be some $i$ such that $e(v)=e_i(v)$, which implies $e(v)=v$ (since $e_i$ is an embedding).

Finally, let $(v_1,p,v_2)$ be a triple in $Q$. Since ${\cal D}_Q$ is a decomposition of $Q$, $(v_1,p,v_2)$ belongs to some $Q_i$. Since $e_i$ is a total embedding of $Q_i$ in an s-segment of $G$, it is also a total embedding of $Q_i$ in  $G$. Thus,
 $(e_i(v_1),p,e_i(v_2)) \in G$, which implies $(e(v_1),p,e(v_2)) \in G$.
\qed
\end{pf}

The above theorem suggests the following strategy for the evaluation of a query $Q$ on a data graph $G$, called \emph{QE-with-Redundancy}.
\emph{QE-with-Redundancy} strategy assumes a star-oriented decomposition of the data graph $G$.
To obtain such a decomposition we assume an arbitrary partition
${\cal N_P}=(N_1, \dots, N_m)$, with $m \geq 1$ of the nodes in ${\cal N}(G)- L$.
Then we decompose the data graph $G$ into a tuple of graph segments ${\cal D}_G = (G_1, \dots, G_m)$,
such that ${\cal D}_G$ is a star-oriented decomposition of $G$ (as defined in Definition~\ref{def:stargraphpartition}).

The \emph{QE-with-Redundancy} strategy consists of the following steps:

\begin{quote}
\begin{itemize}
\item [\textbf{Step 1:}] Decompose the query $Q$ into a tuple of queries ${\cal D}_Q = (Q_1, \dots, Q_n)$,
with $n \geq 1$,  such that  each query in ${\cal D}_Q$ is a subject-object star query.

\item [\textbf{Step 2:}]  Compute all possible embeddings of each subquery in ${\cal D}_Q$ on every segment in ${\cal D}_G$.

\item [\textbf{Step 3:}] Compute the embeddings of $Q$ on $G$ by joining compatible embeddings of the subqueries $Q_1, \dots, Q_n$.
\end{itemize}
\end{quote}

It is important to note that the algorithm is independent of the choice of the specific partition ${\cal N_P}$ of the nodes in ${\cal N}(G)- L$, used for the data graph decomposition, as well as of the specific query decompotition strategy (employed in Step 1).


\subsection{Query decomposition algorithms}
\label{subsec:query-decomp-algorithms}

In this section, we present and analyze algorithms for decomposing queries into a set of so-subqueries. In the previous subsections, we assumed that the queries are decomposed into a set of subqueries, but we have not typically discussed any algorithm for finding such a decomposition, so far. Although the algorithms presented in Sections~\ref{subsec:QEJPE-algorithm},	\ref{subsec:QE-using-STARS}, and \ref{subsec:QE-redundancy} can be used to evaluate a query over a single machine, they are designed to be efficiently applied on a distributed environment, as we will see in the next sections. In this context, we focus on decomposition algorithms that can boost parallelization. Furthermore, although the QEJPE algorithm (Sections~\ref{subsec:QEJPE-algorithm}) is quite generic and can support every query decomposition, the algorithms presented in this section aim to take advantage of the special, so-queries  decomposition, which can be used in the evaluation algorithms eval-STARS (Section~\ref{subsec:QE-using-STARS}) and QE-with-Redundancy (Section~\ref{subsec:QE-redundancy}).

Intuitively, the decomposition approach followed can affect the efficiency of the overall query evaluation process, since an appropriate algorithm can significantly reduce the amount of the data transferred through the network (i.e., the amount of intermediate results). For example, in the extreme scenario that we decompose the query in Figure~\ref{fig:rdf-embedding} so that each edge defines a different subquery, it is easy to see that all the 6 edges with predicate $``hasAuthor"$ are mapped by the edge-subquery $\{(?A$, $hasAuthor$, $?W)\}$; hence, 6 embeddings are resulted by  Step 2 of the QE-with-Redundancy algorithm and passed to Step 3. If, however, we decompose the query $Q$ in such a way that at least one constant (i.e., non-variable node) is included in each subquery, the number of embeddings found in Step 2 and used in Step 3 is significantly reduced; e.g., consider the query $\{(?A,hasAuthor,?W)$, $(?A,year,2008)\}$, or $Q$ itself. Practically, the more the number of constants each query has, the less the embeddings that are found, since the constants filter out useless embeddings (i.e., partial embeddings that surely cannot be used in Step 3 to construct a total embedding). The Steps 3 and 4 of the eval-STARS algorithm operate similarly. The following proposition proves this statement.

\begin{prop}
\label{prop:star-cont}
Let $Q_1$ and $Q_2$ be two generalized star queries, such that $Q_2\subseteq Q_1$ and each triple in the set $(Q_1-Q_2)$ is either of the form $(C,p,c)$ or of the form $(c,p,C)$, where $C=C(Q_1)=C(Q_2)$, $p$ is a predicate, and $c$ is not a variable. Then, for every data graph $G$ the set of answers of $Q_1$ over $G$ is a subset of the set of answers of $Q_2$ over $G$, and $n^e_1\leq n^e_2$, where $n^e_i$ is the number of embeddings of $Q_i$ over $G$, with $i=1,\; 2$.
\end{prop}	

The proof of the previous proposition is straightforwardly given by expressing both queries as conjunctive queries and checking containment of the corresponding conjunctive queries \cite{CMSTOC77, AC2019Book}.

To maximize the number of constants in each subquery, one could come up with the following simple decomposition algorithm (called \emph{naive algorithm}).
Let $Q$ be a query.
\begin{description}
	\item[Step 1:] For each node $n$ in $\CN(Q)$, we construct the star query $Q_n$ such that $Q_n$ includes all the edges in $Q$ of either the form $(n,p,m)$ or the form $(m,p,n)$, where $m\in\CN(Q)$. Let $S_Q$ be the set including all the subqueries constructed by this process.	
	\item[Step 2:] We, then, remove from $S_Q$ the subqueries that are not so-queries.
\end{description}
It is easy to see that the remaining subqueries in $S_Q$ form a decomposition $\CD_Q$ of $Q$ that can be used in both QE-with-Redundancy and eval-STARS algorithms.

\begin{prop}
	\label{prop:naive-decomp}
	Considering a query $Q$, the naive algorithm results a decomposition $\CD_Q$ of $Q$ such that each query in $\CD_Q$ is an so-query.
\end{prop}

The proof of the Proposition~\ref{prop:naive-decomp} is straightforward since each edge of $Q$ will be at least in the subquery centered by its subject. Furthermore, it is easy to see that the naive algorithm results a query decomposition that maximizes the number of constants in each star subquery, since each subquery is constructed by a query node along with all of its adjacent edges.
%
This algorithm, however, results a quite large number of subqueries
as at the worst-case scenario one subquery for each query edge is obtained (the subject of each query edge, may introduce a new subquery),
%
and does not limit the number of variables in each subquery, which might impact the overall performance of the query evaluation, as we will see in the next sections.

In the following sections, we study two additional parameters, the number of variables into each subquery and the number of subqueries in the decomposition, in order to provide an effective decomposition approach. In particular, in Section~\ref{section:min-res-alg}, we present an algorithm that decomposes a query in a way that the number of variables do not exceed a given threshold. Section~\ref{section:non-redund-max-alg} discusses multiple algorithms that aim to reduce the number of subqueries into the decomposition.

\subsubsection{Subqueries with a limited number of variables}
\label{section:min-res-alg}
In this section, we present a decomposition algorithm that aims to keep the number of the variables in each subquery as low as possible. As Proposition~\ref{prop:star-cont} shows, by decomposing into so-subqueries with large number of constants we can achieve significant improvement in the overall performance of both QE-with-Redundancy and eval-STARS algorithms. However, we might not have the same result if we choose star subqueries with large number of variables. To see which is the impact of the number of variables into the overall query evaluation process we start our analysis with an example.

Consider the simple  query $Q=\{(c,p_1,?X),(c,p_2,?Y),(?X,p_3,?Y)\}$ and the data graph $G=\{(c,p_1,c_{11})$, $(c,p_1,c_{12})$, $(c,p_1,c_{13})$, $(c,p_2,c_{21})$, $(c,p_2,c_{22})$, $(c,p_2,c_{23})\}$, where $p_1$, $p_2$, $p_3$ are two predicates and $c$, $c_{ij}$ are either URIs or literals. Suppose now a decomposition ${\cal D}=\{Q_1,Q_2\}$ of $Q$, such that $Q_1=\{(c,p_1,?X),(c,p_2,?Y)\}$ and $Q_2=\{(?X,p_3,?Y)\}$. Notice that although the answers of both $Q$ and $Q_2$ are empty, there are 9 total embeddings from $Q_1$ to $G$, giving 9 answers. Looking at the algorithms eval-STARS (Steps 4 and 5) and QE-with-Redundancy (Steps 3 and 4), it is worth further decomposing $Q_1$ into two subqueries $Q_{11}$, $Q_{12}$, one for each edge, instead  of keeping $Q_1$ into ${\cal D}$. In particular, if we replace $Q_1$ in ${\cal D}$ with $Q_{11}$ and $Q_{12}$, the number of embeddings found in Step 4 of eval-STARS and passed to Step 5 (resp., found in Step 3 of QE-with-Redundancy and passed to Step 4) is 6, instead of 9 in the case we use $Q_1$.

In the previous example, we saw that the presence of multiple variables in a subquery might increase the number of embeddings of this subquery in both QE-with-Redundancy and eval-STARS algorithms. Especially, if we apply these algorithms into a distributed environment, as we will see in the next sections, we might have significant impact on the performance of each algorithm, since the communication cost might be increased tremendously from the large number of embeddings transferred through the network.

In this context, we present the \emph{min-res} decomposition algorithm, which finds a decomposition by keeping the number of variables into each subquery at most 2. One could wonder how we come up with the threshold number 2. Typically, we want to keep the number of variables in each subquery as low as possible. If we set such a threshold to one variable, we miss edges consisting of two variables, i.e., we cannot find a valid decomposition of any given query.

Min-res algorithm decomposes a query $Q$ into a set of so-subqueries, such that
each subquery has at most two variables. It also allows replication of triples that contains at most one variable, and maximizes the number of ``constraints'' (triples that do not increase the number of variables in the query) in each subquery containing variables. As for the subqueries that do not contain any variable, the algorithm constructs maximal subqueries without redundant constraints. The \emph{min-res} decomposition algorithm, in detail, is given as follows.
 
%
%
{\scriptsize \sl
\begin{tabbing}
aa \= aa \= aa \= aa \= aa \= aa \= aa\kill
\textbf{min-res}(Q)\\
// $Q$ a query.\\
// The min-res function returns a decomposition $R$ of $Q$ consisting of so-subqueries of $Q$\\

\textbf{begin}\\
\>  $T_{sub-obj} = \{t \in Q \mid  t = (s, p, o)$ and $s, o \in {\cal V}(Q)\}$; // subject and object are variables \\
\>  $T_{sub} = \{t \in Q \mid  t = (s, p, o)$ and $s \in {\cal V}(Q)$ and $o \not\in {\cal V}(Q)\}$; // only subject is variable \\
\>  $T_{obj} = \{t \in Q \mid  t = (s, p, o)$ and $o \in {\cal V}(Q)$ and $s \not\in {\cal V}(Q)\}$; // only object is variable \\
\>  $T_{c} = \{t \in Q \mid  t = (s, p, o)$ and $s, o \not\in {\cal V}(Q)\}$; // subject and object are nonvariables \\
\>  $R = \emptyset$;\\

\>  \textbf{foreach} $t = (s, p, o) \in T_{sub-obj}$ \textbf{do} // select a maximal so-query centered at subject
of t\\
\>  \>  \textbf{begin} \hskip 3cm  // by adding triples that do not add variables\\
\>  \>  \>  $Q_s = \{t\} \cup \{t' \mid  t' \in T_{sub}$ and $t' = (s, p', c)\} \cup \{t'' \mid  t'' \in T_{obj}$ and $t'' = (c', p'', s)\}$;\\
\>  \>  \>  \> // $Q_s$ is an so-query with central node s \\
\>  \>  \> $S = \{t' \mid  t' \in T_{sub}$ and $t' = (o, p', c)\}$; \\
\>  \>  \> \textbf{If} $S = \emptyset$ \textbf{then} $Q_o =  \emptyset$  \textbf{else} \\
\>  \>  \> \> $Q_o = \{t\} \cup S \cup \{t'' \mid  t'' \in T_{obj}$ and $t'' = (c', p'', o)\}$;\\
\>  \>  \>  \>  \> // $Q_o$ is an so-query with central node o\\
\>  \>  \>  \textbf{If} $|Q_s| \geq |Q_o|$ \textbf{then} $Q' = Q_s$ \textbf{else} $Q' = Q_o;$\\
\>  \>  \>  $R = R \cup \{Q'\}$;\\
\> \>  \textbf{end}\\

\> $T_{sub} = T_{sub} - \{t \mid  t \in Q'$ and $Q' \in R\}$;   // Remove from $T_{sub}$ the triples used so far\\
\>  $T_{obj} = T_{obj} - \{t \mid  t \in Q'$ and $Q' \in R\}$; // Remove from $T_{obj}$ the triples used so far\\

\>  \textbf{while} $T_{sub} \neq \emptyset$ \textbf{do}  //For each member of $T_{sub}$ construct an so-query  \\
\>  \>  \textbf{begin}\\
\>  \>  \>  extract a triple $t = (s, p, o)$ from $T_{sub}$;\\
\>  \>  \>  $Q' = \{t\} \cup \{t' \mid  t' \in T_{sub}$ and $t' = (s, p', c)\} \cup \{t'' \mid  t'' \in T_{obj}$ and $t'' = (c', p'', s)\}$;\\
\>  \>  \>  \textbf{If} $|Q'| = 1$ \textbf{then} // No other triple has s as object or subject\\
\>  \>  \>   \>  \textbf{begin}\\
\>  \>  \>  \>  \> $S = \{t' \mid  t' \in T_{c}$ and $t' = (o, p', c)\}$;\\
\>  \>  \>  \>  \> \textbf{If} $S \neq \emptyset$ \textbf{then} \\
\>  \>  \>   \>  \>   \>$Q' = \{t\} \cup S \cup \{t'' \mid  t'' \in T_{c}$ and $t'' = (c', p'', o)\}$;\\
\>  \>  \>   \>  \textbf{end}\\
\>  \>  \>  $T_{sub} = T_{sub} - \{t \mid  t \in Q'\}$; // Remove from $T_{sub}$ the triples used in $Q'$\\
\>  \>  \>  $T_{obj} = T_{obj} - \{t \mid  t \in Q'\}$; // Remove from $T_{obj}$ the triples used in $Q'$\\
\>  \>  \>  $R = R \cup \{Q'\}$;\\
\> \>  \textbf{end}\\

\>  \textbf{foreach} $t = (s, p, o) \in T_{obj}$  \textbf{do} //For each member of $T_{obj}$ construct an so-query  \\
\>  \>  \textbf{begin}\\
 \>  \>   \>    $Q' = \{t\} \cup \{t' \mid  t' \in T_{c}$ and $t' = (s, p', c)\} \cup \{t'' \mid  t'' \in T_{c}$ and $t'' = (c', p'', s)\}$;\\
%
\>  \>  \>  $R = R \cup \{Q'\}$;\\
\> \>  \textbf{end}\\

\> $T_{c} = T_{c} - \{t \mid  t \in Q'$ and $Q' \in R\}$;\\

\>  \textbf{while} $T_{c} \neq \emptyset$  \textbf{do}  // select maximal so-query centered at subject or object\\
\>  \>  \textbf{begin}\\
\>  \>  \>     extract a triple $t = (s, p, o)$ from $T_{c}$;\\
\>  \>  \>  $Q_s = \{t\} \cup \{t' \mid  t' \in T_{c}$ and $t' = (s, p', c)\} \cup \{t'' \mid  t'' \in T_{c}$  and $t'' = (c', p'', s)\}$;\\
\>  \>  \>   $S = \{t' \mid  t' \in T_{c}$ and $t' = (o, p', c)\}$; \\
\>  \>  \>  \textbf{If} $S = \emptyset$ \textbf{then} $Q_o =  \emptyset$  \textbf{else} \\
\>  \>  \>   \>  \> \>  $Q_o = \{t\} \cup S  \cup \{t'' \mid  t'' \in T_{c}$ and $t'' = (c', p'', o)\}$;\\
\>  \>  \>     \textbf{If} $|Q_s| \geq |Q_o|$ \textbf{then} $Q' = Q_s$ \textbf{else} $Q' = Q_o;$\\
\>  \>  \>  $R = R \cup \{Q'\}$;\\
\>  \>  \>  $T_{c} = T_{c} - \{t \mid  t \in Q'\}$; // Remove from $T_c$ the triples used in $Q'$\\
\> \>  \textbf{end}\\
\> \>  \textbf{return} $R$;\\
\> \textbf{end.}
\end{tabbing}.
}

Intuitively, the algorithm performs as follows. Let $Q$ be a query. Initially, for each edge $t$ of two variables in $Q$, it constructs an so-query $Q_s$  having the subject $s$ of $t$ as central node. All the adjacent edges of $s$ in $Q$ such that $s$ is their only variable are added into the subquery.
In each construction step, the possibility to get an so-query $Q_o$, whose central node is the object of $t$, is also considered,
and the query with maximum number of edges between $Q_s$ and $Q_o$ is finally selected.
It is easy to see that the subqueries constructed in this step include 2 variables. Next, the algorithm constructs the subqueries that have a single variable as central code. These subqueries have at least one edge whose subject is the central variable. Then, the remaining query triples give so-subqueries whose central node is not a variable. Each of the subqueries constructed in this step have at most a single variable which is not the central node. Notice here that the min-res algorithm constructs subqueries with two variables only if those variables are used by an edge in $Q$. Note also that in each of the aforementioned steps, we build an so-subquery by initially selecting an edge from a set (e.g., the set of edges having two variables). The order the edges are selected might give different decompositions. Here, we consider an arbitrary ordering of the edges included in each set. The following proposition shows that the min-res algorithm results a decomposition of a query into a set of so-queries.

\begin{prop}
	\label{prop:min-res-decomp}
	Considering a query $Q$, the min-res algorithm results a decomposition $\CD_Q$ of $Q$ such that each query in $\CD_Q$ is an so-query.
\end{prop}

\begin{pf}
	Consider the sets  $T_{sub-obj}$, $T_{sub}$, $T_{obj}$, $T_{c}$ of edges as defined in the min-res algorithm. To prove that $R$ is a decomposition of $Q$, we need to show that (1) each query $Q_i$ in $R$ is a subquery of $Q$, and (2) $\bigcup_{Q_i\in R}Q_i = Q$. The first condition is straightforward since each edge of $Q_i$ is an edge in $T_{sub-obj}\cup T_{sub}\cup T_{obj}\cup T_{c}$, which equals $Q$. To show the second condition, we need to prove that each edge $t$ of $Q$ is included in at least one subquery in $R$. Since the algorithm uses all the edges in $T_{sub-obj}\cup T_{sub}\cup T_{obj}\cup T_{c}$ to construct the subqueries, we have that $t$ is included in at least one subquery in $R$.
Besides, it is easy to see that, by construction, all queries in $\CD_Q$  are so-queries.\qed
	\end{pf}

\begin{figure*}[htb]
\centering
\includegraphics[width=0.80\textwidth]{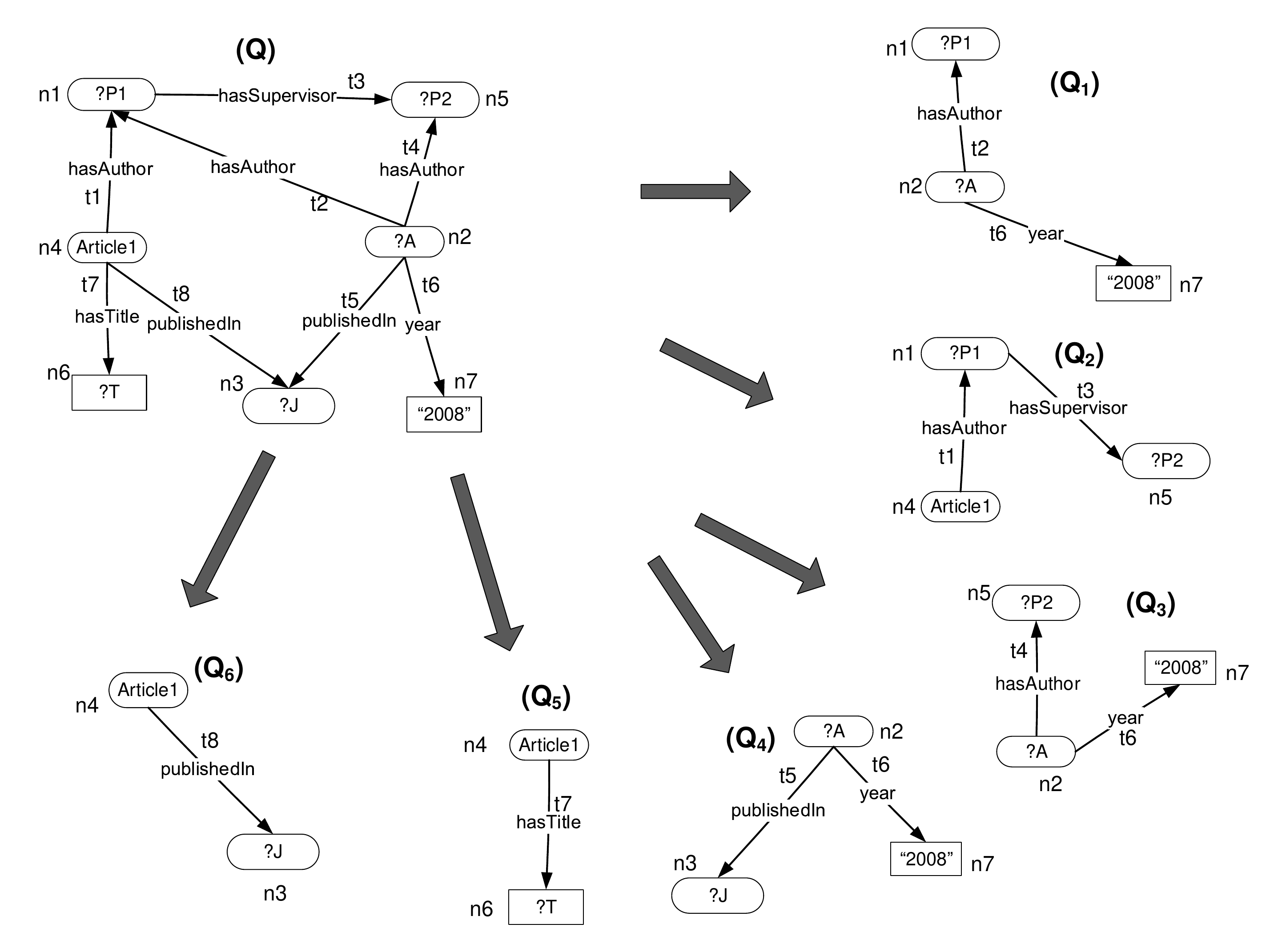}
\caption{Min-res Query decomposition.\label{fig:min-res-decomp}}
\end{figure*}

\begin{exmpl}
	\label{exam:min-res-decomp}
	Consider the query $Q$ depicted in Figure~\ref{fig:rdf-query-star-decomp}. Figure~\ref{fig:min-res-decomp} illustrates a decomposition of $Q$ resulted by the min-res algorithm. In particular, we initially select the edge $t2$ and construct $Q_1$. Similarly, the queries $Q_2$, $Q_3$ and $Q_4$ are given by selecting the edges $t3$, $t4$ and $t5$, respectively, having two variables, as well.  $Q_5$ and $Q_6$ are then constructed by selecting the corresponding edges of $Q$. Note here that the edge $t6$ is replicated to multiple subqueries, as min-res algorithm requires, since it can reduce the number of intermediate answers through the constant $``2008"$.
\hfill$\Box$
\end{exmpl}

\subsubsection{Reducing the number of subqueries}
\label{section:non-redund-max-alg}
Unlike the min-res algorithm which minimizes the number of variables in each subquery, in this section, we investigate
algorithms that keep the number of so-subqueries as low as possible, as well as select so-subqueries with high degree. As we will see in the following, there are settings where the number of queries in the decomposition affects the overall performance of the query evaluation, since large number of subqueries might increase the amount of intermediate results. The following example presents such a case.

\begin{figure*}[htb]
	\centering
	\includegraphics[width=0.90\textwidth]{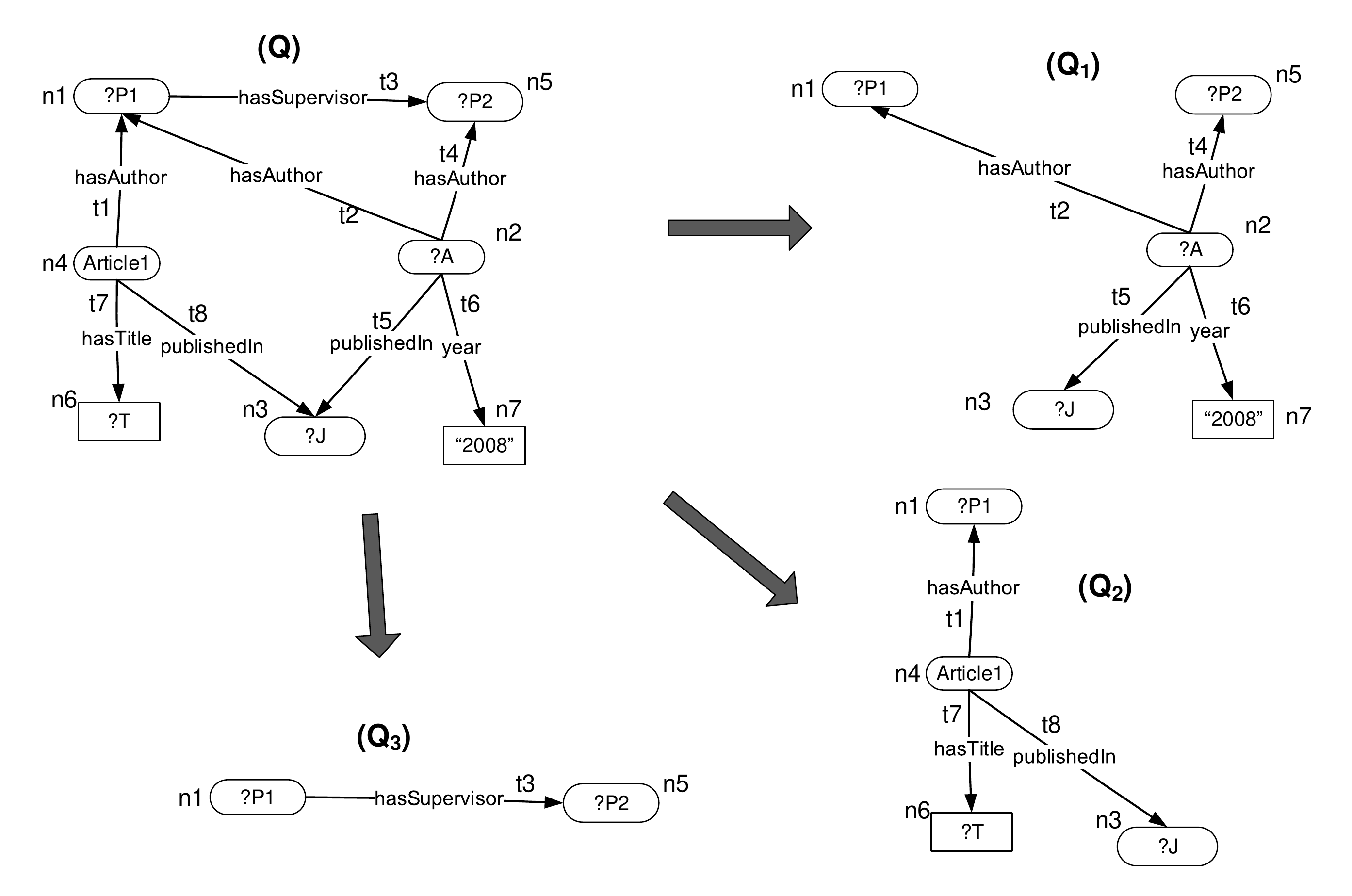}
	\caption{Min-subquery decomposition.\label{fig:min-subqueries-decomp}}
\end{figure*}

\begin{exmpl}
\label{exam:min-subqueries-decomp}
Consider the query $Q$ and the data graph $G$ depicted in Figure~\ref{fig:rdf-query-star-decomp} and Figure~\ref{fig:rdf-embedding}, respectively. It is easy to see that there is a single total embedding from $Q$ to $G$. Suppose two decompositions $\CD_Q^1$ and $\CD_Q^2$ illustrated in Figure~\ref{fig:min-res-decomp} and Figure~\ref{fig:min-subqueries-decomp}, respectively. As we saw in Example~\ref{exam:min-res-decomp}, $\CD_Q^1$ is resulted by min-res algorithm. Counting now the embeddings found for the subqueries of each decomposition over $G$, we have that there are $12$ embeddings, in total, from  queries in $\CD_Q^1$ to $G$, while $\CD_Q^2$ gives $10$ embeddings. Hence, we can see that although each subquery in $\CD_Q^1$ has minimum number of variables, the total number of embeddings is high, due to the large number of subqueries.
\hfill$Box$
\end{exmpl}

To construct a decomposition with minimum number of subqueries, we follow an approach based on the naive algorithm. In particular, considering a query $Q$, a simple algorithm, called \emph{min-subquery} decomposition algorithm, computing a decomposition with minimum number of so-subqueries is given as follows.
\begin{description}
	\item[Step 1:] We initially apply the naive algorithm and get a decomposition $\CD_N$.
	\item[Step 2:] Then, we construct the set $\CS$ including all the subsets $\CD$ of $\CD_N$ such that the queries in $\CD$ cover all the edges of $Q$; i.e., $\bigcup_{Q_i\in\CD}(Q_i)=Q$.
	\item[Step 3:] Finally, we find the sets in $\CS$ with the minimum number of subqueries and output one of them.
\end{description}
It is easy to see that the min-subquery algorithm returns a decomposition with minimum number of so-subqueries. Note that there might be multiple decompositions that minimize the number of so-subqueries.

\begin{prop}
	\label{prop:min-subquery-decomp}
	Considering a query $Q$, the min-subquery algorithm results a decomposition $\CD_Q$ of $Q$ such that each query in $\CD_Q$ is an so-query and $\CD_Q$ has the minimum number of so-subqueries, among all the decompositions of $Q$ including so-subqueries.
\end{prop}

The proof of the previous proposition follows by the Proposition~\ref{prop:naive-decomp}. The last step of the algorithm also ensures that the output has the minimum number of so-subqueries.

Although the min-subquery algorithm returns a minimal decomposition, it applies an exhaustive search over the search space and the resulted decomposition has high redundancy (i.e., there are triples that are included in two subqueries). Especially, if the replicated edges include variables, as we saw in the previous section, the amount of the intermediate results could affect the overall evaluation time. To overcome these issues, we focus on an efficient approach that constructs a decomposition based on the nodes' degree. In particular, we focus on selecting first the subqueries containing as many triples as possible. In addition, each query triple is included in a unique so-subquery (i.e. redundancy is not allowed in query decomposition). The decomposition algorithm, called \emph{max-degree}, that follows this approach is given below.


{\scriptsize \sl
	\begin{tabbing}
		aa \= aa \= aa \= aa \= aa \= aa  \= aa \= aa  \= aa \= aa \kill
		\textbf{max-degree}(Q)\\
		// $Q$ a query.\\
		// max-degree function returns a decomposition $R$ of $Q$ consisting of so-subqueries \\
		
		\textbf{begin}\\
		\> $R = \emptyset$;\\
		\> $N = {\cal N}(Q) - L$; // The non-literal nodes.\\
		
		\> $S_Q = FindInitMaxSoQueries(N)$\\
		\> $T_{Covered} = \emptyset$;\\
		\> \textbf{while} $S_Q \neq \emptyset$ \textbf{do}\\
		\> \>\textbf{begin}\\
		\> \> \> $S_Q^{old}$, $R$, $T_{Covered}$ = $ReconstructS_Q(S_Q,\; R,\; T_{Covered})$;\\
		\> \> \> $S_Q =  \emptyset$;\\
		\>  \> \> \textbf{foreach} $(m,S) \in S_Q^{old}$ \textbf{do}\\
		\>  \> \> \> \textbf{begin}\\
		\> \> \> \> \> $S' = S - T_{Covered}$;\\
		\> \> \> \> \> \textbf{if}  there is a triple with subject $m$ in $S'$ \textbf{then} $S_Q = S_Q \cup \{(m,S')\}$;\\
		\> \> \> \> \textbf{end}\\
		\> \> \textbf{end}\\
		\>  \textbf{return} $R$;\\
		\textbf{end}.
	\end{tabbing}
}

{\scriptsize \sl
	\begin{tabbing}
		aa \= aa \= aa \= aa \= aa \= aa  \= aa \= aa  \= aa \= aa \kill
		\textbf{$FindInitMaxSoQueries$}(N)\\
		// $N$ is a set of the non-literal nodes of a query\\
		
		\textbf{begin}\\
		\> $S_Q = \emptyset$;\\
		\> \textbf{foreach} $n \in N$ \textbf{do} // $S_Q$ contains all pairs $(n, S(n))$ where $n \in N$ and\\
		\> \> \textbf{begin} \hskip 2cm // S(n) is the maximal so-query with n as central node.\\
		\> \> \> \textbf{if} there is a triple $(n, p, o) \in Q$ \textbf{then} \\
		\> \> \> \> \textbf{begin}\\
		\> \> \> \> \>    $S(n) = \{t \mid  t \in Q$ and $(t = (n, p, o)$ or $t = (s, p', n)) \}$;\\
		\> \> \> \> \> $S_Q = S_Q \cup \{(n, S(n))\}$;\\
		\> \> \> \> \textbf{end}\\
		\> \>\textbf{end}\\
		\> \textbf{return} $S_Q$;\\
		\textbf{end}.
	\end{tabbing}
}

{\scriptsize \sl
	\begin{tabbing}
		aa \= aa \= aa \= aa \= aa \= aa  \= aa \= aa  \= aa \= aa \kill
		\textbf{$ReconstructS_Q$}($S_Q$, $R$, $T_{Covered}$)\\
		//Find next so-subquery and update both the result $R$ and the set $T_{Covered}$ of covered edges. \\
		\textbf{begin}\\
		\>  //select a maximal so-query in $S_Q$\\
		\>  select a $(n, Q') \in S_Q$ such that $|Q'|$ is maximal among all elements in $S_Q$. \\
		\> $R = R \cup \{Q'\}$; // ... add $Q'$ to the result and ...\\
		\> $T_{Covered} = T_{Covered} \cup Q'$; // ... add its triples to $T_{Covered}$.\\
		\>  $S_Q^{old} =  S_Q - \{(n,Q')\}$;\\
		\> \textbf{return} $S_Q^{old}$, $R$, $T_{Covered}$;\\
		\textbf{end}.
	\end{tabbing}
}

\begin{figure*}[htb]
	\centering
	\includegraphics[width=0.90\textwidth]{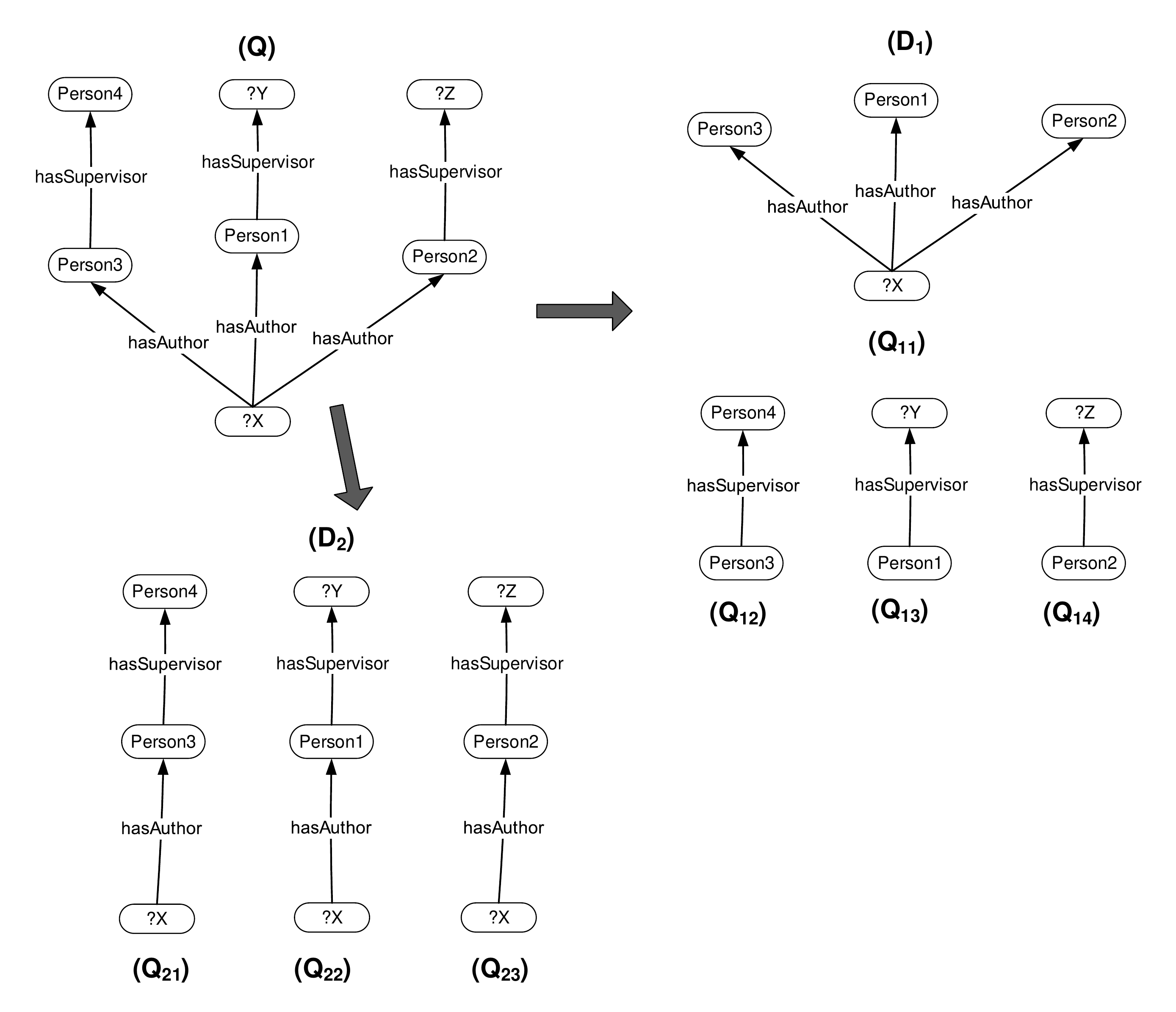}
	\caption{Min-subquery vs. Max-degree decomposition.\label{fig:non-redundant-max-decomp}}
\end{figure*}

Intuitively, the max-degree algorithm performs similarly to the min-subquery algorithm. In particular, in each step, it finds an so-query with max degree and removes its edges from the remaining so-stars. The algorithm stops once all the query edges are covered. Note that the max-degree algorithm does not aim to minimize the number of subqueries in the decomposition.
An example describing such a case is illustrated in Figure~\ref{fig:non-redundant-max-decomp}. Notice that $D_1$ and $D_2$ are two decompositions of $Q$, where $D_1$ is the result of the max-degree algorithm and $D_2$ is the result of min-subquery algorithm. On the other hand, there are queries where the results of both algorithms match. Such an example is illustrated in Figure~\ref{fig:min-subqueries-decomp}.

As we mentioned above,  the max-degree decomposition algorithm does not apply any edge replication (no redundant edges are allowed). Lack of replication might improve the performance of finding the embeddings of a subquery, since less edges are checked in order to find an embedding. However, as we saw in Proposition~\ref{prop:star-cont}, replicating edges that add more constraints in the subquery might decrease the total number of embeddings of subqueries; hence, it may also decrease the communication cost in a distributed execution. Taking this into account, we present a modification of the max-degree algorithm, called \emph{max-degree-with-redundancy}, which replicates edges with constants.
The max-degree-with-redundancy algorithm is given by replacing the function $ReconstructS_Q$ in max-degree algorithm with the function $ReconstructS_QRedundancy$, which is defined as follows.

{\scriptsize \sl
	\begin{tabbing}
		aa \= aa \= aa \= aa \= aa \= aa  \= aa \= aa  \= aa \= aa \kill
		\textbf{$ReconstructS_QRedundancy$}($S_Q$, $R$, $T_{Covered}$)\\
		//Find next so-subquery and update both the result $R$ and the set $T_{Covered}$ of covered edges. \\
		\textbf{begin}\\
		\>  //select a maximal so-query in $S_Q$\\
		\>  select a $(n, Q') \in S_Q$ such that $|Q'|$ is maximal among all elements in $S_Q$. \\
		\>  //add triples that have already covered and do not add any new variable to the subquery found\\
		\>  $Q'' = Q'\cup\{t=(n_1,p,n_2)|t\in (Q-Q')$, and either $n_1=n$ and $n_2\notin\CV(Q)$ or $n_2=n$ and $n_1\notin\CV(Q)\}$;\\
		\> $R = R \cup \{Q''\}$; // ... add $Q''$ to the result and ...\\
		\> $T_{Covered} = T_{Covered} \cup Q'$; // ... add its triples to $T_{Covered}$.\\
		\>  $S_Q^{old} =  S_Q - \{(n,Q')\}$;\\
		\> \textbf{return} $S_Q^{old}$, $R$, $T_{Covered}$;\\
		\textbf{end}.
	\end{tabbing}
}

Comparing the max-degree and max-degree-with-redundancy algorithms,
we can easily see that the function $ReconstructS_QRedundancy$ used in the max-degree-with-redundancy algorithm to construct each subquery and add it into the resulting set $R$, constructs each subquery $Q''$ from $Q'$ (which is similar to the query $Q'$ constructed by $ReconstructS_Q$ in the max-degree algorithm) and all the query triples in $Q$ that either start or end to the central node of $Q'$ and do not include a variable in the other node; i.e., the triples having constants in the non-central node are replicated and reused. In the contrary, the max-degree algorithm (function
$ReconstructS_Q$) does not replicate any edge during construction of the result.

\begin{prop}
\label{prop:max-degree-decomp}
Considering a query $Q$, the results of both the max-degree and max-degree-with-redundancy algorithms are decompositions of $Q$ that include only so-queries.
\end{prop}

\begin{pf}
By construction all subqueries produced in both algorithms are so-queries. Besides, as all query triples are used the algorithms produce decompositions of $Q$. \qed
\end{pf}

As we mentioned above the main difference between max-degree and max-degree-with-redundancy algorithms is that in the latter, we replicate edges that have constants in the adjacent nodes of the central node. As Proposition~\ref{prop:max-degree-vs-max-degree-with-redundancy} shows, the decomposition resulted by the max-degree-with-redundancy algorithm might reduce the number of embeddings exchanged between the last two steps of the evaluation algorithms eval-STARS and QE-with-Redundancy, comparing to the corresponding decomposition resulted by the max-degree algorithm.

\begin{prop}
\label{prop:max-degree-vs-max-degree-with-redundancy}
Let $Q$  be a query and $\CD_M$ be a decomposition of $Q$ resulted by the max-degree. Then, there is a decomposition $\CD_R$ of $Q$ resulted by the max-degree-with-redundancy algorithm such that the following hold:
\begin{itemize}
\item there is an one-to-one mapping $\mu$ from $\CD_M$ to $\CD_R$ such that $\mu(P_R)=P_M$, if $P_M\subseteq P_R$, where $P_R\in\CD_R$ and $P_M\in\CD_M$; and
\item for each data graph $G$ and every query $P$ in $\CD_R$, the number of embeddings of $P$ over $G$ is less than or equal to the  number of embeddings of $\mu(P)$ over $G$.
\end{itemize}
\end{prop}

\begin{pf} Let $Q$  be a query and $\CD_M$ be a decomposition of $Q$ resulted by the max-degree. We now need to prove that the max-degree-with-redundancy algorithm can result a decomposition of $Q$ which satisfies the aforementioned properties.

Each subquery $Q'$ in $\CD_M$ is constructed by the function $ReconstructS_Q$, and specifically, once it is constructed it is inserted into the resulting set $R$ (which eventually equals $\CD_M$). Let's consider that instead of returning the subquery $Q'$ into the result $R$, we return the subquery $Q'' = Q'\cup\{t=(n_1,p,n_2)|t\in (Q-Q')$, and either $n_1=n$ and $n_2\notin\CV(Q)$ or $n_2=n$ and $n_1\notin\CV(Q)\}$, where $n$ is the central node of $Q'$. Since $R$ simply stores the resulting subqueries and is not used in any other step of the algorithm, such a modification does not affect the construction of the subqueries. It is easy to verify that the modified function is given by the function $ReconstructS_QRedundancy$, and the modified algorithm is the max-degree-with-redundancy. Let also $\CD_R$ be the result of the modified algorithm (i.e., the final set $R$ returned by the algorithm); hence, $\CD_R$ is the result of the max-degree-with-redundancy.

According to the previous modification, for each query $Q'$ in $\CD_M$, there is a query $Q''$ in $\CD_R$, such that
$Q'' = Q'\cup\{t=(n_1,p,n_2)|t\in (Q-Q')$, and either $n_1=n$ and $n_2\notin\CV(Q)$ or $n_2=n$ and $n_1\notin\CV(Q)\}$, where $n$ is the central node of $Q'$. Hence, there is an one-to-one mapping $\mu$ from $\CD_M$ to $\CD_R$ such that $Q''=\mu(Q')$ and $Q'\subseteq Q''$; which proves the first condition of the proposition.

Furthermore, Proposition~\ref{prop:star-cont} and the construction of $Q''$ from $Q'$ imply that for each data graph $G$, the number of embeddings of $Q''$ over $G$ is less than or equal to the number of embeddings of $Q'$ over $G$; which means that the second property is also satisfied. Hence, the decomposition $\CD_R$ satisfies both properties of the Proposition~\ref{prop:max-degree-vs-max-degree-with-redundancy}.
\qed
\end{pf}


As we have seen,
both max-degree and max-degree-with-redundancy algorithms iterate over the maximal so-subqueries found by the  $FindInitMaxSoQueries$ function, from the queries of max degree to the queries with min degree, and remove triples covered in the previous iterations. In each iteration, if the query resulted by removing the covered triples is not an so-query, then both the algorithms ignore this query and continue to the next iteration. Let $Q_R$ be the set of the remaining triples, in such cases. Note that $Q_R$ will be covered in the next iterations, but the number of iterations might increase due to the triples that do not form an so-query in some iterations. To reduce the number of iterations, we can construct so-queries from the triples in $Q_R$ by adding to $Q_R$ a triple that makes it so-query. Such a triple $t$ is found in the set of covered triples. In addition, to avoid replicating triples that add variables to a query, we remove the triple $t$ from the so-query that was constructed in the previous iterations. Such an approach might reshape the so-queries constructed in the previous iterations. A decomposition algorithm following this approach is given as follows, and called \emph{max-degree-with-reshaping}.

{\scriptsize \sl
\begin{tabbing}
aa \= aa \= aa \= aa \= aa \= aa  \= aa \= aa  \= aa \= aa \kill
\textbf{max-degree-with-reshaping}(Q)\\
// $Q$ a query.\\
// The function returns a decomposition $R$ of $Q$ consisting of so-subqueries of $Q$\\

\textbf{begin}\\
\> $R = \emptyset$;\\
\> $N = {\cal N}(Q) - L$; // The non-literal nodes.\\

\> $S_Q = FindInitMaxSoQueries(N);$\\
\> $T_{Covered} = \emptyset$;\\
\> \textbf{while} $S_Q \neq \emptyset$ \textbf{do}\\
\> \>\textbf{begin}\\
\> \> \>select a $(n, Q') \in S_Q$ s.t. $\forall (m, Q'') \in (S_Q - \{(n,Q')\})$ it holds $NC(Q') \geq NC(Q'')$; \\
\hskip 3cm // $NC(Q)$ function returns the number of Not Covered triples in $Q$\\
\> \> \> $S' = Q' - \{t \mid  t = (s, p, n) \in Q'$ and  $t \in T_{Covered}$ and $s \in {\cal V}(Q')\}$;\\
\hskip 3cm //i.e. remove covered triples whose object is $n$ that add variable in $Q'$\\
\> \> \> $T = \{t \mid  t = (n, p, o) \in Q'$ and $t \in T_{Covered}$ and $o \in {\cal V}(Q')\}$;\\
\> \> \>\textbf{If} $S' - T$ is so-query \textbf{then} $S = S' - T$ \textbf{else} \\
\> \> \> \> \textbf{begin}\\
\> \> \> \> \> $S = S' - T \cup \{t'\}$ where $t'$ is a triple in $T$;\\
\> \> \> \> \> replace $F$ by $F - \{t'\}$ in $R$ where $F$ is the query in $R$ containing $t'$;\\
\> \> \> \> \> // notice that $F - \{t'\}$ is also an so-query\\
\> \> \> \> \textbf{end}\\
\\

\> \> \> $R = R \cup \{S\}$; // ... add $S$ to the set of subqueries of $Q$ ...\\
\> \> \> $T_{Covered} = T_{Covered} \cup S$; // ... add the triples of $S$ to $T_{Covered}$.\\
\> \> \> $S_Q^{old} =  S_Q - \{(n,Q')\}$;\\
\> \> \> $S_Q =  \emptyset$;\\
\> \> \> \textbf{foreach} $(m,S) \in S_Q^{old}$ \textbf{do} //Reconstruct $S_Q$ by removing the queries whose triples  are ... \\
\> \> \> \> \textbf{begin} //... completely covered by the so-queries already constructed ...\\
\> \> \> \> \> \textbf{if} $S - T_{Covered} \neq \emptyset$  \textbf{then} $S_Q = S_Q \cup \{(m,S)\}$;\\
\> \> \> \> \textbf{end}\\
\> \> \textbf{end}\\
\> \textbf{return} R;\\
\textbf{end}.
\end{tabbing}
}

\begin{prop}
\label{prop:max-degree-vs-max-degree-with-reshaping}
Considering a query $Q$, the results of max-degree-with-reshaping algorithm are decompositions of $Q$ that include only so-queries.
\end{prop}

\begin{pf}
	By construction all subqueries produced in the algorithm are so-queries. Besides, as all query triples are used the algorithm produces decompositions of Q.\qed
\end{pf}

\section{Distributed query evaluation algorithms using  MapReduce}
\label{sec:implementations}

In this section, we present a set of distributed algorithms implementing the query evaluation approaches presented in Section 4. These algorithms take advantage of the commutation power provided by the MapReduce computation framework.

\subsection{The MapReduce framework}
\label{sec:MapReduce}

MapReduce is a programming model for processing large datasets in a distributed manner.
It is based on the definition of two functions, the \textit{Map} and the \textit{Reduce} function.
The storage layer for the MapReduce framework is a Distributed File System (DFS), such as Hadoop Distributed File System (HDFS), and is characterized by the block/chuck size (the chunk size, which is larger than the chuck size in conventional file systems, is typically 16-128MB in most of DFSs) and the replication of chunks in relatively independent locations to ensure availability.
Creating a MapReduce job is straightforward. Briefly, the user defines the functions, which run in each cluster node, in isolation.
The map function is applied on one or more files, in DFS, and results \textsf{[key,value]} pair. This process is called \textit{Map process/task}.
The nodes that run the Map processes are called \textit{Mappers}, and may run multiple tasks over different input files.
The \textit{master controller} is responsible to route the pairs to the \textit{Reducers} (i.e., the nodes that apply the reduce function on the pairs) so that all pairs with the same key initialize a single reduce process, called \textit{reduce task}.
The reduce tasks apply the reduce function on the input pairs and result \textsf{[key,value]} pairs; which are stored in the DFS. This procedure describes one \textit{MapReduce step}. Furthermore, the output of the reducer can be set as the input of a map function, which gives to the user the flexibility to create pipelines of multiple steps.

\subsection{Overall methodology}
\label{c-proproc}

Before describing the query MapReduce query evaluation algorithms, we focus on presenting the main patterns used to construct these algorithms. In particular, the algorithms presented in the upcoming sections are based on the following patterns:

\begin{enumerate}
\item \textbf{Data graph decomposition:} The  data graph $G$ is decomposed into a set of data segments according to a given decomposition approach. The data graph segments are stored in the nodes of a cluster of commodity computers.
\item \textbf{Storage of the data graph segments:}
A generic methodology for storing the data graph segment is used. Such an approach focuses on storing the RDF data into simple text files in N-triple format. Each file also includes the set of border nodes of the segment represented by the triples in the file. Although the segments are stored in simple text files, relational, NoSQL and graph databases could be used, instead, for storing the corresponding segments. Especially the use of multiple relational databases to store the data segments
can facilitate the implementation of certain algorithms, but it has a significant impact on the scalability, and fault tolerance.

\item \textbf{Query graph decomposition:} The query graph $Q$ is decomposed into a tuple of subqueries $(Q_1, \dots, Q_n)$, with $n\geq 1$, according to the principles specified in the definition of the corresponding algorithm.

\item \textbf{Implementing the query evaluation algorithm:} The proposed query evaluation algorithms are implemented in the MapReduce programming framework.
In general, the implementation of each algorithm consists of \emph{a preprocessing phase} followed by \emph{two MapReduce phases} (see next section).
\end{enumerate}

\subsection{Preprocessing Phase}
\label{subsec:preproc}

As mentioned earlier, all the query evaluation algorithms presented in the subsequent sections consider a \emph{preprocessing phase}, where the setting is prepared.
In particular, the pre-processing phase accepts a query $Q$ which is posed by the user
and decomposes it into a tuple of subqueries $(Q_1, \dots, Q_n)$, with $n\geq 1$,
following the decomposition principles determined by the specific query evaluation algorithm.
These subqueries broadcasted or distributed to the mappers of the first MapReduce phase of the query evaluation algorithm.

Preprocessing phase also constructs some auxiliary structures and emits them to the mappers/reducers that implement the algorithm.
To define these structures we assume an enumeration $n_1$, $n_2$, $\dots$, $n_{|{\cal N}(Q)|}$ of the nodes of the query $Q$, so that
$n_1, n_2, \dots, n_{|{\cal B}(Q)|}$ are the border nodes of $Q$ and $n_{|{\cal B}(Q)|+1}$, $\dots$, $n_{|{\cal N}(Q)|}$ are the non-border nodes of $Q$.
We denote by $I$ the function that gives the index of a
node in ${\cal N}(Q)$ with respect to the above enumeration (that is, for
every $x \in {\cal N}(Q)$ it holds $x = n_{I(x)}$). We also denote by
$I_{nb}$ the function from ${\cal N}(Q) - {\cal B}(Q)$ to $\{1,\dots, |{\cal N}(Q) -
{\cal B}(Q)|\}$, with $I_{nb}(x)=I(x)-|{\cal B}(Q)|$. Similarly, we assume
an enumeration $t_1, t_2, \dots, t_{|Q|}$ of the triples in $Q$.
Using the above enumeration functions we now define the concept of \emph{query prototype}.
A query prototype is a triple of tuples of the form:

$(BorderNodeFlags, NonBorderNodeFlags, TripleFlags)$

where $BorderNodeFlags$ is a tuple of $|{\cal B}(Q)|$ items, one item for each border node in ${\cal B}(Q)$. Similarly, the  $NonBorderNodeFlags$ is a tuple of $|{\cal N}(Q) - {\cal B}(Q)|$ items, one for each non border node in ${\cal N}(Q) - {\cal B}(Q)$. Finally, the tuple $TripleFlag$ has $|Q|$ items, one for each triple in $Q$.

Consider now that a query prototype is assigned to each (sub)query $Q_i$.
Each item in the tuples of the prototype has either the value '+' to denote the presence of the corresponding border node/non-border node/triple,  in $Q_i$, or the value '-' to denote the absence of that node or triple.

We also construct a set\footnote{In the algorithms presented in this section we represent the MBN set as list.} called \emph{Missing Border Nodes (MBN)} as follows: $MBN$ = $\{$($b_i$, $Q_j$)$|$ $b_i \in {\cal B}(Q)$ and $b_i \not\in {\cal N}(Q_j)\}$.
An element $(b_i, Q_j)$ in MBN denotes that the border node $b_i$ of $Q$ does not appear among the nodes of the subquery $Q_j$ of $Q$.

Based on the idea of query prototype we can represent a partial or total embedding $e$ of a (sub)query in a similar way; i.e. as a triple of tuples of the form $(BorderNodeValues, NonBorderNodeValues, TriplesMatched)$.
More specifically, $BorderNodeValues$  stores the images of the border nodes of the query through the (partial)embedding, while $NonBorderNodeValues$  stores the images of non border nodes of the query.
The star symbol ('*') is placed in the corresponding node place if no image of that node is defined in $e$.
Finally, $TriplesMatched$ keeps track of the triples of the query that have images on the data graph through the  (partial) embedding $e$ (by putting a '+' sign or a '-' sign in the corresponding place of $TriplesMatched$).

\subsection{QEJPE-algorithm}
\label{subsec:QEJPE-algorithm-impl}


In this section we present an implementation of the query evaluation algorithm (QEJPE-algorithm) presented in Subection~\ref{subsec:QEJPE-algorithm}. The implementation is based on  the MapReduce programming framework.
Besides the general assumption on which all algorithms are based, we have the following specific assumptions of the present algorithm:

\begin{enumerate}
\item In this algorithm both the decomposition of the data graph $G$ and the query graph $Q$ may be redundant or non-redundant.
The query graph $Q$ is decomposed into a tuple of arbitrary subqueries $(Q_1, \dots, Q_n)$, with $n\geq 1$.

\item The implementation of the algorithm
consists of a preprocessing phase followed by two map-reduce phases:

\begin{enumerate}
\item In the first map-reduce phase the subqeries are applied to each graph segment, in isolation, and intermediate results
are computed. More specifically, the mappers of phase 1 compute useful (partial or total)
embeddings of the subqueries, by applying each subquery to each specific graph segment.
Then the reducers of phase 1 combine (i.e. join) the partial embeddings to compute the total embeddings of each subquery.
Notice that the total embeddings of the subqueries are, in general, partial embeddings of the query $Q$ to the graph $G$.

\item In the second map-reduce phase, the embeddings of the subqueries are combined appropriately  to produce the embedding of the query $Q$ on the graph $G$.
More specifically, the mapper of phase 2 fills the missing border nodes in each sub-query embeddings
using the values obtained from the embeddings of the other subqueries.
Then, reducers of phase 2 construct the embeddings of the query $Q$ by combining compatible embeddings, one for each subquery.
\end{enumerate}
\end{enumerate}

\subsubsection{The preprocessing phase}

In the preprocessing phase the users' query $Q$ is decomposed into a tuple of subqueries  $(Q_1, \dots, Q_n)$,
with $n\geq 1$ and the auxiliary structures presented in Subsection~\ref{c-proproc} are constructed.
Preprocessing phase emits these structures with key the pair $(subqueryID, SegmentID)$ to the mappers of Phase 1, except of $MBN$ list,
which is emitted directly to the reducers of Phase 1.

\begin{exmpl}
\label{ex:preproc1}
Consider the query $Q$ appearing in Fig.~\ref{fig:rdf-query-decomp} and assume that the subqueries $Q_1$, $Q_2$, and $Q_3$ are constructed in the preprocessing phase. Assume also that the numbering functions has numbered the nodes and the edges of $Q$ as shown in Fig.~\ref{fig:rdf-query-decomp}. Then, it is easy to see that
${\cal B}(Q) = \{n1, n2, n3\}$ while ${\cal N}(Q) - {\cal B}(Q) = \{n4, n5\}$.
Finally, the lists of triples is $(t1, t2, t3, t4, t5)$.
It is thus easy to see that the query prototypes for the subqueries $Q_1$, $Q_2$ and $Q_3$ are:

{\small
\textsf{Q1: ($<$+,+,\_$>$, $<$\_,+$>$, $<$+,\_,\_,\_,+$>$)}

\textsf{Q2: ($<$\_,+,+$>$, $<$+,\_$>$, $<$\_,\_,+,+,\_$>$)}

\textsf{Q3: ($<$+,\_,+$>$, $<$\_,\_$>$, $<$\_,+,\_,\_,\_$>$)}
}

\noindent
while the list of missing border nodes is $MBN$ = $[(n1,Q2)$, $(n2,Q3)$, $(n3,Q1)]$.
\hfill$\Box$
\end{exmpl}

\subsubsection{Phase 1 of the QEJPE-algorithm}


The \emph{mapper of phase 1} gets as input a subquery $Q_i$ and a graph segment $G_j$ and evaluates $Q_i$ on $G_j$ obtaining in this way
all useful (total and partial) embeddings.  These embeddings  are emitted
to the reducers  of Phase 1 with key the subquery ID $Q_i$.
The  procedure for the Mapper of Phase 1 is given below:

{\small \sl
\begin{tabbing}
aa \= aa \= aa \= aa \kill
\textbf{mapper1} (($Q_i$, $G_j$), ($GjData$, $subqueryInfo$))\\
//($Q_i$,$G_j$): $Q_i$ is the ID of a subquery, $G_j$ is the ID of a data segment\\
// $GjData$: the content of the data graph segment $G_j$\\
// $SubqueryInfo$: prototypes/border \& non-border nodes/triples of $Q$\\

\textbf{begin}\\
\>  compute $E = \{e\mid e$ is a useful partial embedding of $Q_i$ in GjData$\}$;\\
\>  \textbf{for each} $e \in E$ \textbf{do} emit $([Q_i, e])$;\\
\textbf{end.}
\end{tabbing}
}

\begin{exmpl} (Continued from Example~\ref{ex:preproc1}\label{ex:mapper11}).
Some embeddings of the subqueries
$Q_1$, $Q_2$ and $Q_3$ (see Fig.~\ref{fig:rdf-query-decomp}) in the segments $G_1$, $G_2$ and $G_3$ (see Fig.~\ref{fig:rdf-apartition})
computed by the corresponding mappers and emitted with key the subquery ID, appear below.
More specifically, a total embedding evaluated and emitted by \textit{the mapper working on $(Q_1, G_1)$} is\footnote{Notice that
we can check if an embedding is total or partial by comparing it with the corresponding subquery prototype (see Example~\ref{ex:preproc1}).
An embedding is total if it has images for all (border and non-border) nodes  and triples of the subquery (i.e. for all nodes and triples of the subquery the '+' sign appears in the corresponding place of the query prototype.)}:

{\small
\textsf{(1) \ \ \  key = Q1, value = ($<$Person4,Article1,*$>$, $<$*,"Title1"$>$, $<$+,\_,\_,\_,+$>$)}
%
%
%
}

\noindent
The \textit{Mapper working on $(Q_2,G_1)$} computes and emits the partial embedding:

{\small
\textsf{(2)\ \ \  key = Q2, value = ($<$*,Article1,Person4$>$, $<$*,*$>$, $<$\_,\_,+,\_,\_$>$)}
}

\noindent
Among the embeddings obtained and emitted by \textit{the Mapper working on $(Q_1,G_2)$} is the (partial) embedding:

{\small
\textsf{(3)\ \ \  key = Q1, value = ($<$Person2,Article2,*$>$, $<$*,*$>$, $<$+,\_,\_,\_,\_$>$)}
}

\noindent
Among the embeddings obtained and emitted by the \textit{Mapper working on
$(Q_2,G_2)$} are the (partial) embeddings:

{\small
\textsf{(4)\ \ \  key = Q2, value = ($<$*,Article1,Person1$>$, $<$*,*$>$, $<$\_,\_,+,\_,\_$>$) }

\textsf{(5)\ \ \  key = Q2, value = ($<$*,Article2,Person3$>$, $<$*,*$>$, $<$\_,\_,+,\_,\_$>$) }
}

\noindent
The \textit{Mapper working on $(Q_3,G_2)$} computes and emits the total embeddings:

{\small
\textsf{(6)\ \ \  key = Q3, value = ($<$Person4,*,Person1$>$, $<$*,*$>$, $<$\_,+,\_,\_,\_$>$) }

\textsf{(7)\ \ \  key = Q3, value = ($<$Person2,*,Person3$>$, $<$*,*$>$, $<$\_,+,\_,\_,\_$>$) }
}

\noindent
The \textit{Mapper working on $(Q_1,G_3)$} emits the partial embedding:

{\small
\textsf{(8)\ \ \  key = Q1, value = ($<$*,Article2,*$>$, $<$*,"Title2"$>$, $<$\_,\_,\_,\_,+$>$) }
}

\noindent
The \textit{Mapper working on $(Q_2,G_3)$} emits the partial embeddings:

{\small
\textsf{(9)\ \ \  key = Q2, value = ($<$*,Article1,*$>$, $<$Journal1,*$>$, $<$\_,\_,\_,+,\_$>$)}

\textsf{(10)\ \ \  key = Q2, value = ($<$*,Article2,*$>$, $<$Journal1,*$>$, $<$\_,\_,\_,+,\_$>$)}
}

\noindent
Finally,
the mappers working on $(Q_3, G_3)$ and $(Q_3, G_1)$ return no (partial or total) embeddings.
\hfill$\Box$
\end{exmpl}

It is important to note that the procedure for the {\sl \textbf{mapper1}} does not determine a specific method for the computation
of the useful (partial) embeddings of the subqueries.  This means that, every algorithm that can compute
all partial embeddings can be used in a specific implementation of the {\sl \textbf{mapper1}}.
Moreover, {\sl \textbf{mapper1}}  is independent of the way the data graph is stored.


A \emph{Reducer of Phase 1 }receives all useful (partial) embeddings
of a subquery $Q_i$ whose ID is the key of the reducer, in all graph segments $G_1, \dots, G_m$ of $G$.
A reducer: (a) computes all total embeddings of $Q_i$
in $G$ and emits them to the mappers of Phase 2 with key the  subquery ID,
and (b) it finds all border node values from the total embeddings of $Q_i$  that are missing from the total embeddings of other subqueries and emits them with the appropriate subquery IDs as keys. The reducer is defined as follows\footnote{In the presentation of the procedures the following abbreviations are used: \emph{bnv} stands for BorderNodeValues,  \emph{nbnv} for NonBorderNodeValues, and \emph{tm} for TriplesMatched.}:

{\small \sl
\begin{tabbing}
aa \= aa \= aa \= aa \= aa \= aa \= aa \= aa \= aa \= aa \kill
\textbf{reducer1}($Q_i, values$)\\
// $Q_i$: a subquery ID. \\
// $values$: contains the list of the embeddings for $Q_i$ and the $MBN$ list\\


\textbf{begin}\\
\> collect in a list $F_i$ the total embeddings of $Q_i$ appearing in values or \\
\> \> obtained by joining compatible partial embeddings in $values$;\\
\> \textbf{if} $F_i$ is empty \textbf{then} EXIT; // there is no solution for the subquery $Q_i$ \\
\>  \>  \>  \>  \>   \>  \> \> // and thus for the original query $Q$\\
\> extract the  MBN list from $values$;\\
\> \textbf{foreach} embedding e = (bnv, nbnv, tm) in $F_i$ \textbf{do}\\
\>  \>   \textbf{begin}\\
\>  \>  \>  emit([$Q_i$, (bnv, nbnv)]);  // emits total embedding with key the subquery ID $Q_i$\\
\>  \>  \>  \textbf{for} i = 1 to $|$bnv$|$ \textbf{do}\\
\>  \>  \>  \>        \textbf{if} (bnv[i] != '*') \textbf{then} \\
\>  \>  \>  \>  \>   \textbf{for each} $(n_i, Q_j)$ in $MBN$ \textbf{do}\\
\>  \>  \>  \>  \>   \>  \>   emit([$Q_j$, $(n_i, bnv[i])]$); \\
\>  \>  \>  \textbf{end}\\
\>  \>       \textbf{end} \\
\textbf{end.}
\end{tabbing}
}

\begin{exmpl} (Continued from Example~\ref{ex:mapper11}\label{ex:reducer11}).
Among the total embeddings of $Q_1$ that constructs and emits \emph{reducer with key $Q1$} are:

{\small
\textsf{(1)$=>$\   key = Q1, value = ($<$Person4,Article1,*$>$, $<$*,"Title1"$>$)}

\textsf{(3)+(8)$=>$\ key = Q1, value = ($<$Person2,Article2,*$>$, $<$*,"Title2"$>$)}
}

\noindent
Taking into account the contents of the MBN list: $MBN = [(n1, Q2)$, $(n2, Q3)$, $(n3, Q1)]$
the reducer also emits the following missing border node values:

{\small
\textsf{key = Q2, value = (1,Person2) \ \ \ \ \  key = Q2, value = (1,Person4) }


\textsf{key = Q3, value = (2,Article1) \ \ \ \ \       key = Q3, value = (2,Article2), ...}

}

\noindent
The \textit{Reducer for key Q2} constructs and emits the total embeddings for $Q_2$:

{\small
\textsf{(4)+(9)$=>$\  key = Q2, value = ($<$*,Article1,Person1$>$, $<$Journal1,*$>$)}

\textsf{(5)+(10)$=>$\  key = Q2, value = ($<$*,Article2,Person3$>$, $<$Journal1,*$>$)}
}

\noindent
and the following values for missing border nodes:

{\small
\textsf{key = Q3, value = (2,Article1) \ \ \ \ \       key = Q3, value = (2,Article2)}

\textsf{key = Q1, value = (3,Person1)  \ \ \ \ \      key = Q1, value = (3,Person3), ...}

}

\noindent
Reducer for key Q3 emits:

{\small
\textsf{(6)$=>$\  key = Q3, value = ($<$Person4,*,Person1$>$, $<$*,*$>$)}

\textsf{(7)$=>$\  key = Q3, value = ($<$Person2,*,Person3$>$, $<$*,*$>$)}

\textsf{key = Q2, value = (1,Person2) \ \ \ \ \        key = Q2, value = (1,Person4)}

\textsf{key = Q1, value = (3,Person1) \ \ \ \ \       key = Q1, value = (3,Person3)}
} \hfill$\Box$
\end{exmpl}

\subsubsection{Phase 2 of the QEJPE-algorithm}
\label{subsec:GlobePhase2}


Each M\emph{apper in Phase2} manipulates the embeddings of a specific subquery.
It fills in their missing border node values using values from the embeddings of other subqueries that have been emitted by the
{\sl \textbf{reducer1}} based in the information in MBN list
and emits the resulted embeddings to the reducers of \emph{Phase 2} (the key is the tuple of the border node values).
The mapper of \emph{Phase 2} is given below:

{\small \sl
\begin{tabbing}
aa \= aa \= aa \= aa \= aa \= aa \kill
\textbf{mapper2}($Q_i$, $values$)\\
// $Q_i$: the ID of a subquery\\
// $values$: a list $E$ of the parts (bnv, nbnv) of the total embeddings of $Q_i$ and\\
//         a list $V$ of pairs $(i,v)$, where $v$ is a candidate value for bnv[i]\\

\textbf{begin}\\
\>  \textbf{for each} embedding e = (bnv, nbnv) in $E$ \textbf{do}\\
\> \>      \textbf{for each} instance bnv' of bnv using the values in $V$ \textbf{do} \\
\> \>  \> \>  \>  \textbf{emit}([bnv', ($Q_i$, nbnv)]);\\
\textbf{end.}
\end{tabbing}
}

\begin{exmpl} (Continued from Example~\ref{ex:reducer11}\label{ex:mapper21}).
\textit{Mapper with key Q1} receives:

{\small
\textsf{E = $[$($<$Person4,Article1,*$>$, $<$*,"Title1"$>$),}
\textsf{($<$Person2,Article2,*$>$, $<$*,"Title2"$>$),...$]$}

\textsf{V = [(3,Person1), (3,Person3),... ]}
}

\noindent
This mapper produces instances of the border node tuples in \textsf{E}
by replacing the '*' in the 3rd place with a value in \textsf{V}.
Among the key-value pairs obtained and emitted in this way are:

{\small
\textsf{key = (Person4,Article1,Person1), value = (Q1, $<$*,"Title1"$>$)}

\textsf{key = (Person2,Article2,Person3), value = (Q1, $<$*,"Title2"$>$)}
}

\noindent
The input of the \textit{Mapper with key Q2} is:

{\small
\textsf{E = [($<$*,Article1,Person1$>$, $<$Journal1,*$>$),}

\textsf{ \ \ \ \    ($<$*,Article2,Person3$>$, $<$Journal1,*$>$), ... ]}

\textsf{V = [(1,Person2), (1,Person4), ...]}
}

\noindent
Some of the instances that this mapper produces and emits are:
%

{\small
\textsf{key = (Person4,Article1,Person1), value = (Q2, $<$Journal1,*$>$)}

\textsf{key = (Person2,Article2,Person3), value = (Q2, $<$Journal1,*$>$)}
}

\noindent
The input of the \textit{Mapper with key Q3} is:

{\small
\textsf{E = [ ($<$Person4,*,Person1$>$, $<$*,*$>$), ($<$Person2,*,Person3$>$, $<$*,*$>$) ]}

\textsf{V = [(2,Article1), (2,Article2)]}
}

\noindent
Some key-value pairs produced and emitted (as above) by this mapper are:

{\small
\textsf{key = (Person4,Article1,Person1), value = (Q3, $<$*,*$>$)}

\textsf{key = (Person2,Article2,Person3), value = (Q3, $<$*,*$>$)}
}
\hfill$\Box$
\end{exmpl}


In each \emph{reducer of phase 2}, the embeddings (one for each subquery in $(Q_1, \dots, Q_n)$) are joined\footnote{Notice that the joined embeddings
are, by construction, compatible.} to construct the final answers of $Q$.
The reducer of phase 2 is given below:

{\small \sl
\begin{tabbing}
aa \= aa \= aa \= aa \kill
\textbf{reducer2}($key$, $values$)\\
// $key$: a tuple of border node values\\
// $values$: pairs of the form ($Q_i$, \emph{partial embedding for non-border nodes})\\

\textbf{begin}\\
\>  \textbf{for each} join of compatible embeddings \\
\> \>  obtained by using one embedding for each subquery \textbf{do} \\
\>      \>    \> Emit the result produced by this join;\\
\textbf{end.}
\end{tabbing}
}

\begin{exmpl} (Continued from Example~\ref{ex:mapper21}\label{ex:reducer21}).
The \textit{Reducer with key (Person4, Article1, Person1)} receives the list:

{\small
\textsf{[(Q1, $<$*,"Title1"$>$), (Q2, $<$Journal1,*$>$), (Q3, $<$*,*$>$)]}
}

\noindent
Combining (i.e. joining) these embeddings the reducer returns  the answer:

\centerline{\small \textsf{$<$Person4,Article1,Person1,Journal1,"Title1"$>$}}

The \textit{reducer with key (Person2,Article2,Person3)} receives  the following list:

{\small
\textsf{[(Q1, $<$*,"Title2"$>$), (Q2, $<$Journal1,*$>$), (Q3, $<$*,*$>$)]}
}

\noindent
 which joins giving the answer:

\centerline{\small \textsf{$<$Person2,Article2,Person3,Journal1,"Title2"$>$}}

\noindent
Notice that no other reducer returns solution (as they do not receive embeddings for all subqueries). This can be verified by considering all possible embeddings  of all subqueries, which do not appear, for space reasons, in the examples of this subsection.  
\hfill$\Box$
\end{exmpl}


\subsubsection{Discussion}

QEJPE-algorithm computes the answers to the given query $Q$ correctly, independently of a) the data graph partitioning,
b) the way the graph segments are stored,
c) the query graph decomposition, and d) the algorithm used for calculating intermediate (partial) results.

Some improvements to the proposed algorithm are as follows :

\begin{enumerate}
\item
Note that, in order to obtain all total embeddings of a subquery $Q_i$ in {\sl \textbf{reducer1}}
it suffices to combine partial embeddings obtained from different data graph segments.
However, such provenance information is not emitted from {\sl \textbf{mapper1}}   in its present form.
It is, however, easy to adapt the QEJPE-algorithm, so as the {\sl \textbf{mapper1}}   emits this information to {\sl \textbf{reducer1}}
and the {\sl \textbf{reducer1}} takes it into account to construct more efficiently total embeddings of the subqueries.

\item
Notice that, as we can see in Example~\ref{ex:reducer11}, several instances of {\sl \textbf{reducer1}}  may emit the same values, either embeddings or missing node values, to the  {\sl \textbf{mapper2}}.  This is due to the fact that the same embedding or the same candidate missing node value may be found and emitted by several reducers.
Thus, a specific instance of {\sl \textbf{mapper2}} may receive  multiple times the same value which may lead in the construction of the same embedding several times.
A possible optimization is to eliminate redundant values from the lists $E$ of embeddings and the list $V$ of candidate values for missing nodes that an instance of a {\sl \textbf{mapper2}} receives before the computation of embeddings that will be emitted to {\sl \textbf{reducer2}} .

%
%
\end{enumerate}

\subsection{eval-STARS algorithm}
\label{subsec:eval-STARS_algorithm-impl}

In this section we present a MapReduce based implementation of the eval-STARS query evaluation algorithm
presented in Subection~\ref{subsec:QE-using-STARS}.
The algorithm is based on similar assumptions on which the QEJPE-algorithm is based.
The main difference is that in eval-STARS algorithm, a queries $Q$ posed by the user is decomposed into
a tuple of queries $(Q_1, \dots, Q_n)$, with $n\geq 1$, of a specific form called \emph{generalized star queries}. The query decomposition may be redundant or non-redundant.

The implementation of the algorithm consists of a preprocessing phase followed by two map-reduce phases:
The first map-reduce phase takes advantage of the generalized  star form of the sub-queries and focuses on
evaluating the generalized star subqueries over the input segments. The results of the sub-queries are emitted to the second phase, which
combines them properly in order to produce the answers of the initial query.

In the \emph{preprocessing phase} the users' query $Q$ is decomposed into a tuple of generalized star subqueries, with $n\geq 1$ and the auxiliary structures presented in Subsection~\ref{c-proproc} are constructed.
Preprocessing phase emits these structures with key the pair $(subqueryID, SegmentID)$ to the mappers of Phase 1.

\begin{exmpl} \label{ex:preproc2}
Consider
the query graph appearing in the left part of Fig. \ref{fig:rdf-query-star-decomp} which is decomposed into three generalized star subqueries $Q_1$, $Q_2$, and $Q_3$
appearing in the right part of Fig. \ref{fig:rdf-query-star-decomp}.
The border nodes are ${\cal B}(Q) = \{n1, n2, n3\}$, while the non-border nodes are  ${\cal N}(Q) - {\cal B}(Q) = \{n4, n5, n6, n7\}$.

\noindent
The query prototypes are the following:

{\small
\textsf{$Q_1$: ($<$+,\_,+$>$, $<$+,\_,+,\_$>$, $<$+,\_,\_,\_,\_,\_,+,+$>$)}

\textsf{$Q_2$: ($<$+,+,+$>$, $<$\_,\_,\_,+$>$, $<$\_,+,\_,\_,+,+,\_,\_$>$)}

\textsf{$Q_3$: ($<$+,+,\_$>$, $<$\_,+,\_,\_$>$, $<$\_,\_,+,+,\_,\_,\_,\_$>$)}
}

\noindent
The list $MBN = [(n_2,Q_1),(n_3,Q_3)]$  is also constructed in preprocessing phase.
\hfill$\Box$
\end{exmpl}

\subsubsection{Phase 1 of the algorithm}

The first phase of the algorithm computes the embeddings of the generalized star subqueries $Q_1, \dots, Q_n$ in $G$.


In Phase 1 each \emph{mapper} gets as input a generalized star subquery $Q_i$, a graph segment $G_j$  and the $MBN$ list.
Let $c_i = C(Q_i)$ be the central node of $Q_i$
(recall that this node appears in every triple of $Q_i$).
The operation of the mapper is divided into two parts.

\underline{Part 1:} The mapper computes the embeddings of each triple of $Q_i$ in $G_j$ that map the central node $c_i$
to a {\em border node} or to a literal, and emits the results to appropriate reducers.
More specifically, let $t=(s,p,o)$ be a triple that belongs to subquery $Q_i$
and let $e$ be an embedding of $t$ into $G_j$ such that $e(c_i) \in {\cal B}(G_j)$.
If the central node of $Q_i$ is $s$ then the mapper emits a pair
$(key,value)$, where $key = (Q_i, e(s))$ and $value = (o,e(o))$.
Otherwise (i.e., if the central node of $Q_i$ is $o$) then $key = (Q_i, e(o))$ and $value = (s,e(s))$.

Notice that embeddings of triples in $Q_i$ that map $c_i$ to different
nodes of $G_j$ are incompatible and cannot be joined to obtain an embedding of $Q_i$.
Since the value of $c_i$ is included in the key, incompatible embeddings of triples are emitted
to different reducers, while compatible embeddings are emitted to the same reducer.

\underline{Part 2:} This part of the {\sl \textbf{mapper1}} computes all embeddings of $Q_i$ into $G_j$ which map the central node of $Q_i$ to a non-border and non-literal node of $G_j$.
Notice that if for some embedding $e$ of $Q_i$ in $G$ the value of $c_i$ is a non-border and non-literal node of $G_j$ (i.e., is $e(c_i) \in ({\cal N}(G_i)- ({\cal B}(G_i) \cup L))$),
then $e(v) \in G_j$ for every node $v \in {\cal N}(Q_i)$. This means that $e$ is an embedding of $Q_i$
into $G_j$ and it can be computed locally i.e. no other data graph segments are needed to compute $e$.

The  computation  of the embeddings of $Q_i$ into $G_j$, which map $c_i$ to a non-border node of $G_j$
can be achieved either by adding an appropriate conjunct to
$Q_i$, or by computing all the embeddings of $Q_i$ in $G_j$ and
then removing those that assign border nodes to $c_i$. The
embedings computed in the second part of the mapper are directly
emitted to the mappers of Phase 2 (rather than to the reducers of
Phase 1). Similarly, the values of missing border nodes are emitted to
the mappers of Phase 2.

{\small \sl
\begin{tabbing}
aa \= aa \= aa \= aa \= aa \= aa \= aa \= aa \kill
\textbf{mapper1}(($Q_i$, $G_j$), (GjData, B(GjData), subqueryInfo, $MBN$))\\
//($Q_i$,$G_j$): $Q_i$ is the ID of a subquery; $G_j$ is the ID of a data segment\\
// GjData: the content of the data graph segment $G_j$\\
// B(GjData): the set of border nodes of $G_j$\\
// SubqueryInfo: prototypes/border \& non-border nodes/triples of $Q$\\
// MBN: the list of missing border nodes \\

\textbf{begin} \\
\>   Let $c_i$ = $C(Q_i)$; \\
\>    \% Part 1 \\
\> \textbf{foreach} triple $t = (c_i,p,o)$ in $Q_i$ \textbf{do}\\
\>  \>   \textbf{begin}\\
\>  \>  \>  compute $E = \{e \mid e$ is an embedding of $t$ in GjData  and $e(c_i) \in {\cal B}(GjData)$ $\}$;\\
\>  \>  \>  \textbf{foreach} embedding $e$ in $E$ \textbf{do} \\
\>  \>  \>  \>  \textbf{emit}([$(Q_i,e(c_i))$,$(o,e(o))$]);\\
\>  \>  \textbf{end} \\
\>  \textbf{foreach} triple $t = (s,p,c_i)$ in $Q_i$ \textbf{do}\\
\>  \>   \textbf{begin}\\
\>  \>  \>  compute $E = \{e \mid e$ is an embedding of $t$ in GjData and $e(c_i) \in ({\cal B}(GjData) \cup L)$ $\}$;\\
\>  \>  \>  \textbf{for each} embedding $e$ in $E$ \textbf{do}\\
\>  \>  \>  \>  \textbf{emit}([$(Q_i,e(c_i))$,$(s,e(s))$]);\\
\>  \>  \textbf{end} \\
\>    \% Part 2 \\
\>   compute $E = \{e \mid e$ is a embedding of $Q_i$ in GjData and $e(c_i) \notin ({\cal B}(GjData) \cup L)$ $\}$;\\
\> \textbf{for each} embedding $e = (bn, nbn)$ in $E$ \textbf{do}\\
\>  \>   \textbf{begin}\\
\>  \>  \>  emitToSecondPhase([$Q_i$, (bnv, nbnv)]); // i.e. to the mapper of phase 2 \\
\>  \>  \>  \textbf{for} k = 1 to $|bnv|$ \textbf{do}\\
\>  \>  \>  \>       \textbf{if} (bnv[k] != '*') \textbf{then} \\
\>  \>  \>  \>  \>   \textbf{for each} $(n_k, Q_j)$ in $MBN$ \textbf{do} \\
\>  \>  \>  \>  \>  \>  emitToSecondPhase$([Q_j, (n_k$, bnv[k])]); \\
\>  \>  \textbf{end} \\
\textbf{end.}
\end{tabbing}
}

\begin{exmpl} (Continued from Example~\ref{ex:preproc2}\label{ex:mapper12}).
In this example, we assume that the query graph $Q$ and its generalized star subquries are those appearing in Fig.~\ref{fig:rdf-query-star-decomp},
while the data graph $G$ and the graph segments obtained by decomposing $G$  are those appearing in Fig.~\ref{fig:rdf-apartition}.
Below, we see the application of {\sl \textbf{mapper1}} on the pairs of subqueries and graph segments:

\noindent
Applying {\sl \textbf{mapper1}} on ($Q_1$, $G_1$) results in emission (see Part 1 of the procedure for {\sl \textbf{mapper1}})
of the following ($key, value$) pairs to the {\sl \textbf{reducer1}}:

{\small
\textsf{key = (Q1, Article1), value = (n1, Person4)} (embedding of $t1$)

\textsf{key = (Q1, Article1), value = (n6, ``Title1")} (embedding of $t7$)
}

\noindent
No key value pairs are emitted to Phase 2 (see Part 2 of the procedure for {\sl \textbf{mapper1}}).

\noindent
Applying {\sl \textbf{mapper1}}  on ($Q_2$, $G_1$) results in emission (see Part 1)
of the following $key, value$ pair to the {\sl \textbf{reducer1}}:

{\small
\textsf{key = (Q2, Article1), value = (n1, Person4)} (embedding of $t2$)
}

\noindent
Besides, the following $key, value$ pairs are emitted directly to the {\sl \textbf{mapper2}} (Mapper of Phase 2) (see Part 2):

{\small
\textsf{key = Q2, value = ($<$Person4,Article3,Journal2$>$, $<$*,*,*,``2008"$>$)}

\textsf{key = Q1, value = (n2, Article3)}

\textsf{key = Q3, value = (n3, Journal2)}
}

\noindent
Notice that the last two emissions are conducted by the MBN list which, as we have seen in Example~\ref{ex:preproc2},
is  $MBN = [(n_2,Q_1),(n_3,Q_3)]$.

\noindent
Applying {\sl \textbf{mapper1}} on ($Q_3$, $G_1$) results in emission (see Part 1)
of the following ($key, value$) pairs to {\sl \textbf{reducer1}}:

{\small
\textsf{key = (Q3, Person4), value = (n2, Article1)} (embedding of $t4$)

\textsf{key = (Q3, Person4), value = (n2, Article3)} (embedding of $t4$)
}

\noindent
No key value pairs are emitted to Phase 2.

\noindent
Applying {\sl \textbf{mapper1}} on ($Q_1$, $G_2$) results in emission (see Part 1)
of the following ($key, value$) pairs to {\sl \textbf{reducer1}}:

{\small
\textsf{key = (Q1, Article1), value = (n1, Person1)} (embedding of $t1$)

\textsf{key =(Q1, Article1), value = (n1, Person2)} (embedding of $t1$)
}

\noindent
No key value pairs are emitted to Phase 2.

\noindent
Applying {\sl \textbf{mapper1}} on ($Q_2$, $G_2$) results in emission (see Part 1)
of the following $key, value$ pairs to the {\sl \textbf{reducer1}}:

{\small
\textsf{key = (Q2, Article1),  value = (n1, Person1)} (embedding of $t2$)

\textsf{key = (Q2, Article1),  value = (n1, Person2)} (embedding of $t2$)

\textsf{key = (Q2, Article2),  value = (n1, Person2)} (embedding of $t2$)

\textsf{key = (Q2, Article2),  value = (n1, Person3)} (embedding of $t2$)
}

\noindent
No key value pairs are emitted to Phase 2.

\noindent
Applying {\sl \textbf{mapper1}} on ($Q_3$, $G_2$)
results in no emission of any ($key, value$) pair to {\sl \textbf{reducer1}} (see Part1).
However, the following ($key, value$) pairs are emitted (see Part 2) to {\sl \textbf{mapper2}}:

{\small
\textsf{key = Q3, value = ($<$Person4, Article1, *$>$, $<$*, Person1,* , *$>$)}

\textsf{key = Q3, value = ($<$Person2, Article2, *$>$, $<$*, Person3, *, *$>$)}

\textsf{key = Q1, value = (n2, Article1)}

\textsf{key = Q1, value = (n2, Article2)}
}

\noindent
Applying {\sl \textbf{mapper1}} on ($Q_1$, $G_3$) results in emission of
the following ($key, value$) pair to {\sl \textbf{reducer1}}:

{\small
\textsf{key = (Q1, Article1),  value = (n3, Journal1)} ($t8$)
}

\noindent
No key value pairs are emitted to Phase 2  (see Part2).

\noindent
Applying {\sl \textbf{mapper1}} on ($Q_2$, $G_3$) results in emission of
the following ($key, value$) pair to {\sl \textbf{reducer1}}:

{\small
\textsf{key = (Q2, Article1), value = (n3, Journal1)} ($t5$)

\textsf{key = (Q2, Article2), value = (n3, Journal1)} ($t5$)

\textsf{key = (Q2, Article2), value = (n7, ``2008")} ($t6$)
}

\noindent
No key value pairs are emitted to Phase 2.

\noindent
Applying {\sl \textbf{mapper1}} on ($Q_3$, $G_3$) results in no emission
of any ($key, value$) pair.\hfill$\Box$
\end{exmpl}


Concerning the  Reducer of Phase 1
For each key $(Q_i, v)$ the corresponding reducer computes all the embeddings of $Q_i$ that map the
central node $c_i$ of $Q_i$ to $v$.
The input to this reducer is a list of pairs of the form $(n_k, u)$, where
$n_k$ is a node of $Q_i$ different from $c_i$ and $u$ is a possible value for
$n_k$ in an embedding of $Q_i$ in $G$.
Suppose that $n_{k_1}, n_{k_2}, \dots, n_{k_m}$ are the non-central nodes in $Q_i$.
Then, for every $j=1, \dots, m$,
the reducer constructs a set $L[k_j]$ of all possible values for node $n_{k_j}$.
More specifically, for each element $(x_1, x_2, \dots, x_m)$ of the cartesian product $L[k_1]\times L[k_2]\times \dots \times L[k_m]$,
it constructs an embedding e = (bnv,nbnv) of $Q_i$ in $G$, such that $e(c_i) = v$ and $e(n_{k_j}) = x_j$ and emits $(Q_i$, (bnv,nbnv)) (see Subsection~\ref{subsec:preproc} for the representation of an embedding).
Moreover, if at least one embedding of $Q_i$ has been found,
Reducer 1 emits the values of missing border nodes.

{\small \sl
\begin{tabbing}
aa \= aa \= aa \= aa \= aa \= aa \= aa \= aa \= aa \= aa \= aa \kill
\textbf{reducer1}($(Q_i,v), values$)\\
// $Q_i$: a subquery ID \\
// $v$: the value of the central node of $Q_i$ \\
// values: contains (i) a list of pairs $(x,u)$, with $x \neq C(Q_i)$ and \\
\>  \> $u$ is a candidate image of $x$  and (ii) the $MBN$ list. \\

\textbf{begin}\\
\% Part 1\\
\>    $allNonEmpty$ = $true$; \\
\>   \textbf{foreach} non-central node $x$ in $Q_i$ \textbf{do}\\
\>  \>   \textbf{begin}\\
\>  \>  \>  $L[I(x)]$ = $\{u \mid (x,u) \in values \}$; \\
\>  \>  \>  \textbf{if} $L[I(x)]$ = $\emptyset$ \textbf{then} $allNonEmpty$ = $false$; \\
\>  \>   \textbf{end}\\

\% Part 2\\
\>  \textbf{if} $allNonEmpty$ = $true$ \textbf{then} // i.e. there are values for all non-central nodes of $Q_i$\\
\>  \>   \textbf{begin}\\
\>  \>  \>  create an embedding with undefined values;\\
\>  \>  \>  \> (bnv,nbnv)= $(\langle *,\dots,*\rangle ,\langle *,\dots,*\rangle )$; \\
\>  \>  \>  $c_i$ = C($Q_i$); \\
\>  \>  \>  $L[I(c_i)]$ = $\{v\}$; \\
\>  \>  \>  \textbf{if} $c_i$ is a border node \textbf{then} \\
\>  \>  \>  \>  \> bnv$[I(c_i)]$ = $v$; \\
\>  \>  \>  \>  \textbf{else} nbnv$[I_{nb}(c_i)]$ = $v$;\\
\>  \>  \>  $E$ = $\{$(bnv,nbnv)$\}$;\\
\>  \>  \>  \textbf{for each} non-central node $x$ in $Q_i$ \textbf{do}\\
\>  \>  \>  \>   \textbf{begin}\\
\>  \>  \>  \>  \>  $E'$ = $\emptyset$;\\
\>  \>  \>  \>  \>  \textbf{foreach} $e$ in $E$ \textbf{do}\\
\>  \>  \>  \>  \>  \>  \textbf{foreach} $u$ in $L[I(x)]$ \textbf{do}\\
\>  \>  \>  \>  \>  \>  \>  \textbf{begin}\\
\>  \>  \>  \>  \>  \>  \>  \> create a copy e'=(bnv',nbnv') of $e$; \\
\>  \>  \>  \>  \>  \>  \>  \> \textbf{if} $x$ is a border node \textbf{then}\\
\>  \>  \>  \>  \>  \>  \>  \>  \>  \> bnv'$[I(x)]$ = $u$; \\
\>  \>  \>  \>  \>  \>  \>  \>  \> \textbf{else} nbnv'$[I_{nb}(x)]$ = $u$; \\
\>  \>  \>  \>  \>  \>  \>  \> insert (bnv',nbnv') in $E'$; \\
\>  \>  \>  \>  \>  \>  \>  \textbf{end}\\
\>  \>  \>  \>  \>  $E$ = $E'$; \\
\>  \>  \>  \>   \textbf{end}\\

\>  \>  \>  \textbf{foreach} embedding e = (bnv, nbnv) in $E$ \textbf{do}\\
\>  \>  \>  \> \textbf{emit}([$Q_i$, (bnv, nbnv)]); \\

\>  \>  \>  \textbf{foreach} $(x, Q_j)$ in $MBN$ \textbf{do} \\
\>  \>  \>  \>  \textbf{if} $x$ is a node in $Q_i$ \textbf{then} \\
\>  \>  \>  \>  \>  \textbf{foreach} $u$ in $L[I(x)]$ \textbf{do} \textbf{emit}$([Q_j, (x, u)])$; \\
\>  \>   \textbf{end}\\
\textbf{end.}
\end{tabbing}
}

\begin{exmpl} (Continued from Example~\ref{ex:mapper12}\label{ex:reducer12}).

\noindent
The \emph{reducer with key $(Q_1, Article1)$} receives the following list of values: \\
$[ (n_1, Person4), (n_6, ``Title1"), (n_1, Person1), (n_1, Person2),$ $(n_3, Journal1)]$. \\
Notice that, as we can conclude from the sub-query prototypes appearing  in Example~\ref{ex:preproc2},
the border nodes of $Q$ is ${\cal B}(Q) = \{n1, n2, n3\}$, while the non-border are ${\cal N}(Q) - {\cal B}(Q) = \{n4, n5, n6, n7\}$. Besides, from Fig.~\ref{fig:rdf-query-star-decomp}, we see that the central node of $Q_1$ is $n_4$ while its non-central nodes are $n_1$, $n_3$ and $n_6$.
Finally, the  $MBN$ list is $MBN = [(n_2,Q_1),(n_3,Q_3)]$.
Taking into account the above, the {\sl \textbf{reducer1}} with key $(Q_1, Article1)$, concludes by applying Part1 of the procedure that it has received values for all non-central nodes of $Q_1$.
More  specifically, {\sl \textbf{reducer1}}  constructs the following lists:

{\small
\textsf{L[1] = [Person1, Person2, Person4]}

\textsf{L[3] = [Journal1]}

\textsf{L[6] = [``Title1"]}
}

\noindent
which contain the values for the non-central nodes $n_1$, $n_3$ and $n_6$ respectively.
Combining these values, as well as the value $Article1$ of the central node $n_4$, {\sl \textbf{reducer1}} in Part 2  constructs and emits the following $(key, value)$ pairs (that represent embeddings of $Q_1$):

{\small
\textsf{key = Q1, value = ($<$Person1,*,Journal1$>$, $<$Article1,*,Title1,*$>$)}

\textsf{key = Q1, value = ($<$Person2,*,Journal1$>$, $<$Article1,*,Title1,*$>$)}

\textsf{key = Q1, value = ($<$Person4,*,Journal1$>$, $<$Article1,*,Title1,*$>$)}
}

\noindent
Besides, {\sl \textbf{reducer1}}, based on the MBN list, emits the following:

{\small
\textsf{Q3, (n3,Journal1)}
}

\noindent
The \emph{reducer with key $(Q_2, Article1)$} receives the following list of values:

{\small
\textsf{[(n1, Person4), (n1, Person1) (n1, Person2), (n3, Journal1)]}.
}

\noindent
Based on this values it constructs the following lists (corresponding to the values of the
non-central nodes $n_1$, $n_3$ and $n_7$ of the subquery $Q_2$):

{\small
\textsf{L[1]  = [Person1, Person2, Person4]}

\textsf{L[3]  = [Journal1]}

\textsf{L[7] = [ ]}
}

\noindent
From the above we see that the list for the non-central node $n_7$ is empty. Thus, these values cannot be user to construct a valid embedding for the query $Q_2$. Therefore, nothing is emitted to the next phase from this reducer.

\noindent
The \emph{reducer with key $(Q_2, Article2)$} receives the following list of values:

{\small
\textsf{[(n1, Person2), (n1, Person3),  (n3, Journal1), (n7, ``2008")]}.
}

\noindent
It constructs the lists:

{\small
\textsf{L[1] = [Person2, Person3]}

\textsf{L[3]  = [Journal1]}

\textsf{L[7]  = [``2008"]}
}

\noindent
Combining these values, as well as the value $Article2$ of the central node $n_2$, Part 2 of {\sl \textbf{reducer1}}  constructs and emits the following $(key, value)$ pairs (that represent embeddings of $Q_2$):

{\small
\textsf{key = Q2, value = ($<$Person2,Article2,Journal1$>$, $<$*,*,*,``2008"$>$)}

\textsf{key = Q2, value = ($<$Person3,Article2,Journal1$>$, $<$*,*,*,``2008"$>$)}
}

\noindent
Besides, {\sl \textbf{reducer1}}, based on the MBN list, emits the following:

{\small
\textsf{key = Q1, value = (n2,Article2)}

\textsf{key = Q3, value = (n3,Journal1)}
}

\noindent
The \emph{reducer with key $(Q_3, Person4)$} receives the following list of values:

{\small
\textsf{[(n2, Article1), (n2, Article3)]}.
}

\noindent
Based on this values it constructs the following lists (corresponding to the values of the
non-central nodes $n_1$ and $n_2$ of the subquery $Q_3$):

{\small
\textsf{L[1] = [ ]}

\textsf{L[2] = [Article1, Article3]}
}

\noindent
From the above we see that the list for the non-central node $n_1$ is empty. Thus, these values cannot be used to construct a valid embedding for the query $Q_3$. Therefore, nothing is emitted to the next phase from this reducer.
\hfill$\Box$
\end{exmpl}

\subsubsection{Phase 2 of the algorithm}


Phase 2 of the algorithm is similar to the Phase 2 of the QEJPE-algorithm presented in Subsection~\ref{subsec:GlobePhase2}.
The input of each \emph{mapper} of Phase 2, consist of all the embeddings of a specific subquery $Q_i$. Besides,
for each border node that does not occur in $Q_i$, mapper gets  as input,  the values
assigned to this node by the embeddings of the other queries.
These values are sent by the mappers and reducers of Phase 1 based on the MBN list.
Based on its input, the mappers of Phase 2 fills in their missing border node values using the corresponding input values, and emits the resulted embeddings to the reducers of Phase 2 using as key the tuple of the border node values.
This means that two embeddings are emitted to the same reducer if and only if they are compatible.

{\small \sl
\begin{tabbing}
aa \= aa \= aa \= aa \kill
\textbf{mapper2}($Q_i$, values)\\
// $Q_i$: the ID of a subquery\\
// values: a set $E$ of the parts (bnv, nbnv) of the total embeddings of $Q_i$ \\
//  \ \ \ \ \ and a set $V$ of pairs $(n_k,v)$, where $v$ is a candidate value for bnv[k]\\

\textbf{begin}\\
\>  \textbf{foreach} embedding e = (bnv, nbnv) in $E$ \textbf{do}\\
\> \>       \textbf{foreach} instance $bnv'$ of $bnv$ using the values in $V$ \textbf{do}\\
\> \> \> \textbf{emit}([bnv', ($Q_i$, nbnv)]);\\
\textbf{end.}
\end{tabbing}
}

\begin{exmpl} (Continued from Example~\ref{ex:reducer12}\label{ex:mapper22}).
\noindent
The mapper that works for the subquery $Q_1$ (i.e. the key is $Q_1$),
gets a list of values that contain the embeddings of $Q_1$ in $G$:

{\small
\textsf{($<$Person1,*,Journal1$>$, $<$Article1,*,Title1,*$>$)}

\textsf{($<$Person2,*,Journal1$>$, $<$Article1,*,Title1,*$>$)}

\textsf{($<$Person4,*,Journal1$>$, $<$Article1,*,Title1,*$>$)}
}

\noindent
and the values of missing border nodes emitted by the mappers and reducers of the Phase 1 directly to the mappers of Phase 2 (see Example~\ref{ex:mapper12}):

{\small
\textsf{(n2, Article1), (n2, Article2), (n2, Article3)}
}

\noindent
The {\sl \textbf{mapper2}}  emits the following ($key, value$) pairs to the {\sl \textbf{reducer2}} (reducer of Phase 2) by completing the missing node values:

{\small
\textsf{key = ($<$Person1,Article1,Journal1$>$), value = (Q1, $<$Article1,*,Title1,*$>$)}

\textsf{key = ($<$Person1,Article2,Journal1$>$),  value = (Q1, $<$Article1,*,Title1,*$>$)}

\textsf{key = ($<$Person1,Article3,Journal1$>$),  value = (Q1, $<$Article1,*,Title1,*$>$)}

\textsf{key = ($<$Person2,Article1,Journal1$>$),  value = (Q1, $<$Article1,*,Title1,*$>$)}

\textsf{key = ($<$Person2,Article2,Journal1$>$),  value = (Q1, $<$Article1,*,Title1,*$>$)}

\textsf{key = ($<$Person2,Article3,Journal1$>$),  value = (Q1, $<$Article1,*,Title1,*$>$)}

\textsf{key = ($<$Person4,Article1,Journal1$>$),  value = (Q1, $<$Article1,*,Title1,*$>$)}

\textsf{key = ($<$Person4,Article2,Journal1$>$),  value = (Q1, $<$Article1,*,Title1,*$>$)}

\textsf{key = ($<$Person4,Article3,Journal1$>$),  value = (Q1, $<$Article1,*,Title1,*$>$)}
}

\noindent
The mapper that works for the subquery $Q_2$ (i.e. the key is $Q_2$),
receives a list of values containing the following embeddings of $Q_2$ in $G$:

{\small
\textsf{($<$Person4,Article3,Journal2$>$, $<$*,*,*,``2008"$>$)}

\textsf{($<$Person2,Article2,Journal1$>$, $<$*,*,*,``2008"$>$)}

\textsf{($<$Person3,Article2,Journal1$>$, $<$*,*,*,``2008"$>$)}
}

\noindent
Notice that $Q_2$ has no missing border nodes.
The mapper emits the following ($key, value$) pairs to the reducers of Phase 2:

{\small
\textsf{key = ($<$Person4,Article3,Journal2$>$),  value = (Q2, $<$*,*,*,``2008"$>$)}

\textsf{key = ($<$Person2,Article2,Journal1$>$),  value = (Q2, $<$*,*,*,``2008"$>$)}

\textsf{key = ($<$Person3,Article2,Journal1$>$),  value = (Q2, $<$*,*,*,``2008"$>$)}
}

\noindent
The mapper that works for the subquery $Q_3$ (i.e. the key is $Q_3$),
receives a list of values that contain the embeddings of $Q_3$ in $G$:

{\small
\textsf{($<$Person4,Article1,*$>$, $<$*,Person1,*,*$>$)}

\textsf{($<$Person2,Article2,*$>$, $<$*,Person3,*,*$>$)}
}

\noindent
and the values of missing border nodes:

{\small
\textsf{(n3, Journal1)$, $(n3, Journal2)}
}

\noindent
The {\sl \textbf{mapper2}}  emits the following ($key, value$) pairs to the {\sl \textbf{reducer2}} (reducer of Phase 2) by completing the missing node values:

{\small
\textsf{key = ($<$Person4,Article1,Journal1$>$), value =  (Q3, $<$*,Person1,*,*$>$)}

\textsf{key = ($<$Person4,Article1,Journal2$>$), value =  (Q3, $<$*,Person1,*,*$>$)}

\textsf{key = ($<$Person2,Article2,Journal1$>$), value =  (Q3, $<$*,Person3,*,*$>$)}

\textsf{key = ($<$Person2,Article2,Journal2$>$), value =  (Q3, $<$*,Person3,*,*$>$)}
}
\hfill$\Box$
\end{exmpl}


Concerning the reducers of Phase 2, each reducer gets as input embeddings for each sub-query that are compatible
(each one of them assigns the values in the key of the reducer to the border nodes of the query).
The embeddings (one for each subquery in $(Q_1, \dots, Q_n)$)
are joined to construct the final answers of $Q$:

{\small \sl
\begin{tabbing}
aa \= aa \= aa \= aa \kill
\textbf{reducer2}(key, values)\\
// key: a tuple of border node values\\
// values: pairs of the form ($Q_i$, \emph{partial embedding for non-border nodes})\\

\textbf{begin}\\
\>   \textbf{foreach} join obtained by using one embedding for each subquery \textbf{do} \\
\>      \>     Emit the result produced by this join;\\
\textbf{end.}
\end{tabbing}
}

\begin{exmpl} (Continued from Example~\ref{ex:mapper22}\label{ex:reducer22}).

The reducer with key $<$Person2,Article2,Journal1$>$
receives the following list of values:

{\small
\textsf{[(Q1, $<$Article1,*,Title1,*$>$),}

\textsf{(Q2, $<$*,*,*,"2008"$>$), }

\textsf{(Q3, $<$*,Person3,*,*$>$)]}
}

\noindent
and constructs the unique embedding of $Q$ in $G$:

{\small
\textsf{($<$Person2, Article2, Journal1$>$,  $<$Article1, Person3, Title1, ``2008"$>$)}
}

\noindent
The remaining 11 reducers do not return any answer (they don't receive values
for at least one subquery).
\hfill$\Box$
\end{exmpl}

\subsubsection{Discussion}

Due to the specific form in which the user query $Q$ is decomposed, namely the  generalized star queries, the eval-STARS algorithm can compute embeddings  more efficiently than QEJPE-algorithm computes partial embeddings of the subqueries of  $Q$.

Notice also that {\sl \textbf{mapper1}}  computes and emits directly to {\sl \textbf{mapper2}} total embeddings of the subqueries that map their central nodes to non-border nodes of the data graph segment. This is also an advantage of the eval-STARS algorithm compared with the QEJPE-algorithm.

\subsection{QE-with-Redundancy algorithm}
\label{subsec:QE-with-Redundancy algorithm-impl}


In this section we present an implementation of the QE-with-Redundancy query evaluation algorithm  presented in Subsection~\ref{subsec:QE-redundancy} based on  the MapReduce programming framework.
Recall that, for the implementation of the algorithm we assume a star-oriented decomposition (s-decomposition) of the data graph $G$ and a (possibly redundant) decomposition of the query $Q$ posed by the user into a set of \emph{subject-object star subqueries} $\{Q_1, \dots, Q_n\}$, with $n\geq 1$.
The implementation of the algorithm consists of a preprocessing phase followed by
one and a half Map-Reduce phase.
The first phase of our algorithm takes advantage of the star form of the sub-queries and focuses on
evaluating the star subqueries over the input segments. The results of the sub-queries are emitted to the second phase, which
combines them properly in order to produce the answers of the initial query.

%

\subsubsection{The preprocessing phase}
\label{subsubsec:preproc3}

In the preprocessing phase the users' query $Q$ is decomposed into a set of so-queries $\{Q_1, \dots, Q_n\}$, with $n\geq 1$, and the auxiliary structures presented in Subsection~\ref{c-proproc} are constructed.
Preprocessing phase emits the above to the mappers of Phase 1 with key the pair
$(subqueryID, SegmentID)$.

\begin{exmpl} \label{ex:preproc-SO}
To present the QE-with-redundancy algorithm, we will use again the query $Q$ and its decomposition into three so-queries presented in Fig. \ref{fig:rdf-query-star-decomp}. The query prototypes, the MBN list and the tuple of common border nodes of appearing in these subqueries are the same as in Example~\ref{ex:preproc2}.

Concerning the data graph decomposition, to present the algorithm we will use the data graph segments (s-segments) obtained by decomposing the data graph $G$ as presented in Fig.~\ref{fig:rdfStarpartition}.
\hfill$\Box$
\end{exmpl}

\subsubsection{Phase 1 of the algorithm}

The first phase of the algorithm computes the embeddings of the so-queries $Q_1, \dots, Q_n$ in $G$ locally in each star graph segment $G_j$ of $G$.


Each \emph{mapper} in phase 1 gets as input an s-graph segment $G_j$, an so-query $Q_i$, the $MBN$ list, and the tuple ${\cal CB}(Q)$ and computes the embeddings of $Q_i$ into $G_j$.
The embeddings computed are directly
emitted to the mappers of Phase 2. Similarly, the values of the nodes in MBN  are emitted to
the mappers of Phase 2. Notice that the instances of the nodes in ${\cal CB}(Q)$ take part in the keys of the (key, value) pairs emitted to the mappers of Phase 2.

{\small \sl
\begin{tabbing}
aa \= aa \= aa \= aa \= aa \= aa \= aa \= aa \kill
\textbf{mapper1}(($Q_i$, $G_j$), (GjData, SubqueryInfo, $MBN$, ${\cal CB}(Q)$))\\
//($Q_i$,$G_j$): $Q_i$/$G_j$ is the ID of a subquery/data segment\\
// GjData: the content of the data graph segment $G_j$\\
// SubqueryInfo: prototypes of the subqueries of $Q$\\
// MBN: the list of missing border nodes \\
// ${\cal CB}(Q)$ is the tuple of common border nodes of $Q$\\

\textbf{begin} \\
\>   compute $E = \{e \mid e$ is an embedding of $Q_i$ in GjData$\}$\\
\> \textbf{for each} embedding e = (bnv, nbnv) in $E$ \textbf{do}\\
\>  \>   \textbf{begin}\\
\>  \>  \>  \textbf{if} ($MBN != [ ]$) \textbf{then} \\
\>  \>  \>  \> emitToMapper2([($Q_i$,e(${\cal CB}(Q)$)), (bnv, nbnv)]);  \\
\>  \>  \>  \> \textbf{for} k = 1 to $|$bnv$|$ \textbf{do}\\
\>  \>  \>  \> \>       \textbf{if} (bnv[k] != '*') \textbf{then} \\
\>  \>  \>  \>  \> \>   \textbf{foreach} $(n_k, Q_j)$ in $MBN$ \textbf{do} \\
\>  \>  \>  \>  \>  \> \>   emitToMapper2$([(Q_j,e({\cal CB}(Q))), (n_k, bnv[k])])$; \\
\>  \>  \>  \textbf{else} \\
\> \> \> \> emitToReducer2$([bnv, (Q_i, nbnv)])$;\\
\>  \>  \>  \textbf{end} \\
\>  \>  \textbf{end} \\
\textbf{end.}
\end{tabbing}
}

\begin{exmpl} (Continued from Example~\ref{ex:preproc-SO}\label{ex:mapper1}).
This example shows the results obtained by the application of  {\sl \textbf{mapper1}}  on  the pairs $(Q_i, G_j)$, where $Q_i$ is an so-query and $G_j$ is a graph segment.%

\noindent
The following three embeddings of $Q_1$ into $G_1$ are computed by the algorithm:

{\small
\textsf{e1 = ($<$Person1, *, Journal1$>$, $<$Article1, *, Title1, *$>$)}

\textsf{e2 = ($<$Person2, *, Journal1$>$, $<$Article1, *, Title1, *$>$)}

\textsf{e3 = ($<$Person4, *, Journal1$>$, $<$Article1, *, Title1, *$>$)}
}

\noindent
For $e_1$ the algorithm emits the following ($key, value$) pair to {\sl \textbf{mapper2}}:

{\small
\textsf{key = (Q, Person1), value = ($<$Person1, *, Journal1$>$, $<$Article1, *, Title1, *$>$)}
}

\noindent
Besides, based on the MBN list and the ${\cal CB}(Q)$, which, in the preprocessing phase have been computed to  $MBN = [(n_2,Q_1), (n_3,Q_3)]$ and ${\cal CB}(Q) = \{n_1\}$, {\sl \textbf{mapper1}}  also emits to {\sl \textbf{mapper2}}  the following key value pair:

{\small
\textsf{key = (Q3, Person1), value = (n3, Journal1)}
}

\noindent
Similarly, {\sl \textbf{mapper1}} also emits the following (key, value) pais based on the embeddings $e_2$ and $e_3$:

{\small
\textsf{key = (Q1, Person2), value = ($<$Person2, *, Journal1$>$, $<$Article1, *, Title1, *$>$)}

\textsf{key = (Q3, Person2), value = (n3, Journal1)}

\textsf{key = (Q1, Person4), value = ($<$Person4, *, Journal1$>$, $<$Article1, *, Title1, *$>$)}

\textsf{key = (Q3, Person4), value = (n3, Journal1)}
}

\noindent
Concerning query $Q_1$ there are no embeddings in segments $G_2$ and $G_3$. Thus nothing is emitted by the corresponding mappers.

\noindent
The following embedding of $Q_2$ into $G_1$ is computed (among others) by the algorithm:

{\small
\textsf{e'1 = ($<$Person4, Article3, Journal2$>$, $<$*, *, *, ``2008"$>$)}
}

\noindent
For $e'_1$ the algorithm emits the following($key, value$) pair to {\sl \textbf{mapper2}}:

{\small
\textsf{key = (Q2, Person4), value = ($<$Person4, Article3, Journal2$>$, $<$*, *, *, ``2008"$>$)}   
}

\noindent
Besides, based on the MBN list and the ${\cal CB}(Q)$, {\sl \textbf{mapper1}} also emits to {\sl \textbf{mapper2}} the following key value pair:

{\small
\textsf{key = (Q1, Person4), value = (n2, Article3)}

\textsf{key = (Q3, Person4), value = (n3, Journal2)}
}

\noindent
Query $Q_2$ has no embeddings in segment $G_2$; hence nothing is emitted in this case.

\noindent
The following two embedding of $Q_2$ into $G_3$ are computed by the algorithm:

{\small
\textsf{e'2 = ($<$Person2, Article2, Journal1$>$, $<$*, *, *, ``2008"$>$)}

\textsf{e'3 = ($<$Person3, Article2, Journal1$>$, $<$*, *, *, ``2008"$>$)}
}

\noindent
As above, based on these embeddings as well as on the content of the MBL list and the ${\cal CB}(Q)$,
{\sl \textbf{mapper1}} also emits to {\sl \textbf{mapper2}} the following key value pairs:

{\small
\textsf{key = (Q2, Person2), value = ($<$Person2, Article2, Journal1$>$, $<$*, *, *, ``2008"$>$)}

\textsf{key = (Q1, Person2), value = (n2, Article2)}

\textsf{key = (Q3, Person2), value = (n3, Journal1)}

\textsf{key = (Q2, Person3), value = ($<$Person3, Article2, Journal1$>$, $<$*, *, *, ``2008"$>$)}

\textsf{key = (Q1, Person3), value = (n2, Article2)}

\textsf{key = (Q3, Person3), value = (n3, Journal1)}
}

\noindent
The following embedding of $Q_3$ into $G_1$ is computed by the algorithm:

{\small
\textsf{e''1 = ($<$Person4, Article1, *$>$, $<$*, Person1, *, *$>$)}
}

\noindent
For $e''_1$ the algorithm emits the following($key, value$) pair to {\sl \textbf{mapper2}}:

{\small
\textsf{key = (Q3, Person4), value = ($<$Person4, Article1, *$>$, $<$*, Person1, *, *$>$)}
}

\noindent
Besides, based on the MBN list and the ${\cal CB}(Q)$, {\sl \textbf{mapper1}} also emits to {\sl \textbf{mapper2}} the following key value pair:

{\small
\textsf{key = (Q1, Person4), value = (n2, Article1)}
}

\noindent
The following embeddings of $Q_3$ into $G_2$ are computed by the algorithm:

{\small
\textsf{e'''1 = ($<$Person4, Article1, *$>$, $<$*, Person1, *, *$>$)}

\textsf{e'''2 = ($<$Person2, Article2, *$>$, $<$*, Person3, *, *$>$)}
}

\noindent
As above, based on these embeddings as as well as on the content of the MBL list and the ${\cal CB}(Q)$,
{\sl \textbf{mapper1}} also emits to {\sl \textbf{mapper2}} the following key value pairs:

{\small
\textsf{key = (Q3, Person4), value = ($<$Person4, Article1, *$>$, $<$*, Person1, *, *$>$)}

\textsf{key = (Q1, Person4), value = (n2, Article1)}

\textsf{key = (Q3, Person2), value = ($<$Person2, Article2, *$>$, $<$*, Person3, *, *$>$)}

\textsf{key = (Q1, Person2), value = (n2, Article2)}
}

\noindent
Query $Q_3$ has no embeddings in segment $G_3$; hence nothing is emitted by this mapper.
\hfill$\Box$
\end{exmpl}

\subsubsection{Phase 2 of the algorithm}

Phase 2 of the algorithm is similar to the Phase 2 of the
eval-STARS algorithm.


Each \emph{mapper} in Phase 2 gets as input all the embeddings of a specific subquery $Q_i$
which have the same values for the nodes in ${\cal CB}(Q)$; moreover
for each border node that does not occur in $Q_i$ it gets as input the values
assigned to this node by the embeddings of the other subqueries.
It fills in their missing border node values using the corresponding values in the input,
and emits the resulted embeddings to the reducers of Phase 2.
The key is the tuple of the border node values, which implies that
two embeddings are emitted to the same reducer if and only if they are compatible.

{\small \sl
\begin{tabbing}
aa \= aa \= aa \= aa \kill
\textbf{mapper2}(($Q_i$,e(${\cal CB}(Q)$)), values)\\
// $Q_i$: the ID of a subquery\\
// values: a set $E$ of the parts (bnv, nbnv) of the embeddings \\
//  \ \ \ \ \ of $Q_i$,e(${\cal CB}(Q)$) and a set $V$ of pairs $(n_k,v)$, \\
//  \ \ \ \ \ where $v$ is a candidate value for $bnv[k]$\\

\textbf{begin}\\
\>  \textbf{foreach} embedding $e = (bnv, nbnv)$ in $E$ \textbf{do}\\
\> \>     \textbf{foreach} ground instance $bnv'$ of $bnv$ using the values in $V$ \textbf{do}\\
\> \> \>   \textbf{emit}$([bnv', (Q_i, nbnv)])$;\\
\textbf{end.}
\end{tabbing}
}

\begin{exmpl} (Continued from Example~\ref{ex:mapper1}\label{ex:mapper2}).
%
This example shows the application of {\sl \textbf{mapper2}}.
%

\noindent
The mapper applied for the key $(Q_1, Person1)$ gets the  value:

{\small
\textsf{($<$Person1, *, Journal1$>$, $<$Article1, *, Title1, *$>$)}
}

\noindent
but it does not get any value for the missing border nodes.
As $V = \emptyset$ no ground instances of
{\small
\textsf{($<$Person1, *, Journal1$>$)}}
can be found.
Therefore this mapper does not emit (key, value) pairs to {\sl \textbf{reducer2}}.\\
The mapper applied for the key $(Q_1, Person2)$ gets the  value:

{\small
\textsf{($<$Person2, *, Journal1$>$, $<$Article1, *, Title1, *$>$)}

\textsf{(n2, Article2)}
}

\noindent
This mapper constructs the instance {\small
\textsf{$<$Person2, Article2, Journal1$>$}} of $bnv$  and emits the following (key, value) pair to {\sl \textbf{reducer2}}:

{\small
\textsf{key = $<$Person2, Article2, Journal1$>$, value = (Q1, $<$Article1, *, Title1, *$>$)}
}

\noindent
The mapper applied for the key $(Q_1, Person3)$ gets the  values:

{\small
\textsf{(n2, Article2)}
}

\noindent
As $E = \emptyset$ this mapper emits nothing to {\sl \textbf{reducer2}}.\\
The mapper applied for the key $(Q_1, Person4)$ gets (after eliminating duplicates) the  values:

{\small
\textsf{($<$Person4, *, Journal1$>$, $<$Article1, *, Title1, *$>$)}

\textsf{(n2, Article3)}

\textsf{(n2, Article1)}
}

\noindent
This mapper emits the following (key, value) pairs to {\sl \textbf{reducer2}}:

{\small
\textsf{key = $<$Person4, Article3, Journal1$>$, value = (Q1, $<$Article1, *, Title1, *$>$)}

\textsf{key = $<$Person4, Article1, Journal1$>$, value = (Q1, $<$Article1, *, Title1, *$>$)}
}

\noindent
The mapper applied for the key $(Q_2, Person2)$ gets the  value:

{\small
\textsf{($<$Person2, Article2, Journal1$>$, $<$*, *, *, ``2008"$>$)}
}

\noindent
This mapper emits the following (key, value) pair to {\sl \textbf{reducer2}}:

{\small
\textsf{key = $<$Person2, Article2, Journal1$>$, value = (Q2, $<$*, *, *, ``2008"$>$)}
}

\noindent
The mapper applied for the key $(Q_2, Person3)$ gets the  value:

{\small
\textsf{($<$Person3, Article2, Journal1$>$, $<$*, *, *, ``2008"$>$)}
}

\noindent
This mapper emits the following (key, value) pair to {\sl \textbf{reducer2}}:

{\small
\textsf{key = $<$Person3, Article2, Journal1$>$, value = (Q2, $<$*, *, *, ``2008"$>$)}
}

\noindent
The mapper applied for the key $(Q_2, Person4)$ gets the  values:

{\small
\textsf{(Q2, Person4)}

\textsf{($<$Person4, Article3, Journal2$>$, $<$*, *, *, ``2008"$>$)}  
}

\noindent
This mapper emits the following (key, value) pair to {\sl \textbf{reducer2}}:

{\small
\textsf{key = $<$Person4, Article3, Journal2$>$, value = (Q2, $<$*, *, *, ``2008"$>$)}
}

The mapper applied for the key $(Q_3, Person1)$ gets the  value:

{\small
\textsf{(n3, Journal1)}
}

\noindent
As $E = \emptyset$ this mapper emits nothing to {\sl \textbf{reducer2}}.

%
%
%
\noindent
The mapper applied for the key $(Q_3, Person2)$ gets the  value:

{\small
\textsf{($<$Person2, Article2, *$>$, $<$*, Person3, *, *$>$)}

\textsf{(n3, Journal1)}
}

\noindent
This mapper emits the following (key, value) pair to {\sl \textbf{reducer2}}:

{\small
\textsf{key = $<$Person2, Article2, Journal1$>$, value = (Q3, $<$*, Person3, *, *$>$)}
}

\noindent
The mapper applied for the key $(Q_3, Person3)$ gets the  value:

{\small
\textsf{(n3, Journal1)}
}

\noindent
As $E = \emptyset$ this mapper emits nothing to {\sl \textbf{reducer2}}.

\noindent
The mapper applied for the key $(Q_3, Person4)$ gets the  values:

{\small
\textsf{($<$Person4, Article1, *$>$, $<$*, Person1, *, *$>$)}

\textsf{(n3, Journal1)}

\textsf{(n3, Journal2)}
}

\noindent
This mapper emits the following (key, value) pairs to {\sl \textbf{reducer2}}:

{\small
\textsf{key = $<$Person4, Article1, Journal1$>$, value = (Q3, $<$*, Person1, *, *$>$)}

\textsf{key = $<$Person4, Article1, Journal2$>$, value = (Q3, $<$*, Person1, *, *$>$)}
}
\hfill$\Box$
\end{exmpl}

Concerning the \emph{Reducer of Phase 2},
each reducer gets as input embeddings for each subquery that are compatible
as the key for the reducer is a tuple of values for all border nodes of the query $Q$.
The embeddings (one for each subquery $Q_1, \dots, Q_n$) are joined to construct the final answers of $Q$:

{\small \sl
\begin{tabbing}
aa \= aa \= aa \= aa \kill
\textbf{reducer2}(key, values)\\
// key: a tuple of values for the border nodes of Q \\
// values: pairs of the form ($Q_i$, partial embeddings of non-border nodes of $Q_i$)\\

\textbf{begin}\\
\>   \textbf{foreach} join obtained by using one  embedding for each subquery \textbf{do} \\
\>      \>    \>   Emit the result produced by this join;\\
\textbf{end.}
\end{tabbing}
}

\begin{exmpl} (Continued from Example~\ref{ex:mapper2}\label{ex:reducer2}).
The reducer with key:\\
  $<Person2, Article2, Journal1>$ receives the list:

\small{
\textsf{[(Q1, $<$Article1, *, Title1, *$>$), (Q2, $<$*, *, *,``2008"$>$), (Q3, $<$*, Person3, *, *$>$)]}
}

\noindent
As this lists contains  an embedding for each subquery, we join them and obtain the following embedding of the query $Q$:

\small{
\textsf{($<$Person2, Article2, Journal1$>$, $<$Article1, Person3, Title1, ``2008"$>$)}
}

\noindent
This embedding corresponds to the answer:

\small{
\textsf{(?P1, ?A, ?J, ?P2, ?T) = (Person2, Article2, Journal1, Person3, Title1 )}
}

\noindent
Note that, the remaining reducers do not return any answer (they don't receive values
for at least one subquery).
%
%
%
%
%
%
%
%
\hfill$\Box$
\end{exmpl}

\subsubsection{Discussion}

QE-with-Redundancy algorithm has several advantages compared with QEJPE-algorithm and eval-STARS algorithm.
Notice that QE-with-Redundancy algorithm is implemented using one and a half Map-Reduce phases while QEJPE-algorithm
and eval-STARS algorithm are implemented using two Map-Reduce phases. Another advantage of QE-with-Redundancy algorithm is that,
due to the replication of the data triples in the decomposition of the data graph,
and the special form of subqueries in which the user query $Q$ is decomposed, namely subject-object star queries,
all the answers to a subject-object star queries can be obtained from a single data segment.

On the other hand, due to the replication of the data triples in the decomposition of the data graph,
multiple occurrences of the same embedding as well as multiple instances of member of MBL list
may be produced and emitted in Phase 1 of the algorithm.

\section{Experimental results}
\label{sec:experiments}

In this section, we present a set of experiments performed over a cluster of 10 virtual machines, and analyze the outcomes.
Each cluster node has the the following characteristics: Intel(R) Xeon(R) CPU E5-2650 v3 @ 2.30GHz (8 Cores) with 16GB RAM, 60GB HD, Ubuntu 16.04 LTS, 64-bit Operating System.
We used Apache Hadoop v3.1 with HDFS (1 NameNode, 1 Secondary NameMode, 10 DataNodes each one 30GB) and YARN (1 ResourceManager, 10 NodeManagers). The 10 virtual machines were connected through external IP addresses.

To perform the experiments we used four different datasets (D1, D2, D3, D4) in N-Triples format from the Waterloo SPARQL Diversity Test Suite (WatDiv) \cite{watdiv} to evaluate the algorithms proposed in this paper.
The number of triples of each dataset, as well as the scale factors used to generate the datasets, are illustrated in Table~\ref{tbl:datasets}.

\begin{table}[!htb]
	\begin{center}
		\scriptsize
		\begin{tabular}{|c|c|c|c|}
			\hline
			Dataset/Query & Scale Factor & Number of triples & Number of files \\
			\hline
			D1  & 25 & 2,731,510 & 7 \\
			D2  & 50 & 5,486,199 &  13 \\
			D3  & 100 & 10,979,566 &  25 \\
			D4  & 200 & 21,961,070 &  49 \\
			\hline
		\end{tabular}
	\end{center}
	\caption{Description of the Datasets}
	\label{tbl:datasets}
\end{table}

The data graph was partitioned using three approaches, random edge partition, vertex partition, and METIS. In particular, the random edge partition was implemented by randomly adding each edge into a file such that each file had approximately 450,000 triples. We also stored information about the border nodes in each file. The vertex partition essentially describes the implementation of s-decomposition approach defined in Subsection~\ref{subsec:QE-redundancy}.  The last partitioning approach used is METIS \cite{METIS}, in order to minimize the number of border nodes in each file. Note that the random edge and METIS  partitioning approaches were used to evaluate queries using the QEJPE-algorithm and eval-STARS algorithms, while the vertex partition was used to evaluate the QE-with-Redundancy algorithm.

In the implementation of each algorithm, we used the library RDFLib\footnote{RDFLib documentation: \url{https://rdflib.readthedocs.io/}} to pose the subqueries over data segments in each MapReduce task. In particular, in order to find the partial embeddings in each MapReduce task, we parse the data segment and load it into certain structure using RDFLib. Then, we use the query evaluation mechanism of the library to query the loaded segment and find the corresponding partial embeddings. Although the usage of RDFLib facilitates the evaluation of subqueries and provides an efficient evaluation tool in each task, there is an overhead due to loading of each data segment, which is around 40 seconds for each data segment. Note that the loading time does not include the transfer time of each segment as well as the time that takes each task to be initialized.

We conducted several types of experiments to investigate both the performance of the query evaluation algorithms proposed in this paper and the impact of the query decomposition algorithms on the overall query evaluation.  In the following, we initially analyze the scalability of each query evaluation algorithm, in terms of both the size of the dataset and the number of cluster nodes.
Then, we analyze how the query evaluation algorithms perform in widely-used pattern types of queries and different partitions of the data graph (the ones mentioned above). Finally, we analyze how the overall performance of query evaluation is affected by the type of query decomposition selected.

\subsection{Scalability}
In this section, we investigate the scalability of the QEJPE-algorithm, eval-STARS and QE-with-Redundancy algorithms. In particular, we conducted a set of experiments to analyze how the query evaluation algorithms perform in terms of both the size of the input dataset and the  number of compute nodes in the cluster.

\begin{figure}
	\centering
	\begin{subfigure}{0.495\textwidth}
		\centering
		\includegraphics[width=1.0\linewidth]{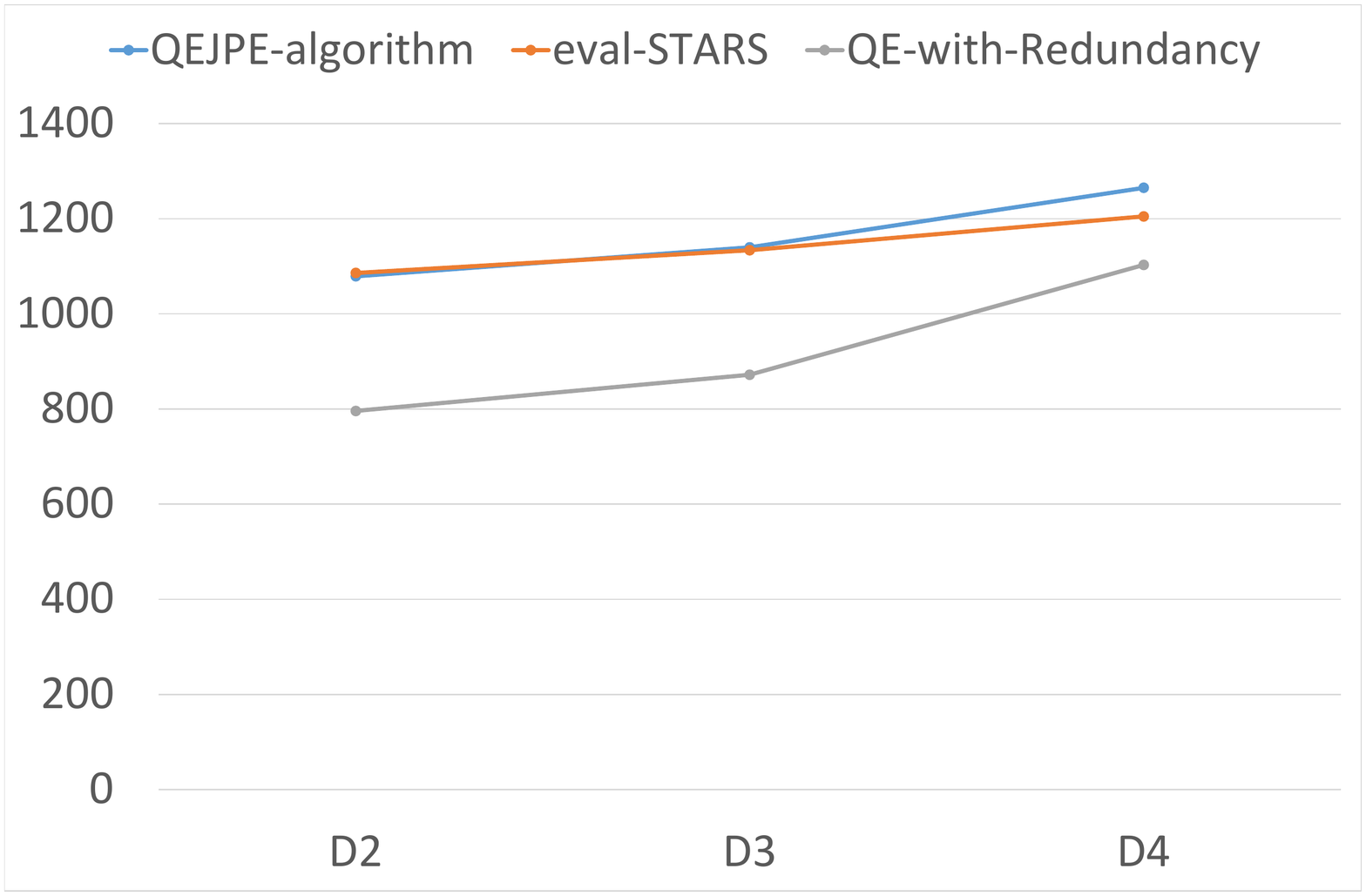}
		\caption{Linear Query}\label{fig:LinearScalability}
	\end{subfigure}%
	\begin{subfigure}{0.495\textwidth}
		\centering
		\includegraphics[width=1.0\linewidth]{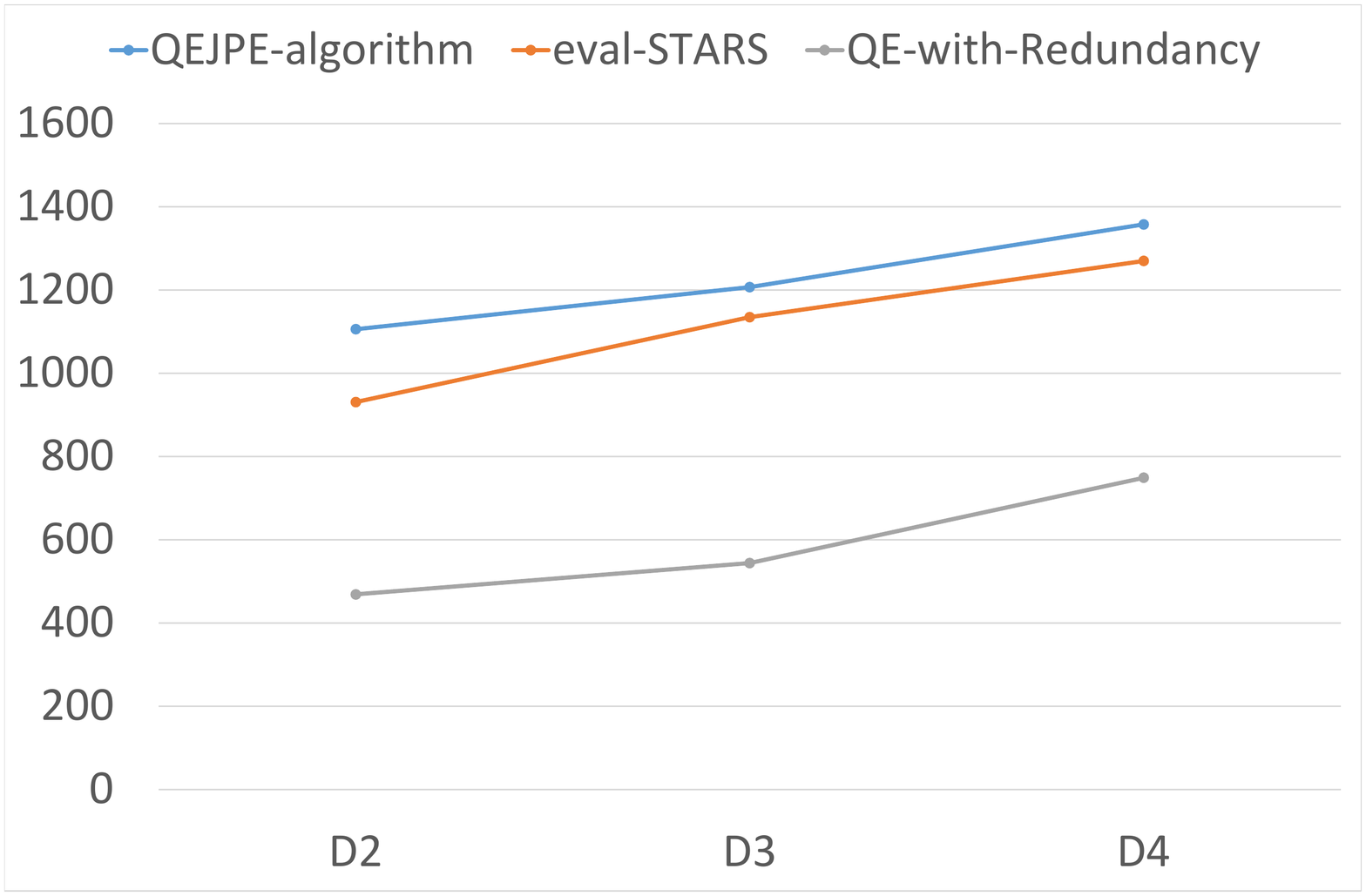}
		\caption{Star Query}\label{fig:StarScalability}
	\end{subfigure}\\
	\begin{subfigure}{0.495\textwidth}
		\centering
		\includegraphics[width=1.0\linewidth]{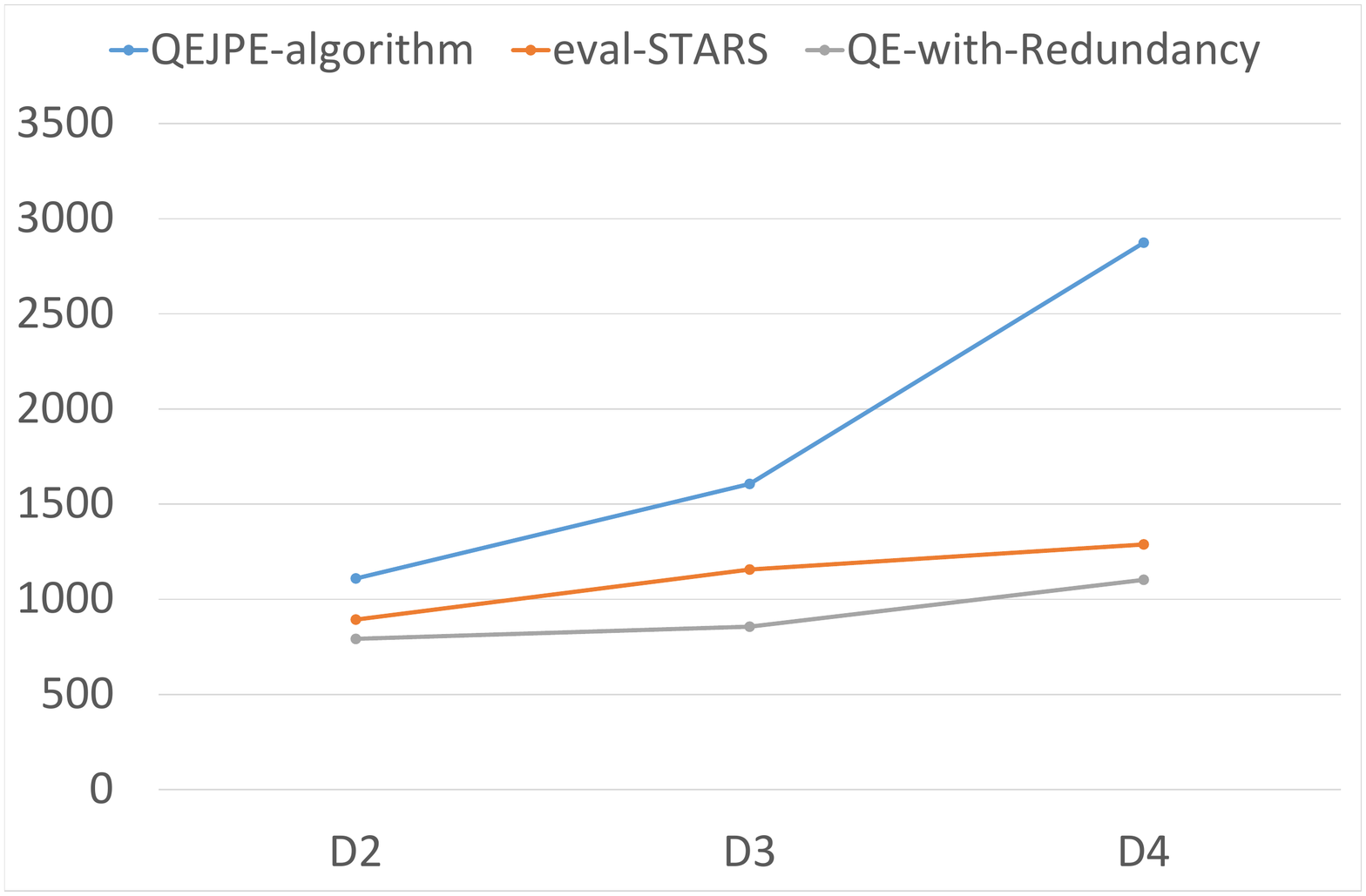}
		\caption{Snowflake Query}\label{fig:SnowFlakeScalability}
	\end{subfigure}
	\begin{subfigure}{0.495\textwidth}
		\centering
		\includegraphics[width=1.0\linewidth]{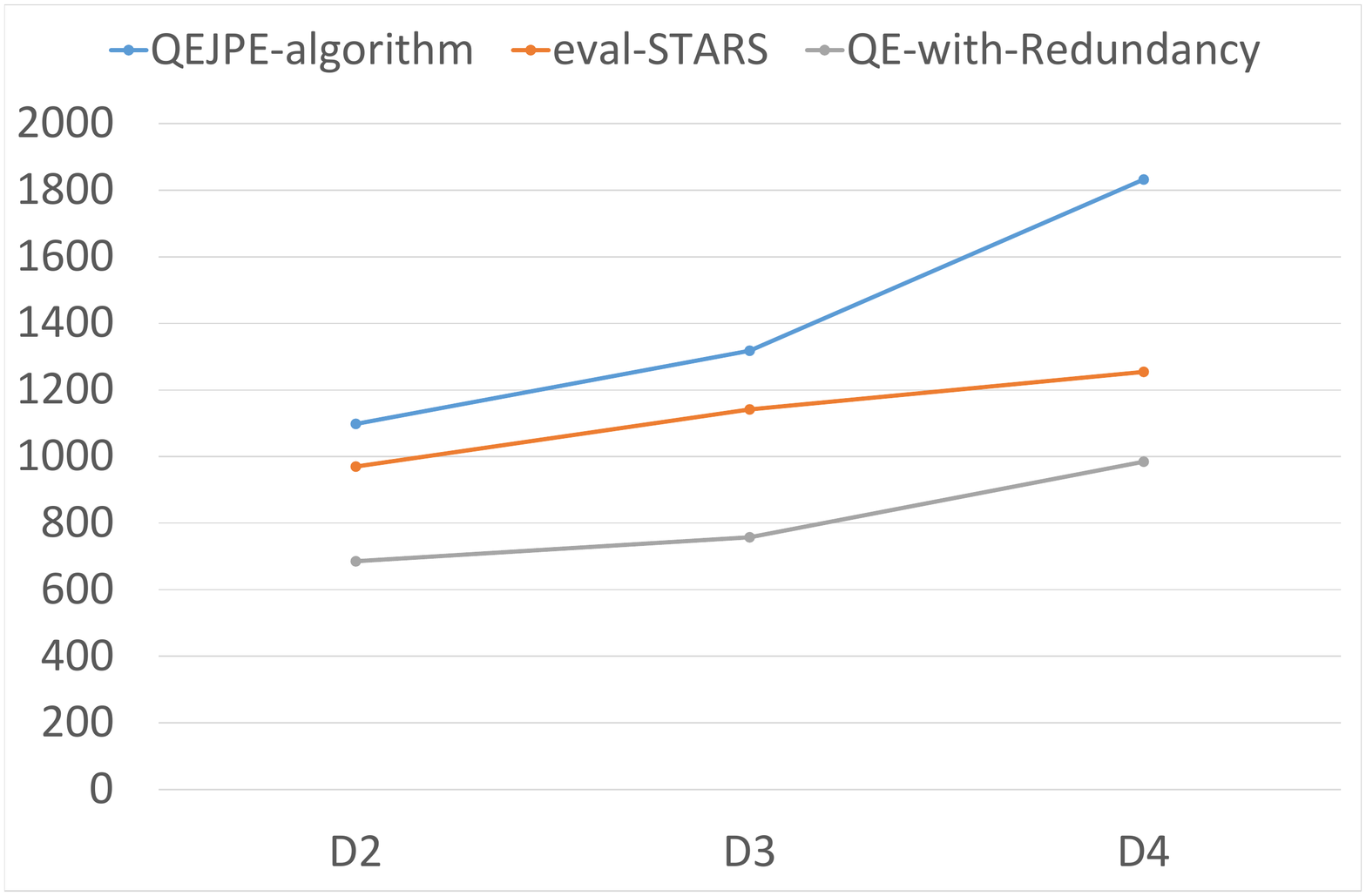}
		\caption{Average per evaluation algorithm}\label{fig:avgDatasetScalability}
	\end{subfigure}
	\caption{Query evaluation in terms of the size of dataset}
\end{figure}

Initially, we selected three queries of different types from the WatDiv Benchmark and evaluated them using each of the algorithms over each of the D2-D4 datasets. The queries selected are illustrated in the Table~\ref{tbl:queries-different-size-datasets}, along with the number of subqueries generated per algorithm. For each query, the type of the query and the number of the resulting tuples for each dataset are included in the table, as well. Table~\ref{tbl:results-different-size-datasets} summarizes the execution time of each query, per evaluation algorithm and dataset, where $L$, $S$ and $F$ represent the Linear, Star and Snowflake queries selected, respectively. Figures~\ref{fig:LinearScalability}, \ref{fig:StarScalability} and \ref{fig:SnowFlakeScalability} graphically show the execution time per dataset and evaluation algorithm, for each query. Figure~\ref{fig:avgDatasetScalability} illustrates the average execution time for each dataset and each algorithm. Looking at the experimental results, we can see that although the amount of the data in each dataset is doubled (i.e., $D3$ and $D4$ have around $100\%$ more triples than $D2$ and $D3$, respectively), the growth rate of the execution time remains less than $30\%$, in average; which shows that each algorithm scales well in terms of the size of the dataset\footnote{Note that the scalability of each algorithm is limited by the capacity of the cluster resources (i.e., memory, disk space).}.

\begin{table}[!htb]
	\begin{center}
		\scriptsize
		\begin{tabular}{|c|c|c|c|c|c|c|}
			\hline
			\multirow{2}{*}{Query type} & \multicolumn{3}{| c |}{Number of subqueries} & \multicolumn{3}{| c |}{Results}
			\\ & QEJPE-algorithm & eval-STARS & QE-with-Redundancy & D2 & D3 & D4 \\
			 Linear (L2) & 2 & 2 & 2 & 36 & 9 & 432 \\
			 Star (S5) & 2 & 1 & 1 & 20 & 33 & 64 \\
			 Snowflake (F1) & 3 & 2 & 2 & 0 & 10 & 4 \\
			\hline
		\end{tabular}
	\end{center}
	\caption{Description of the queries evaluated over datasets of different sizes}
	\label{tbl:queries-different-size-datasets}
\end{table}

\begin{table}[!htb]
	\begin{center}
		\tiny
		\begin{tabular}{|c|c|c|c|c|c|c|c|c|c|}
						\hline
			\multirow{2}{*}{ } & \multicolumn{3}{| c |}{QEJPE-algorithm} &	\multicolumn{3}{| c |}{eval-STARS} & \multicolumn{3}{| c |}{QE-with-Redundancy}\\
			& L	&	S	&	F	& L	&	S	&	F	& L	&	S	&	F \\
						\hline
			D2 & 1079	&	1106	&	1109	&	1086	&	931	&	893	&	796	&	469	&	792 \\
			D3 & 1140	&	1207	&	1606	&	1134	&	1135	&	1156	&	872	&	544	&	856 \\
			D4 & 1265	&	1358	&	2874	&	1205	&	1270	&	1288	&	1103	&	749	&	1102 \\
			\hline
		\end{tabular}
	\end{center}
	\caption{Query evaluation in datasets of different sizes (in seconds)}
	\label{tbl:results-different-size-datasets}
\end{table}

\begin{figure}
	\centering
	\begin{subfigure}{0.495\textwidth}
		\centering
		\includegraphics[width=1.0\linewidth]{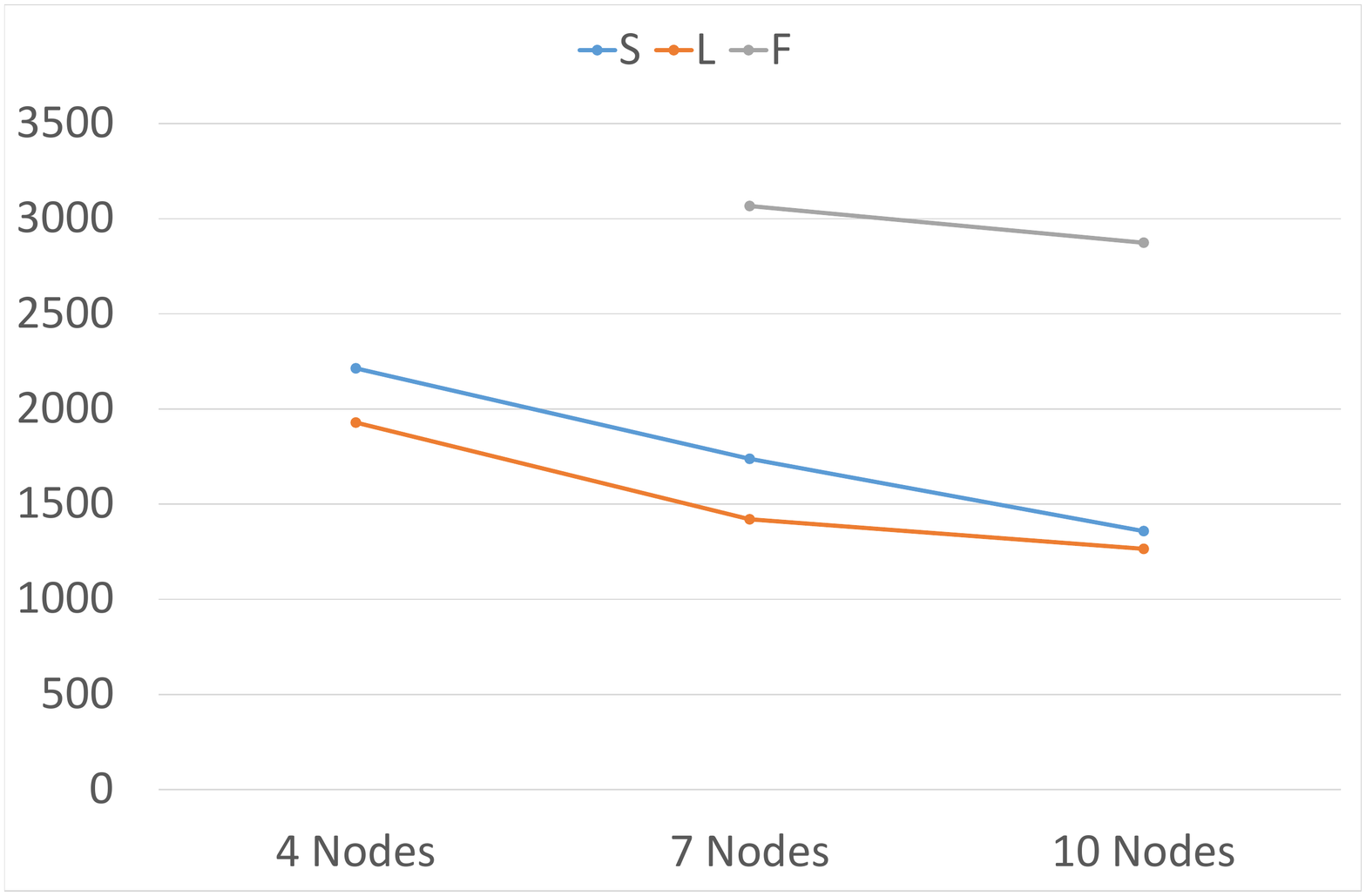}
		\caption{QEJPE-algorithm}\label{fig:NScalabilityQEJPE-algorithm}
	\end{subfigure}%
	\begin{subfigure}{0.495\textwidth}
		\centering
		\includegraphics[width=1.0\linewidth]{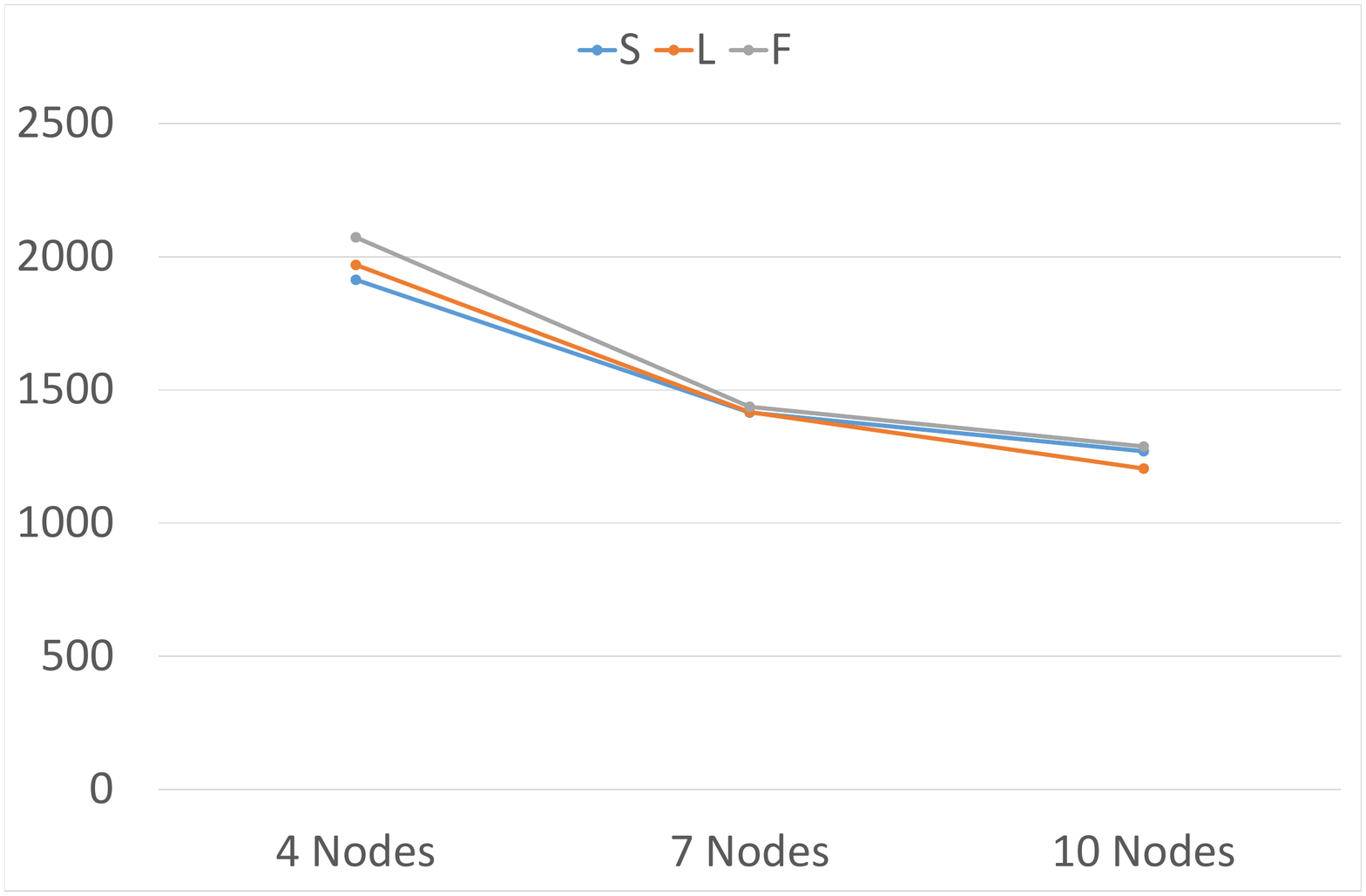}
		\caption{eval-STARS}\label{fig:NScalabilityeval-STARS}
	\end{subfigure}\\
	\begin{subfigure}{0.495\textwidth}
		\centering
		\includegraphics[width=1.0\linewidth]{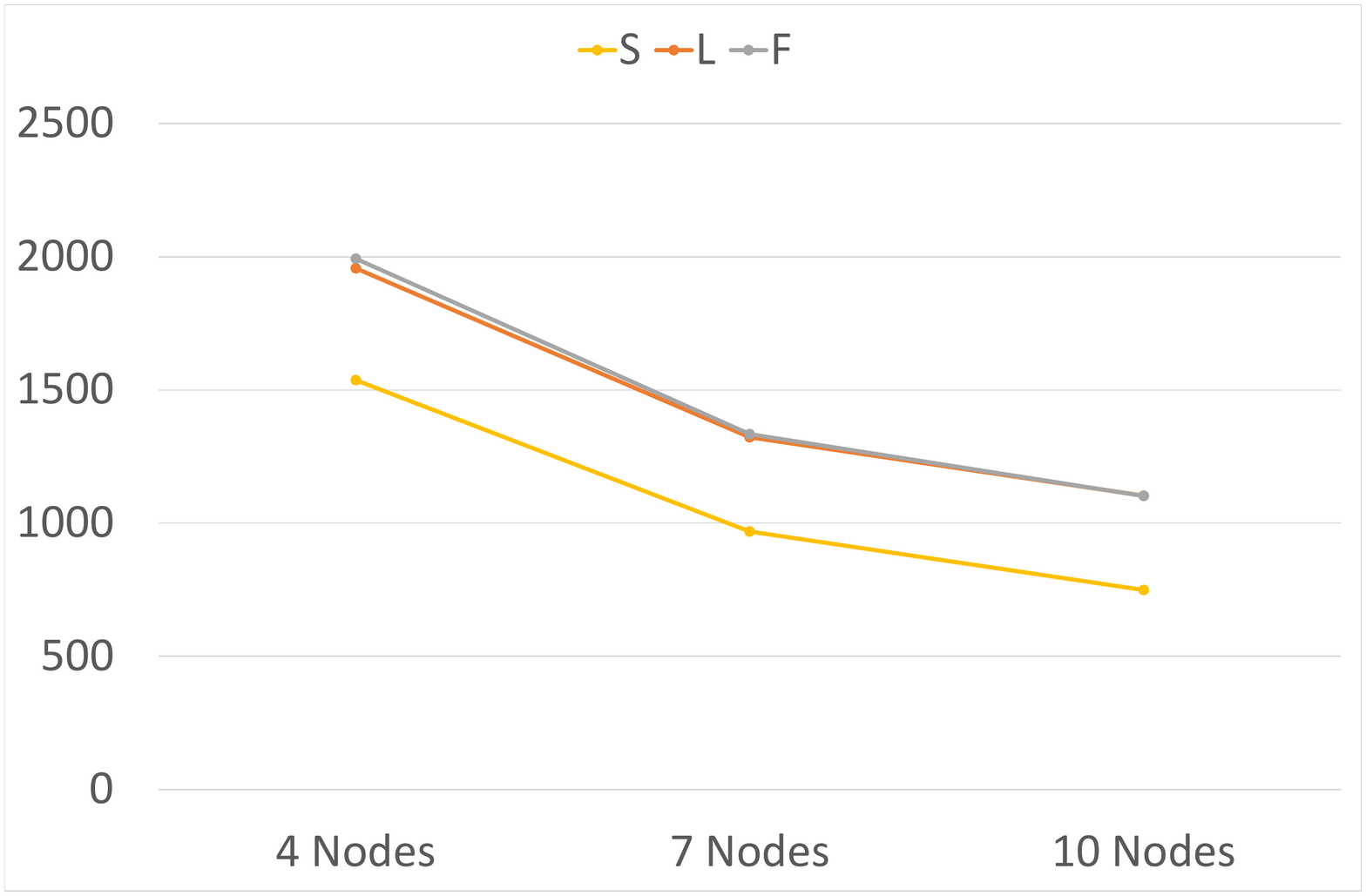}
		\caption{QE-with-Redundancy}\label{fig:NScalabilityQE-with-Redundancy}
	\end{subfigure}
	\caption{Query evaluation in terms of compute nodes size per algorithm}
\end{figure}

To evaluate the scalability in terms of the size of the cluster (i.e., the number of compute nodes), we performed as follows. We evaluated over 3 cluster settings the 3 queries described in Table~\ref{tbl:queries-different-size-datasets} using each algorithm over the dataset D4. In particular, the first setting had 4 compute nodes (NodeManagers), the second had 7 compute nodes and the last one utilized all the 10 available compute nodes. Then, we executed the evaluation algorithms in each setting. The execution times are summarized into the Table~\ref{tbl:results-different-number-nodes}. Figures~\ref{fig:NScalabilityQEJPE-algorithm}, \ref{fig:NScalabilityeval-STARS} and \ref{fig:NScalabilityQE-with-Redundancy}  illustrate the execution time in terms of the size of the cluster per algorithm for each type of query. As we can see, the algorithms scale well in terms of the number of compute nodes; i.e., the execution time is decreasing by increasing the number of compute nodes.

\begin{table}[!htb]
	\begin{center}
		\tiny
		\begin{tabular}{|c|c|c|c|c|c|c|c|c|c|}
						\hline
			\multirow{2}{*}{ } & \multicolumn{3}{| c |}{QEJPE-algorithm} &	\multicolumn{3}{| c |}{eval-STARS} & \multicolumn{3}{| c |}{QE-with-Redundancy}\\
			& L	&	S	&	F	& L	&	S	&	F	& L	&	S	&	F \\
						\hline
			4 Nodes & 1929	&	2214	&	-	&	1970	&	1914	&	2073	&	1957	&	1537	&	1993 \\
			7 Nodes & 1421	&	1738	&	3967	&	1417	&	1415	&	1437	&	1323	&	969	&	1334 \\
			10 Nodes & 1265	&	1358	&	2874	&	1205	&	1270	&	1288	&	1103	&	749	&	1102 \\
			\hline
		\end{tabular}
	\end{center}
	\caption{Query evaluation in D4 dataset for different compute nodes size (in seconds)}
	\label{tbl:results-different-number-nodes}
\end{table}

\subsection{Comparison of query evaluation algorithms}				

In this section, we present the outcomes of the experiments performed in order to compare the performance of the query evaluation algorithms proposed in this work; i.e., the QEJPE-algorithm, eval-STARS and QE-with-Redundancy algorithms. The evaluation performed by applying all the algorithms for a variety of queries over the dataset D4 described in the previous sections. We used multiple queries from  WatDiv Benchmark, from all the proposed query types (Linear, Star, Snowflake, and Complex). Table~\ref{tbl:watdiv-queries} summarizes the queries used in this experiment, along with corresponding characteristics of each query; e.g., number of triples, number of variables, number of resulting tuples over the dataset D4, and the number of subqueries generated by the query decomposition.

\begin{figure}
	\centering
	\begin{subfigure}{0.495\textwidth}
	    \centering
		\includegraphics[width=1.0\textwidth]{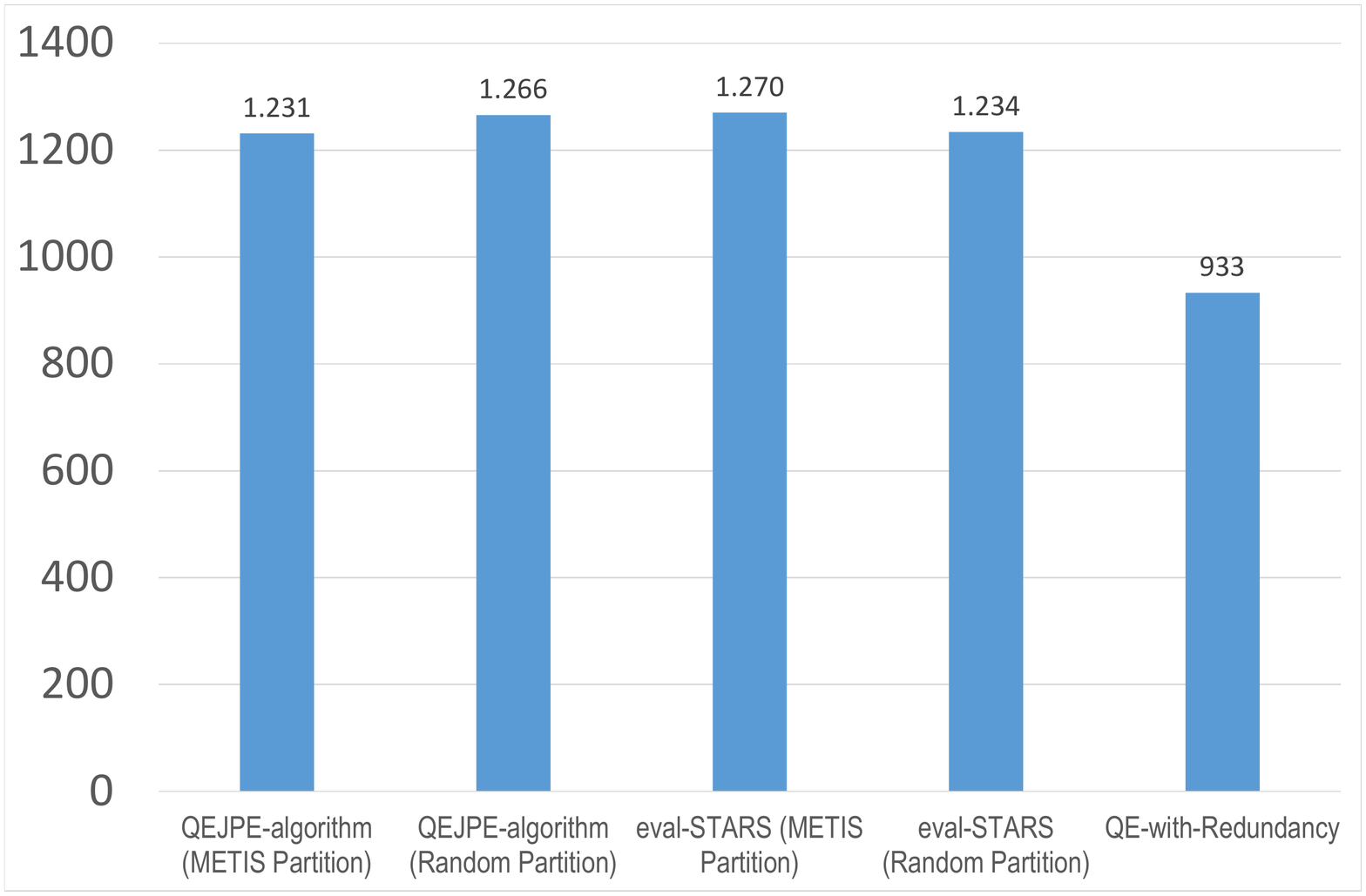}
		\caption{Linear Query Type Evaluation}\label{fig:LinearType}
	\end{subfigure}%
	\begin{subfigure}{0.495\textwidth}
		\centering
		\includegraphics[width=1.0\textwidth]{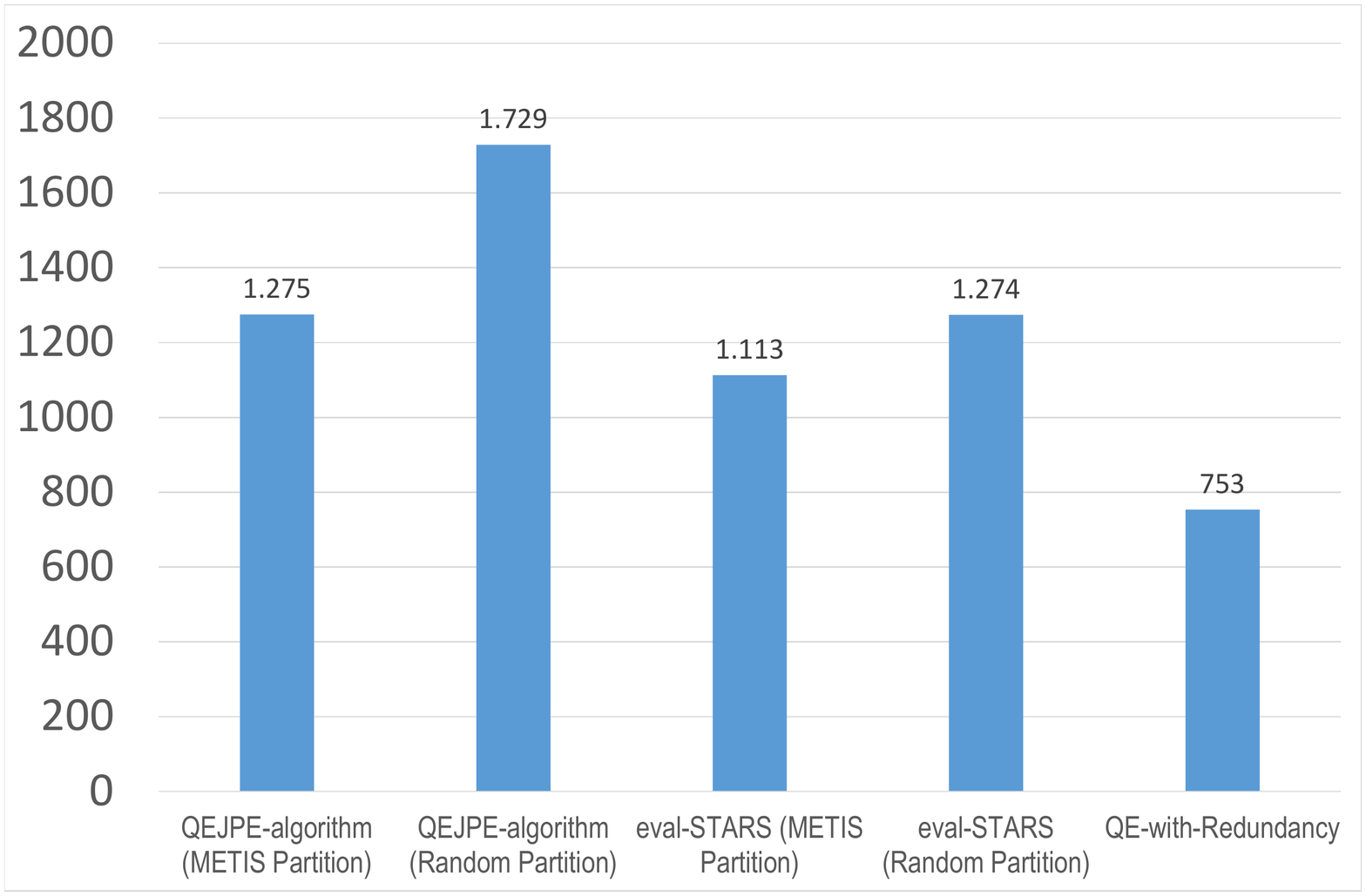}
		\caption{Star Query Type Evaluation}\label{fig:StarType}
	\end{subfigure}\\
	\begin{subfigure}{0.495\textwidth}
		\centering
		\includegraphics[width=1.0\textwidth]{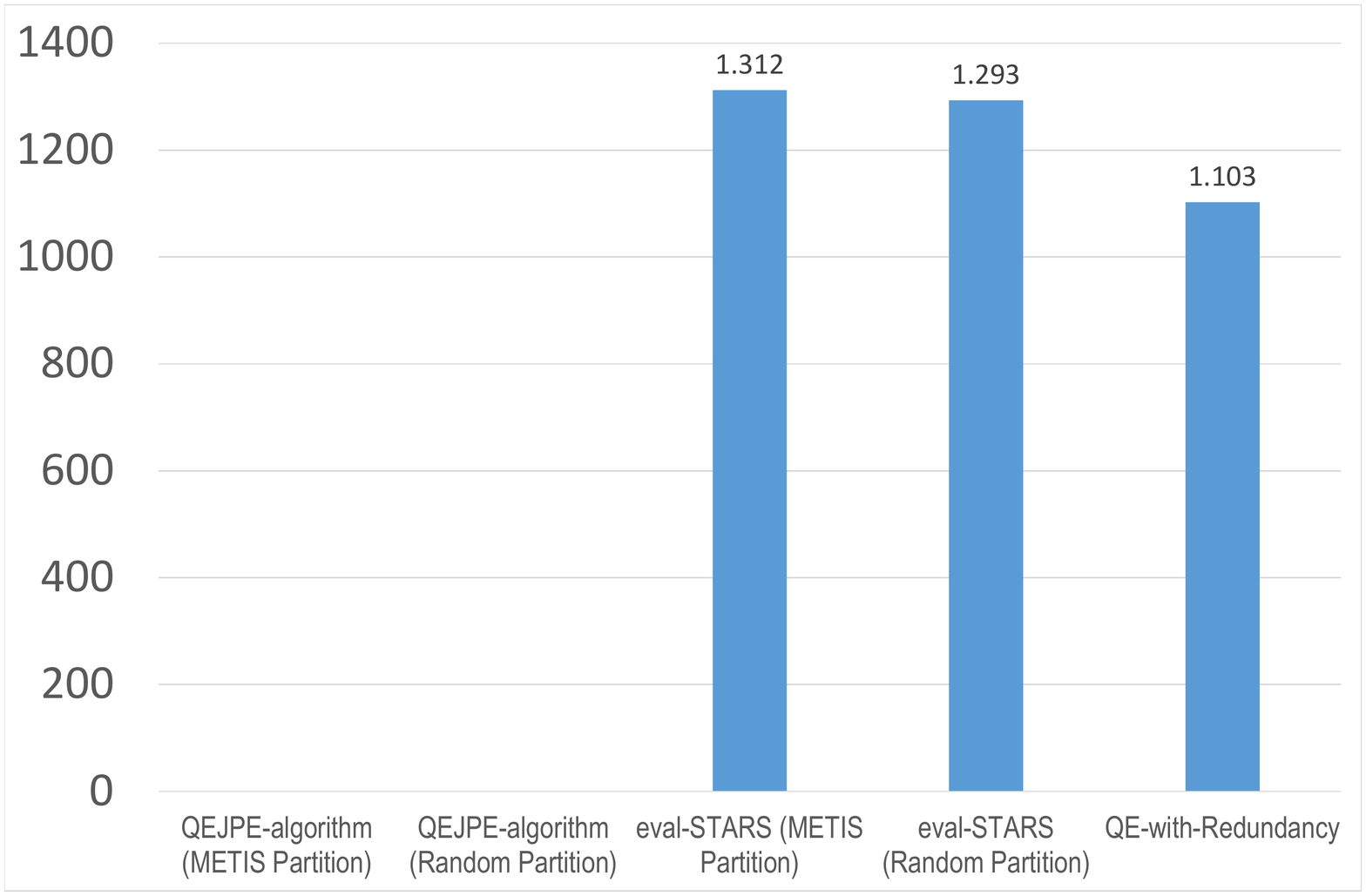}
		\caption{Snowflake Query Type Evaluation}\label{fig:SnowFlakeType}
	\end{subfigure}
	\begin{subfigure}{0.495\textwidth}
		\centering
		\includegraphics[width=1.0\textwidth]{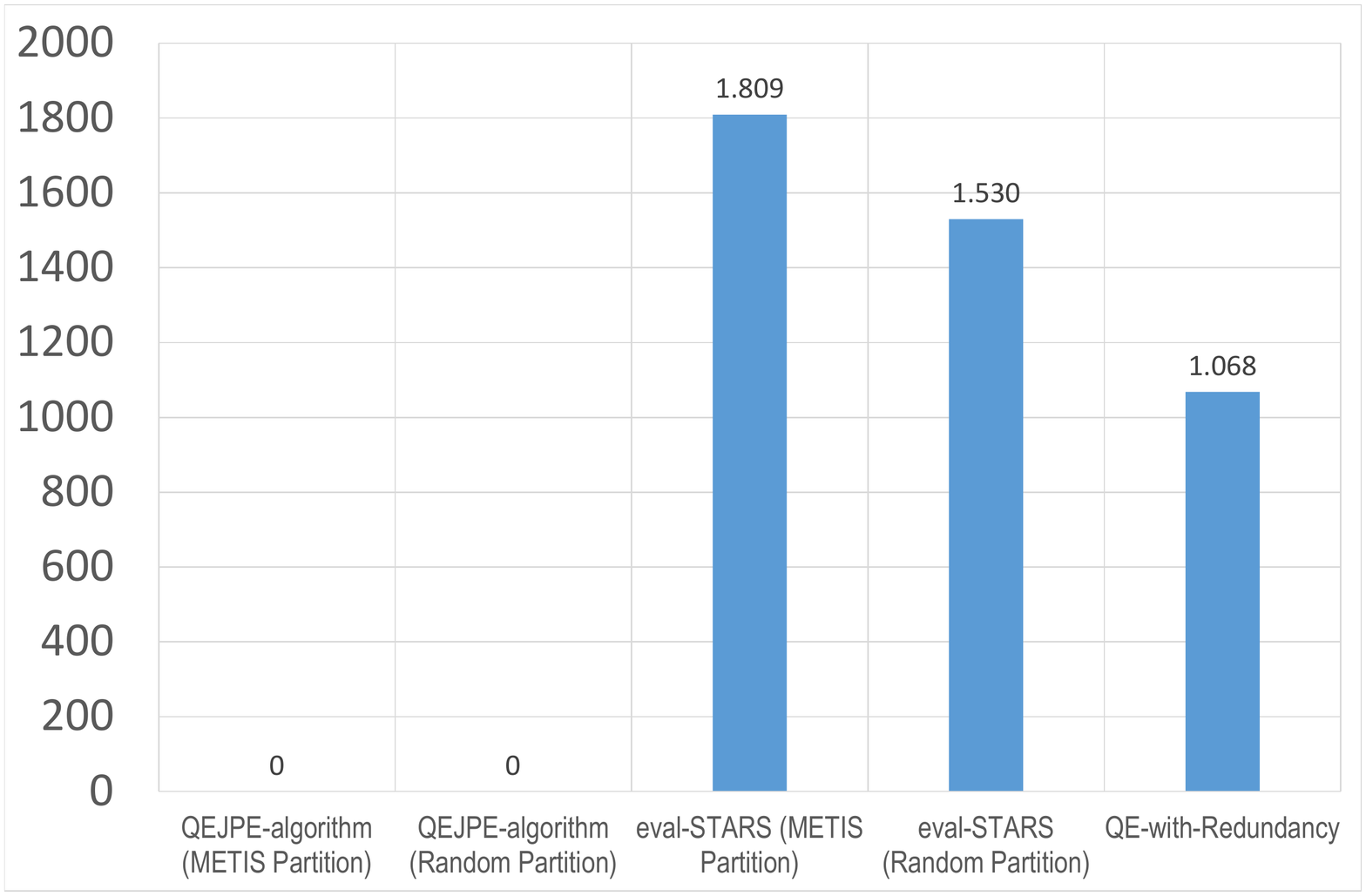}
		\caption{Complex Query Type Evaluation}\label{fig:ComplexType}
	\end{subfigure}
	\caption{Comparison of query evaluation algorithms for a variety of query types}
	\label{fig:comparison-evaluation-algs}
\end{figure}

The QEJPE-algorithm and eval-STARS algorithms were also tested over two data partitioning approaches, random partitioning and METIS, while the QE-with-Redundancy algorithm was only evaluated over vertex-partitioned data (due to the requirements of the algorithm). The execution time for each query is included in the Table~\ref{table:QueryEvaluationperquery}, where the execution time is given in minutes, followed by seconds (i.e., Minutes:Seconds). The average execution time for each algorithm and each query type, per data partitioning approach, is illustrated in Table~\ref{table:QueryEvaluationExperiments} and graphically presented in Figure~\ref{fig:comparison-evaluation-algs}. Note that evaluating the majority of Snowflake and Complex queries using QEJPE-algorithm the cluster reached the memory limits (14.5GB for all YARN containers on a node) and did not manage to provide any result.

\begin{table}[!htb]
	\begin{center}
		\tiny
		\begin{tabular}{|l|c|c|c|c|c|c|c|}
			\hline
			\multirow{2}{*}{Parameters} & \multicolumn{2}{| c |}{Linear} &	\multicolumn{2}{| c |}{Star} & \multicolumn{2}{| c |}{Snowflake} & Complex\\
				&	L2	& L4	&	S3	& S5 & F1 & F4 & C3\\
			\hline
			Triples 	&	4	& 3	&	4	& 4 & 6 & 9 & 6 \\
			Variables 	&	2	& 2	&	4	& 3 & 5 & 8 & 7\\
			Results 	&	432	& 109	&	677	& 64 & 4 & 71 & 763924\\
			Subqueries QEJPE-algorithm 	&	2 & 1	&	2	& 2 & 3 & 5 & 3 \\
			Subqueries eval-STARS and QE-with-Redundancy 	&	2 & 1	&	1	& 1 & 2 & 2 & 1 \\
			Subqueries QE-with-Redundancy 	&	2 & 1	&	1	& 1 & 2 & 2 & 1 \\
			\hline
		\end{tabular}
	\end{center}
	\caption{Description of Watdiv queries}
	\label{tbl:watdiv-queries}
\end{table}

\begin{table}[!htb]
	\tiny
	\begin{center}
		\begin{tabular}{|c|c|c|c|c|c|}
			\hline
			\multirow{2}{*}{Query}	&	\multicolumn{2}{| c |}{QEJPE}	& \multicolumn{2}{| c |}{eval-STARS}	&		\multirow{2}{*}{QE-with-Redundancy}	\\

			&	METIS	&	Random	&	METIS	&	Random	&		\\

			\hline

			L2	&	21:06	&	21:05	&	21:11	&	20:05	&	18:23	\\

			L4	&	19:56	&	21:06	&	21:09	&	21:03	&	12:43	\\

			S3	&	21:47	&	34:59	&	18:51	&	21:18	&	12:37	\\

			S5	&	20:43	&	22:38	&	18:14	&	21:10	&	12:29	\\

			F1	&	23:29	&	47:54	&	22:22	&	21:28	&	18:22	\\

			F4	&	-	&	-	&	21:22	&	21:38	&	18:23	\\

			C3	&	-	&	-	&	30:09	&	25:30	&	17:48	\\
			\hline
		\end{tabular}
		\caption{Execution time per query}\label{table:QueryEvaluationperquery}
	\end{center}
\end{table}

\begin{table}[!htb]
\scriptsize
\begin{center}
	\begin{tabular}{|c|c|c|c|c|c|}
		\hline
		Query Type & Data Partitioning & Linear & Star & Snowflake & Complex
		\\
		\hline
		\multirow{2}{*}{QEJPE-algorithm} & METIS & 1231,0 & 1275,0 & - & -
		\\
		& Random & 1265,5 & 1728,5 & - & -
		\\
		\hline
		\multirow{2}{*}{eval-STARS} & METIS & 1270,0 & 1112,5 & 1312,0 & 1809,0
		\\
		& Random & 1234,0 & 1274,0 & 1293,0 & 1530,0
		\\
		\hline
		QE-with-Redundancy & s-decomposition & 933,0 & 753,0 & 1102,5 & 1068,0
		\\
		\hline
	\end{tabular}
	\caption{Average execution time per query type (seconds)}\label{table:QueryEvaluationExperiments}
\end{center}
\end{table}

As we can see in the experimental results, QE\label{key}JPE is more efficient for Linear queries than Star queries. In addition, METIS outperforms Random partition for both Star and Linear queries.

Queries L2 and L4 from WatDiv Benchmark (Linear query type) evaluated in all the algorithms and the mean execution times in seconds of these queries are presented in Figure~\ref{fig:LinearType} and in Table~\ref{table:QueryEvaluationExperiments}. QE-with-Redundancy algorithm performs better than QEJPE-algorithm and eval-STARS algorithms. eval-STARS algorithm perform better than QEJPE-algorithm using both METIS and Random partition.

Queries S3 and S5 were used to evaluate star type queries. QE-with-Redundancy algorithm performs better than QEJPE-algorithm and eval-STARS algorithms while eval-STARS perform better than QEJPE-algorithm. Both eval-STARS and QEJPE-algorithm perform better for METIS partition than random partition.

In case of Snowflake queries, queries F1 and F4 executed. Experimental results prove that QEJPE-algorithm is not efficient for this type of queries. QE-with-Redundancy algorithm performs again better results from eval-STARS algorithm.  eval-STARS algorithm performed almost the same results for random and METIS partition.

Similar behavior with Snowflake queries had the Complex type queries. C3 query executed and QEJPE-algorithm was not efficient, QE-with-Redundancy algorithm performs better results than eval-STARS algorithm. In this type of query, eval-STARS algorithm performed better using random partition rather than METIS partition.

\subsection{Query Decomposition Algorithms Evaluation}

In this section, we experimentally analyze how the selection of the query decomposition algorithm can affect the overall query evaluation performance. We focus on the three main query decomposition algorithms proposed in Subsection~\ref{subsec:query-decomp-algorithms}; i.e., min-res, max-degree, and max-degree-with-reshaping. To perform this experiment, we decomposed multiple queries using the aforementioned decomposition algorithms and evaluate them using a single evaluation algorithm and over a single dataset.

\begin{figure}[!htb]
	\centering
	\includegraphics[width=0.95\textwidth]{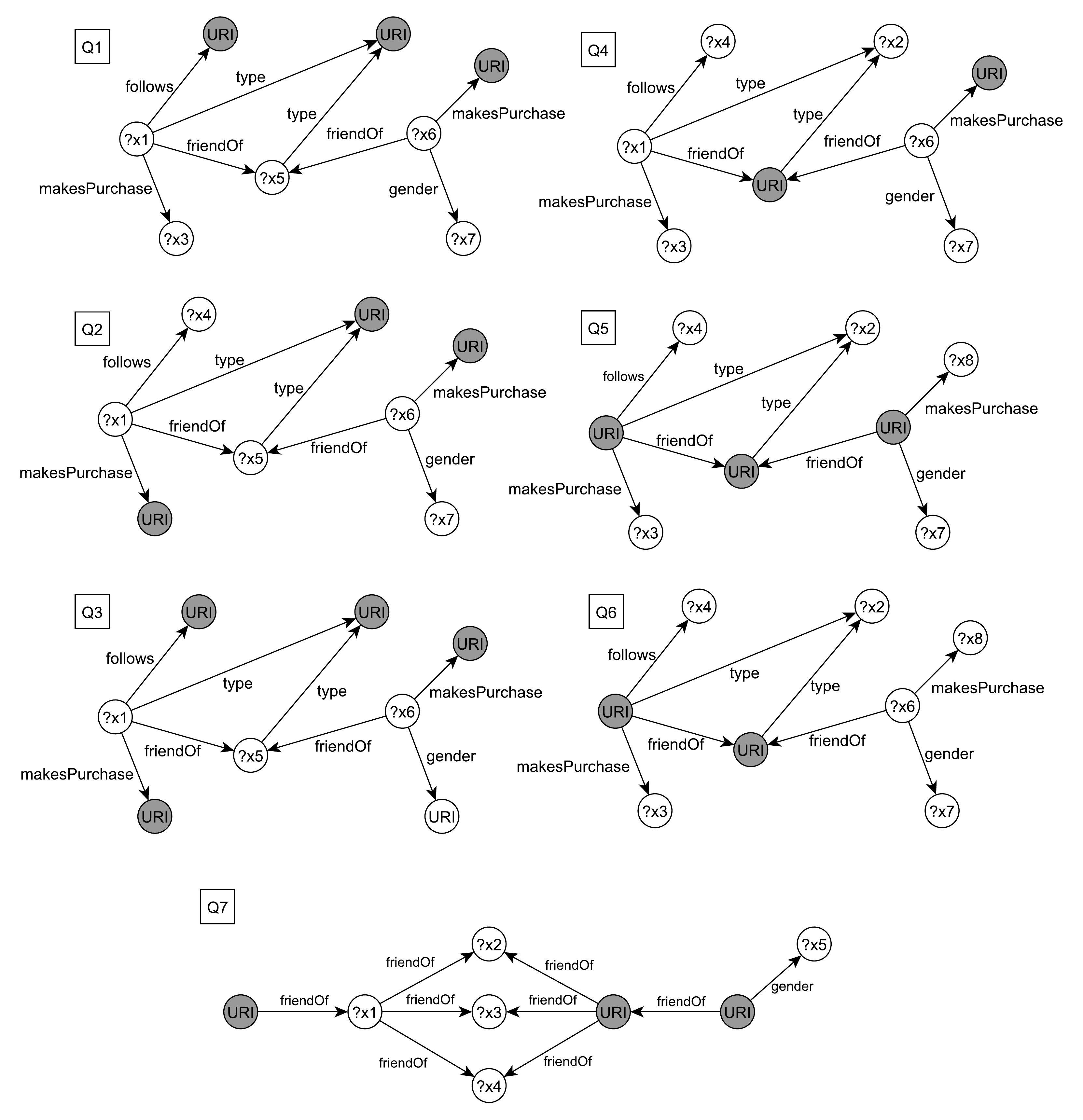}
	\caption{Queries used to compare decomposition algorithms}\label{fig:Q1Q7Decomp}
\end{figure}

\begin{figure}[!htb]
	\centering
	\includegraphics[width=0.80\textwidth]{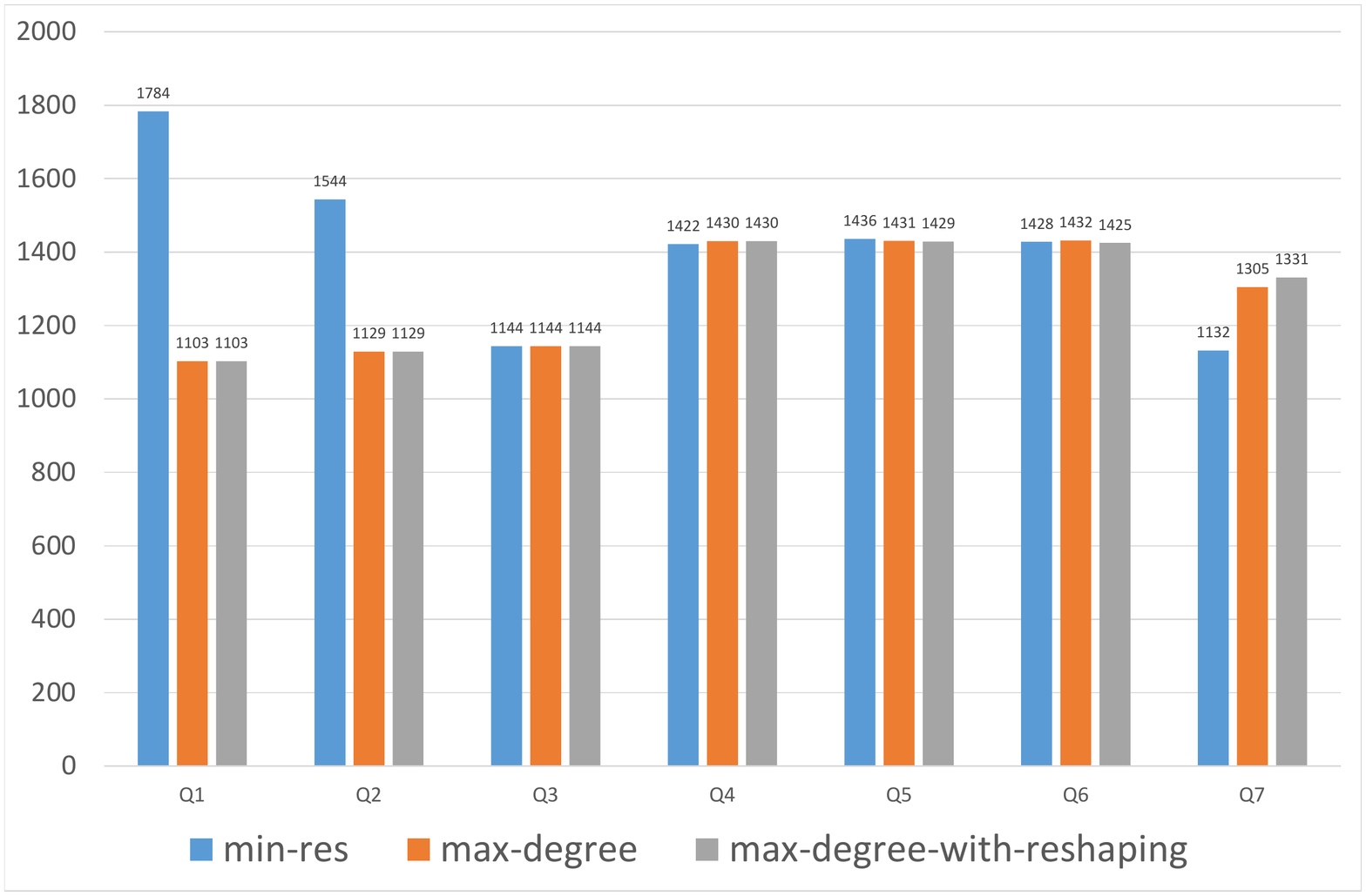}
	\caption{Query Decomposition Algorithms Evaluation}\label{fig:QueryDecompExecutionTime}
\end{figure}

In particular, we initially used a query template (i.e., query graph structure) over the Watdiv data model and generated six different queries by setting variables and constants to the nodes. The queries Q1-Q6 that were constructed are depicted in Figure~\ref{fig:Q1Q7Decomp}, where the white-colored nodes represent the variables and dark-colored nodes represent constants. We also constructed an additional complex query Q7, over the Watdiv data model, asking for certain edges of the data graph multiple times. We then decomposed the queries Q1-Q6 using different decomposition algorithms and evaluated them using the QE-with-Redundancy algorithm and the dataset D4. For Q7 query, the smaller dataset D1 was used to overcome memory limitation due to the large number of results.  The execution time for each query and each decomposition algorithm is illustrated in Table~\ref{table:Querydecompositionresults}, along with the number of subqueries resulted by each decomposition algorithm and the number of resulting tuples. The execution time per query and algorithm is graphically presented in Figure~\ref{fig:QueryDecompExecutionTime}.

\begin{table}[!htb]
	\scriptsize
	\begin{center}
		\begin{tabular}{|c|c|c|c|c|c|c|c|}
			\hline
			\multirow{2}{*}{Queries}	&	\multirow{2}{*}{Results}	&	\multicolumn{2}{| c |}{min-res}		&		\multicolumn{2}{| c |}{max-degree}		&		\multicolumn{2}{| c |}{max-degree-reshaping}\\
				&		&	subqueries	&	time	&	subqueries	&	time	&	subqueries	&	time\\
			\hline
			Q1	&	11	&	5	&	1784	&	3	&	1103	&	3	&	1103	\\
			Q2	&	33	&	5	&	1544	&	3	&	1129	&	3	&	1129	\\
			Q3	&	1	&	3	&	1144	&	3	&	1144	&	3	&	1144	\\
			Q4	&	1580	&	5	&	1422	&	3	&	1430	&	3	&	1430	\\
			Q5	&	5808	&	6	&	1436	&	3	&	1431	&	3	&	1429	\\
			Q6	&	12705	&	6	&	1428	&	3	&	1432	&	3	&	1425	\\
			Q7	&	438976	&	7	&	1132	&	3	&	1305	&	3	&	1331	\\
			\hline
		\end{tabular}
		\caption{Evaluation using QE-with-Redundancy and different query decompositions (Seconds)}\label{table:Querydecompositionresults}
	\end{center}
\end{table}		
		
Analyzing the execution time of the queries per decomposition algorithm (Table~\ref{table:Querydecompositionresults} and Figure~\ref{fig:QueryDecompExecutionTime}), we can easily see that for queries Q1 and Q2, the decompositions of the max-degree and max-degree-with-reshaping perform better than the ones given by min-res. Queries Q4-Q6 perform similarly, for all the three algorithms. This can be explained by the fact that the queries Q1 and Q2 give decompositions with more subqueries using min-res than the decompositions given by the two other algorithms, as well as the results are few. For queries Q4-Q6 the performance of min-res algorithm is improved compared with the max-degree and max-degree-with-reshaping algorithms. Query Q3 is an exception since the decomposition resulted by all the three algorithms is the same.

Comparing now the execution time of the max-degree and the max-degree-with-reshaping, these algorithms resulted similar decompositions. Hence, as we can see  in Table~\ref{table:Querydecompositionresults} and Figure~\ref{fig:QueryDecompExecutionTime}, their execution time for the majority of the queries is very close.

Looking however the execution time of the query Q7, the decomposition resulted by min-res outperforms the decompositions given by max-degree and max-degree-with-reshaping. To analyze this result in more detail, we can easily see that since the node $?x1$ is high degree, gives a subquery with multiple variables in the decompositions given by max-degree and max-degree-with-reshaping. In addition, the variables $?x2$, $?x3$, and $?x4$ in both subqueries map to the same data nodes and increase significantly the number of intermediate results (comparing with the number of the corresponding data edges mapped by these variables in all the embeddings). On the other hand, min-res handles such a case better, since it does not allow subqueries having more than 2 variables to be generated.


\section{Conclusions}
\label{sec:conclusion}

In this paper, we presented a set of distributed query evaluation algorithms that are independent of the storage and data distribution approaches. These algorithms could also be implemented in various distributed processing frameworks. We also presented a set of query decomposition approaches and analysed their advantages and disadvantages. Evaluating the proposed algorithms, we showed that each problem instance (data and query graph) might benefit from different decomposition algorithm and/or evaluation approach.		
		
		As  future  work,  we  aim  to  investigate the proper methods for storing data in order to further improve our algorithms. Investigation of the usage of certain NoSQL databases with the appropriate indices is also considered, in order to optimize the  query plans used to combine the results of the generalized star subqueries in the last approach presented.  Furthermore,  we  aim  to  analyze  additional  query  decomposition  approaches,  focusing  on  finding  an  optimal  query  decomposition  for every different setting.  An additional topic for  further investigation is how our approach could be extended to support query evaluation over dynamic RDF data. Finally, improvements of our algorithms using in-memory processing frameworks, such as Apache Spark and Flink, are also considered for further investigation.

%




\section*{Bibliography}
\bibliographystyle{plain}
\bibliography{RelatedWorkRefs}

\end{document}